\documentclass[12pt,twoside]{article}
\usepackage[dvips]{graphicx}
\usepackage{float}
\usepackage{latexsym}
\usepackage{amsmath,mathrsfs,bm,amssymb,color}

\newtheorem{theorem}{Theorem}[section]
\newtheorem{proposition}[theorem]{Proposition}
\newtheorem{lemma}[theorem]{Lemma}
\newtheorem{corollary}[theorem]{Corollary}
\newtheorem{definition}[theorem]{Definition}
\newtheorem{example}[theorem]{Example}
\newtheorem{remark}[theorem]{Remark}

\newtheorem{assumption}[theorem]{Assumption}

\newcommand{\bd}[1]{\begin{definition}\label{#1}}
\newcommand{\ed}{\end{definition}}
\newcommand{\bl}[1]{\begin{lemma}\label{#1}}
\newcommand{\el}{\end{lemma}}
\newcommand{\bc}[1]{\begin{corollary}\label{#1}}
\newcommand{\ec}{\end{corollary}}
\newcommand{\bt}[1]{\begin{theorem}\label{#1}}
\newcommand{\et}{\end{theorem}}
\newcommand{\bp}[1]{\begin{proposition}\label{#1}}
\newcommand{\ep}{\end{proposition}}
\newcommand{\br}[1]{\begin{remark}\label{#1}}
\newcommand{\er}{\end{remark}}
\newcommand{\eq}[1]{\begin{equation}\label{#1}}
\newcommand{\en}{\end   {equation}}

\newcommand{\ee}{{\rm e}}

\newcommand{\eqn}{\begin{eqnarray*}}
\newcommand{\enn}{\end{eqnarray*}}
\newcommand{\eqnn}{\begin{eqnarray}}
\newcommand{\ennn}{\end{eqnarray}}
\newcommand{\bi}{\begin{description}}
\newcommand{\ei}{\end{description} }

\newcommand{\TTT}[1]{{\bf (#1)}}

\newcommand{\ov}[1]{\overline{#1}}
\newcommand{\f}{^{-1}}
\newcommand{\lk}{\left(\!}
\newcommand{\rk}{\!\right)}
\newcommand{\lkk}{\left\{\!}
\newcommand{\rkk}{\!\right\}}
\newcommand{\lkkk}{\left[\!}
\newcommand{\rkkk}{\!\right]}

\renewcommand{\d}{\displaystyle}
\newcommand{\add}{a^{\dagger}}
\newcommand{\ass}{a^\sharp}
\newcommand{\D}{{\rm D}}

\newcommand{\proof}{{\noindent {\sc Proof}: \,}}
\newcommand{\qed}{\hfill $\Box$ \par\medskip}
\newcommand{\BR}{{{\mathbb  R}^d }}
\newcommand{\BRR}{{{\mathbb  R}^{d+1} }}
\newcommand{\CC}{{{\mathbb  C}}}
\newcommand{\RR}{{\mathbb  R}}
\newcommand{\CCC}{C_0^\infty(\BR)}
\newcommand{\ms}{\mathscr}
\newcommand{\calb}{\ms B}

\newcommand{\limn}{\lim_{n\rightarrow\infty}}
\newcommand{\limm}{\lim_{m\rightarrow\infty}}
\newcommand{\limt}{\lim_{t\rightarrow\infty}}

\newcommand{\wick}[1]{{:\!\! #1 \!\!:}}
\newcommand{\kak}[1]{(\ref{#1})}
\newcommand{\LRZ}{{\cal W}}
\newcommand{\LR}{{L^2(\BR)}}
\newcommand{\hp}{{\rm H}_{\rm p}}
\newcommand{\Ebb}{\mathbb E}
\newcommand{\slim}{{\rm s}\!\!-\!\!\lim}
\newcommand{\slimn}{\slim_{n\to\infty}}
\newcommand{\LRR}{{L^2(\RR^{d+1})}}
\newcommand{\fff}{{\mathscr{F}}}
\newcommand{\ffff}{{\mathscr{F}_{\rm fin}}}
\newcommand{\q}{{\rm q_{\rm M}}}
\newcommand{\qq}{{\rm q_{\rm E}}}
\newcommand{\A}{{\rm  A}}
\renewcommand{\AA}{{\rm  A}_{\rm E}}
\newcommand{\rh}{{\rm h}}
\newcommand{\dG}{{\rm d}\Gamma}
\newcommand{\vk}{\ms V_{\rm Kato}}
\newcommand{\vqf}{\ms V_{\rm qf}}
\newcommand{\vsa}{\ms V_{\rm rel}}

\newcommand{\W}{{\ms M}}
\newcommand{\WW}{{\ms E}}
\newcommand{\EX}{\Ebb_{\P\!\times\!\nu}^{x,0}\!}
\newcommand{\EW}{\Ebb_\msw ^x\!}
\newcommand{\msw}{{\rm W}}

\newcommand{\IIXX}{\int_\BR\!\!\!\dx}

\newcommand{\EZ}{\Ebb_{\P\!\times\!\nu}^{0,0}\!}

\newcommand{\e}{\epsilon}

\newcommand{\jjj}{\sum_{j=1}^{d-1}}
\newcommand{\ob}{\omega}
\newcommand{\PFF}{{\rm H}}
\newcommand{\PFFF}{{\rm H}_{\rm F}}
\newcommand{\PF}{{\rm H}_{\rm qf}}
\newcommand{\PFK}{{\rm H}_{\rm K}}
\newcommand{\T}{{\rm T}_{\rm kin}}
\newcommand{\KKKK}{{\rm I}}
\newcommand{\ima}{{\rm I}_0}

\newcommand{\K}{{\T \,\dot{+}\,\hf}}
\newcommand{\xt}{\Y_t(x)}
\newcommand{\xtt}{\Y_{t\wedge \tau}(x)}
\newcommand{\xs}{\Y_s(x)}
\newcommand{\Y}{{\rm M}}
\newcommand{\pq}{{\rm q}}
\newcommand{\www}{b}
\newcommand{\rmw}{{\rm Z}}

\newcommand{\mmm}{\MMM}

\newcommand{\hhh}{\ms H}
\newcommand{\hhhf}{\ms H_{\rm Fock}}

\newcommand{\SSS}{\ms S_\RR}

\newcommand{\ott}{\bigoplus^d_{\mu=1}}
\newcommand{\ottt}{\bigoplus^d}
\newcommand{\ab}[1]{\langle#1\rangle}

\newcommand{\pro}[1]{(#1_t)_{t\geq 0}}

\newcommand{\Q}  {{\mathscr Q}}
\newcommand{\QE}{{\mathscr Q}_{\rm E}}
\newcommand{\nt}{\frac{t}{2^n}}
\newcommand{\TT}[1]{{T_{t_{#1}}}}
\renewcommand{\tt}[1]{{t_{#1}}}

\newcommand{\tot}{{\rm P}}
\newcommand{\dx}{\!{\rm d} x}
\newcommand{\dy}{\!{\rm d} y}
\newcommand{\dE}{\!{\rm d} E}

\newcommand{\dk }{\!{\rm d} k}
\newcommand{\dB }{\!{\rm d} B}
\newcommand{\dP  }{\!{\rm d P}}
\renewcommand{\P  }{{\rm P}}
\newcommand{\dr}{{\rm d}r}
\newcommand{\ds}{{\rm d}s}

\newcommand{\JJ}{{\rm J}}
\newcommand{\jj}{{\rm j}}
\newcommand{\p}{{\rm p}}

\newcommand{\is}{\inf\sigma}

\def\bbbone{{\mathchoice {\rm 1\mskip-4mu l} {\rm 1\mskip-4mu l}
{\rm 1\mskip-4.5mu l} {\rm 1\mskip-5mu l}}}
\def\one{\bbbone}

\newcommand{\la}{\lambda}

\newcommand{\MMM}[4]
{\left[ \!\!\!\begin{array}{cc}#1&#2\\
#3&#4\end{array}\!\!\!\right] }

\newcommand{\hf}{{\rm H}_{\rm rad}}
\newcommand{\pf}{{{\rm P}_{\rm f}}}
\newcommand{\gr}{\varphi _{\rm g}}
\newcommand{\bou}{\Phi_{\rm b}}
\newcommand{\bouu}{\phi_{\rm b}}

\newcommand{\grt}{\gr^t}
\newcommand{\grtt}{\phi _t}

\newcommand{\grn}{\gr}
\newcommand{\N}{{\rm N}}

\newcommand{\half}{\frac{1}{2}}

\newcommand{\han}{{1/2}}

\newcommand{\s}{\sigma}
\newcommand{\vp}{{\hat \varphi}}

\newcommand{\non}{\nonumber}

\newcommand{\QT}{\QQ_{[-t, t]}}
\newcommand{\QS}{\QQ_{[-s, s]}}
\newcommand{\QR}{\QQ_{[-r, r]}}
\newcommand{\QTT}{\QQ_{[0, t]}}
\newcommand{\QSS}{\QQ_{[0, s]}}
\newcommand{\QQ}{{\rm  J}}

\makeatletter
\@addtoreset{equation}{section}
\makeatother

\title
{Functional integral approach to semi-relativistic Pauli-Fierz models}
\author{
Fumio Hiroshima\thanks{Faculty of Mathematics, Kyushu University, 
Motooka 744, Nishiku, Fukuoka, 819-0395, Japan,  e-mail: hiroshima@ math.kyushu-u.ac.jp}\\
\small \it Dedicated to
 Professor Asao Arai 
 on the occasion of his 60th birthday
}

\begin{document}
\setlength{\baselineskip}{16pt}
\maketitle
\begin{abstract}
By means of  functional integrations  spectral properties of 
  semi-relativistic Pauli-Fierz Hamiltonians 
  $$H=\sqrt{(\p-\alpha\A)^2+m^2}-m+V+\hf$$
in  quantum electrodynamics is considered. 
Here $\p$ is the momentum operator, 
$\A$ a quantized radiation field on which an ultraviolet cutoff is imposed,  
$V$ an external potential, 
$\hf$ the free field Hamiltonian and $m\geq0$
describes the mass of electron.  
Two  self-adjoint extensions of  a semi-relativistic Pauli-Fierz Hamiltonian are defined. The Feynman-Kac type formula of $e^{-tH}$ is given. 
An essential self-adjointness,  a spatial decay of  bound states, 
a Gaussian domination  of the ground state and the existence of a measure associated with the ground state are  shown.
All the results are independent of values of coupling constant $\alpha$, 
and it is emphasized that 
$m=0$ is included.   
\end{abstract}

\newpage{\footnotesize \tableofcontents}\newpage
\section{Introduction}
\subsection{Preliminary}
In the past decade a great deal of work has been devoted to 
studying spectral properties of non-relativistic quantum electrodynamics in the purely mathematical point of view. In this paper 
we are concerned with the  {semi-relativistic} Pauli-Fierz model  (it is  abbreviated as SRPF model)
 in  quantum electrodynamics and its spectral properties by using functional integrations. 
 The SRPF model   describes a minimal interaction between semi-relativistic electrons and 
a massless quantized radiation field $A$ on which 
an  ultraviolet cutoff function is imposed. 
 We assume throughout this paper that 
the electron is spinless and  moves in  $d$  
$(\geq 3)$ dimensional Euclidean space for simplicity. 
In the case where the electron has spin $\han$, the procedure is similar 
  and we shall publish details somewhere. 
A Hamiltonian of semi-relativistic as well as non-relativistic quantum electrodynamics is  usually described as a self-adjoint operator in 
the tensor product of a Hilbert space and a boson Fock space. 
In this paper instead of the boson Fock space  we can formulate 
the Hamiltonian as a self-adjoint operator 
in the known Schr\"odinger representation in  a functional realization of the boson Fock space as a space of  square integrable functions 
with respect to the corresponding Gaussian measure. 
Through the Schr\"odinger representation  a Fyenman-Kac type formula of 
the strongly continuous one parameter semigroup generated by  the SRPF Hamiltonian is given.
A functional integral or a path measure approach is proven to be useful to study 
properties of bound states associated with  embedded eigenvalues in the continuous spectrum. See e.g., 
\cite[Sections 6 and 7]{lhb11} . 
We are interested 
in investigating properties of bound states and ground states of 
the SRPF Hamiltonian  by functional integrations.

\subsection{Self-adjoint extensions and functional integrations}
The SRPF  Hamiltonian can be  realized as 
a self-adjoint operator bounded from below in the   tensor product of 
 $\LR$ and a boson Fock space $\fff$, where $\LR$ denotes the state space of a semi-relativistic  electron and   
$\fff$ that of photons. 
 Then the  decoupled Hamiltonian  is given by 
 \begin{align}
\label{500mile}
(\sqrt{\p^2+m^2}-m+V) \otimes \one +\one\otimes \hf,
 \end{align}
 where 
$\p=(\p_1,...,\p_d)=(-i\partial_{x_1},...,-i\partial_{x_d})$ denotes the momentum operator, 
$m$  electron mass, $V:\BR\to \RR$ an external potntial, 
 and $\hf$ the free  field Hamiltonian on  $\fff$. 
The SRPF  Hamiltonian  is defined by introducing the minimal coupling by the quantized radiation field $A$ with cutoff function $\vp$, i.e.,  
replacing $\p\otimes \one $ with $\p\otimes \one-\alpha A$ and, then 
\begin{align}\label{sa}
\PFF=\sqrt{(\p\otimes\one-\alpha A)^2+m^2}-m+V\otimes \one +\one\otimes \hf,
\end{align}
where $\alpha$ is a real coupling constant. 
In order to investigate  the semigroup $\ee^{-t \PFF}$,  $t\geq0$,  we 
redefine 
$\PFF$ 
on $\LR\otimes L^2(\Q)$ instead of $\LR\otimes\fff$, 
where $L^2(\Q)$ denotes the set of  square integrable functions  
on a  Gaussian probability space $(\Q,\mu)$, and is called a Schr\"odinger representation of $\fff$.

We introduce three classes, $\vqf$, $\vk$ and $\vsa$,  of external potentials. 
The definitions of $\vqf $, $\vk $ and $\vsa$  are given in Definitions \ref{vqf}, \ref{VK}, and \ref{vsa}, respectively.
Note that $\vk $ contains relativistic 
Kato-class potentials (see \kak{rk}), 
$\vsa$ potentials being relatively bounded 
with respect to $\sqrt{\p^2+m^2}-m$, and 
$\vk \subset \vqf$, $\vsa\subset \vqf$ hold.
We show in Theorems 
\ref{essentialselfadjointnesstheorem} and  
\ref{sasa}
that $\PFF$  is 
self-adjoint on $D(|\p|\otimes\one)\cap D(\one\otimes\hf)$
for  $V\in \vsa$. 
For more singular potentials we  
shall construct  two appropriate self-adjoint extensions of $\PFF$, 
which are denoted by $\PF$ and $\PFK$. 
The former is defined for $V\in \vqf$ by the quadratic form sum 
and 
the later for $V\in \vk$ through  Feynman-Kac type formula.
See Definition \ref{vqf} for $\PF$ and 
Definition \ref{vk} for $\PFK$.
Although $\vqf$ is wider  than $\vk$, 
$\PFK$ is defined under weaker condition on cutoff function $\vp$ than that for $\PF$.

In Introduction $\PFF$ stands for $\PF$ or $\PFK$ in what follows. 
We construct the Feynman-Kac type formula of $\ee^{-t\PFF}$
in terms of a composition of 
Euclidean quantum field $\AA(f)$ with test function 
$f\in\WW=\bigoplus^d L_\RR^2(\BRR)$,  $d$-dimensional 
Brownian motion $(B_t)_{t\in\RR}$  on the whole real line $\RR$ defined 
on a probability space 
$(\Omega_\P,\calb _\P, \P^x)$, 
and a subordinator $\pro T$ on  $(\Omega_\nu,\calb _\nu, \nu)$. 
The Euclidean quantum field $\AA(f)$ is Gaussian,  and the 
covariance is given by 
$\Ebb_{\mu_{\rm E}}[\AA(f)\AA(g)]=\qq(f,g)$ with some bilinear form
 $\qq(\cdot,\cdot)$ on $\WW\times\WW$. 
Hence it is driven in Theorem \ref{fkf2} and Corollary \ref{fkfshift} that 
\begin{align}\label{fkfformula}
(F, \ee^{-2t\PFF}G)=
\IIXX  \EX
\lkkk
\lk
\JJ_{-t} F(B_{-T_t}), 
 \ee^{-i\alpha\AA(\KKKK [-t,t]) }\ee^{-\int_{-t}^tV(B_{T_s})\ds }
\JJ_t G(B_{T_t})\rk
\rkkk
\end{align}
for $F, G\in L^2(\BR; L^2(\Q))\cong \LR\otimes L^2(\Q)$. 
Here $\KKKK [-t,t]$ is a  limit of $\WW$-valued stochastic integrals, 
which is formally written as 
\eq{kitu}
\KKKK [-t,t]=\bigoplus_{\mu=1}^d \int_{-T_{t}}^{T_{t}} \jj_{T^\ast_s} \lambda(\cdot-B_s)\dB_s^\mu
\en
with  $\lambda=(\vp/\sqrt\omega\check{)}$.
 Here $T^\ast_s=\inf\{t| T_t=s\}$ is the first hitting time of $\pro T$ at $s$. 
 Notations $\JJ_t$ and $\jj_t$  are defined in Section 2.2 below, and 
 the rigorous  definition of \kak{kitu} is given in Lemma \ref{cauchysequence}, Remarks \ref{remark1} and \ref{remark2}.

\subsection{Main results}
 By using the Feynman-Kac type formula \kak{fkfformula} we study the spectrum of the SRPF Hamiltonian $\PFF$. 
The main results of this paper are (a)-(d) below:
\bi
\item[(a)] Self-adjointness and essential self-adjointness of $\PFF$
(Theorems \ref{essentialselfadjointnesstheorem} and \ref{sasa}).
\item[(b)] Spatial decay  of bound states $\bou$ of $\PFF$ 
(Theorem \ref{falloff1}).
\item[(c)] Gaussian domination  of the  ground state $\gr$ of $\PFF$
(Theorem \ref{gaussiandecay}).
\item[(d)] Existence of a  probability measure $\mu_\infty$  associated with 
 $\gr$
 (Theorem \ref{gibbs}). 
\ei
The spectrum of non-relativistic versions of $\PFF$, which is the so-called Pauli-Fierz model,  have been studied, 
and among other things the existence of a ground state is  proven in \cite{bfs99,gll01}. 
See  also \cite{spo04} and references therein.
 The spectrum of semi-relativistic versions, $\PFF$, is also studied in e.g., 
 \cite{fgs01, hha13, hh13, kms09,kms11,kms12,ms10,ms09} 
 from  an operator-theoretic point of view.
In particular the existence of ground states of $\PFF$ 
are considered  under some conditions in 
\cite{kms09,kms12} for $m>0$ and \cite{hh13} for $m\geq0$. 

Here are outlines of assertions (a)-(d) mentioned above. 

{\bf (a)} 
Following our previous work \cite{hir00b}, we investigate (a).  
This  can be proven  by estimating the scalar product  
$|(K F, \ee^{-t\PFF}G)|$ for 
self-adjoint  operators 
$K=\one\otimes \hf$ and $\p_\mu\otimes\one$.
Let $V=0$. Then a bound  $|(K F, \ee^{-t\PFF}G)|\leq C_{K, G} \|F\| $, $F, G\in \D(\PFF)$,  is shown with some constant $ C_{K, G}$. 
Hence $\ee^{-t\PFF}$ leaves 
$\D(|\p|\otimes\one)\cap \D(\one\otimes\hf)$ invariant for $V=0$ and we can conclude that 
$\PFF$ is essentially self-adjoint on $\D(|\p|\otimes\one)\cap \D(\one\otimes\hf)$ by Proposition \ref{invariant} for $V\in \vsa$
for arbitrary values of $\alpha$. 
This is an extension of that of  a non-relativistic case established in \cite{hir00b}
and \cite[Section 7.4.1]{lhb11}. 
Furthermore the self-adjointness of $\PFF$ is 
shown in Theorem \ref{sasa}.
Examples include a spinless hydrogen like atom (Example \ref{hydrogen}). 
It is noted that our  method is also available to the SRPF Hamiltonian with spin. 
We give a comment on 
known  results.  
Although in \cite{kms11,ms10} the  self-adjointness of the SRPF Hamiltonian with spin $\han$ 
is considered, it is not sure that the method can be available to spinless cases.

{\bf (b)} 
Let 
\eq{hpcite}
\hp =\sqrt{\p^2+m^2}-m +V
\en
be the semi-relativistic Schr\"odinger operator.
Let $\pro z$ be the $d$-dimensional L\'evy   process on a probability space 
$(\Omega_\rmw, \calb _\rmw , \rmw ^x)$ such that 
$\Ebb_\rmw ^x \lkkk \ee^{-iu \cdot z_t} \rkkk =\ee^{-t (\sqrt{|u|^2+m^2}-m)}\ee^{-iu\cdot x}$. 
Hence the self-adjoint generator of $\pro z$ is given by 
$\sqrt{\p^2+m^2}-m$.
The Feynman-Kac type formula for  $\hp $ is thus given by 
\eq{fkfshcrel}
(f, \ee^{-t\hp } g)=\int_\BR \dx 
\Ebb_\rmw ^x\!\lkkk\bar f(z_0)g(z_t)\ee^{-\int_0^t V(z_s) \ds}\rkkk .
\en
Conversely taking  a  potential 
$-V$ such that 
\eq{rk}
\sup_{x\in\BR}\Ebb_\rmw ^x\![\ee^{-\int_0^t 
V(z_s) \ds}]<\infty,
\en
we can define 
the  strongly continuous one-parameter symmetric semigroup 
$s_t$, $t\geq 0$, on $\LR$
by 
\eq{qm}
(s_t f)(x)= \Ebb_\rmw ^x\!\lkkk f(z_t)
\ee^{-\int_0^t V(z_s) \ds}\rkkk.
\en
Thus 
we can define 
the  unique self-adjoint operator 
$\hp ^{\rm K}$ by 
$s_t=\ee^{-t\hp ^{\rm K}}$, $t\geq0$. 
A potential $V$ satisfying 
$
\sup_{x\in\BR}\Ebb_{\rm P}^x
\![\ee^{+\int_0^t V(B_s) \ds}]<\infty
$
is known as 
a Kato-class potential. 
Replacing the Brownian motion $B_t$ 
with L\'evy process $z_t$,  
we call  
a  potential $-V$ satisfying \kak{rk}
a relativistic Kato-class potential. 
The property \kak{rk} is also used in the proofs of Lemmas \ref{sup} and  \ref{bound2}, and Corollary \ref{martingale2}. 
Let 
$V=V_+-V_-$ be such that $V_\pm\geq0$, $V_+\in L_{\rm loc}^1(\BR)$ and $V_-$ is a relativistic Kato-class potential. 
$\vk$ denotes the set of such potentials.  
Furthermore let $\bouu$ be a bound state  of $\hp ^{\rm K}$ with $V\in \vk$, i.e., 
$\hp ^{\rm K}\bouu=E\bouu$ with  some  $E\in\RR$. 
Then 
 the stochastic process 
\eq{stochastic2}
\lk
\ee^{tE}\ee^{-\int_0^t V(z_s+x) \ds}\bouu(z_t+x)\rk_{t\geq0}
\en
is  martingale with respect to the natural filtration
$M_t=\s(z_s,0\leq s\leq t)$.
From martingale property we can derive a spatial decay  
of $\bouu(x)$ (\cite{cms90}). 
Furthermore 
in \cite{hil13}  we can extend these procedures to 
a semi-relativistic Schr\"odinger  operators of the form:
$\sqrt{({\bf \sigma}\cdot(\p-a))^2+m^2}-m +V$ on 
$\CC^2\otimes L^2(\RR^3)$,
where ${\bf \s}=(\s_1,\s_2,\s_3)$ denotes $2\times 2$ Pauli matrices and $a=(a_1,a_2,a_3)$ a vector potential satisfying suitable conditions.

In a similar manner to \kak{qm} we define 
a strongly continuous one-parameter symmetric semigroup and 
define 
the SRPF Hamiltonian with 
$V\in \vk$. 
We can show in Theorem \ref{semigroup}
 that 
the map 
$$(S_t F) (x)=\Ebb_{\P\times\nu}^{x,0}\lkkk \JJ_0^\ast 
\ee^{-i\alpha\AA(\KKKK[0,t])}
\ee^{-\int_0^tV(B_{T_s})ds}\JJ_t 
F(B_{T_t})\rkkk
$$
is 
 the strongly continuous one-parameter symmetric semigroup under the identification 
 $\LR\otimes L^2(\Q)\cong L^2(\BR; L^2(\Q))$.
 Thus  
 we can define the self-adjoint operator 
 $\PFK$ by $S_t=\ee^{-t\PFK}$, $t\geq0$.   
To study (b)
we also show  a martingale property of some stochastic process derived from the Feynman-Kac type formula  \kak{fkfformula}. 
Let $\bou$ be any bound state of $\PFK$, i.e., $\PFK \bou=E\bou$ with some $E\in\RR$.  
We can show in Theorem \ref{martingale} that the $L^2(\QE)$-valued stochastic process 
\begin{align}
( \xt )_{t\geq 0} =
\lk 
\ee^{tE} 
 \ee^{-i\alpha\AA(\KKKK^x [0,t]) }\ee^{-\int_0^tV(B_{T_r}+x)\dr }
\JJ_t \bou(B_{T_t}+x)\rk_{t\geq0},\quad t\geq 0,
\end{align}
  is martingale with respect to 
  a 
filtration $({\cal M}_t)_{t\geq0}$. 
Suppose that $|V(x)|\to0$ as $|x|\to\infty$. 
Then 
we can show  
in Theorem \ref{falloff1} 
that 
$\|\bou(x)\|_{L^2(\Q)}$ spatially decays exponentially in the case of $m>0$ and 
polynomially in the case of $m=0$.
As far as we know a polynomial decay  of bound states  of the SRPF  Hamiltonian with $m=0$ is 
new.

{\bf (c)} 
By the phase factor  
$\ee^{-i\alpha\AA(\KKKK [-t,t]) }$  appeared in the Feynman-Kac type formula \kak{fkfformula}, 
 $(F, \ee^{-t\PFF}G)\in \CC$ for 
 $F,G\geq 0$ in general.  
However it is established in a similar manner to \cite{hir00a}  that 
$(F, \ee^{-i\frac{\pi}{2}\N} \ee^{-t\PFF}\ee^{i\frac{\pi}{2}\N}G)>0$ for 
$F,G\geq 0$ ($F\not\equiv 0, G\not\equiv 0$), where $\N$ denotes 
the number operator. I.e., $\ee^{-i\frac{\pi}{2}\N} \ee^{-t\PFF}\ee^{i\frac{\pi}{2}\N}$ is positivity improving.
 Then 
 the ground state $\gr$  satisfies that 
$\ee^{-i\frac{\pi}{2}\N}\gr>0$. 
This is a key point  to study the ground state of $\PFF$  by path measures. 
By 
 $\ee^{-i\frac{\pi}{2}\N}\gr>0$,  
normalizing sequence  
\eq{ns}
\grt=\ee^{-t\PFF}(\phi\otimes \one) /\|\ee^{-t\PFF}
(\phi\otimes \one)\|
\en
strongly 
 converges to a normalized ground state $\gr$ as $t\to \infty$ for any $0\leq \phi\in\LR$ but $\phi\not\equiv 0$. 

Physically it is  interested in observing  expectation values of 
 some observable ${\cal O}$  
 with respect to $\gr$, i.e., 
 $(\gr, {\cal O}\gr)$. 
Since $\grt\to\gr$ as $t\to\infty$ strongly, we can 
see that 
$
\d (\gr, {\cal O}\gr)=\limt
 (\grt, {\cal O}\grt)$.
Let $\A_\xi$ be the quantized radiation field smeared by 
$\xi\in \bigoplus^d L_\RR^2(\BR)$. 
To show (c) we prove in Lemma \ref{guss1} the bound 
\eq{non}
(\grt, \ee^{\beta \A_\xi^2}\grt)\leq \frac{1}{\sqrt{1-2\beta\qq(\jj_0\xi,\jj_0\xi)^2}}
\en
uniformly in $t$ for some $\beta>0$.
Taking the limit $t\to\infty$ on both sides of \kak{non},  
 we show that $\gr\in \D(\ee^{\beta \A_\xi^2}) $ for some $0<\beta$.

{\bf (d)} 
For  some important observables ${\cal O}$, 
 by \kak{fkfformula}  we can see 
 that 
$
 (\grt, {\cal O}\grt)=\Ebb_{\mu_t}[F_{\cal O}^t]$
 with an 
 integrant $F_{\cal O}^t$ and 
 probability measures (we call this as finite volume Gibbs measure) given by 
 \eq{measure123}
\mu_t^{\rm SRPF}(A)
=\mu_t(A)
=\frac{1}{Z_t}\IIXX 
\EX
\lkkk
\one_A \ee^{-\frac{\alpha^2}{2} \qq(\KKKK [-t,t])}\ee^{-\int_{-t}^tV(B_{T_s})\ds} 
\rkkk, \quad t\geq0,
\en
where $Z_t$ denotes the normalization constant. 
See Definition \ref{64}. 
Furthermore  it is interesting to show  the convergence of measures 
$\mu_t$, $t\geq0$, for its own sake 
in mathematics.    
Formally we have $(\gr, {\cal O}\gr)=\Ebb_{\mu_\infty}[F_{\cal O}^\infty]$. 
Exponent $\qq(\KKKK [-t,t])$ in \kak{measure123} is called a pair interaction associated with $\PFF$, which is formally 
given by 
\eq{srpf}
W^{\rm SRPF}=\qq(\KKKK [-t,t])=\sum_{\mu,\nu=1}^d
\int_{-T_t}^{T_t} \dB_s^\mu 
\int_{-T_t}^{T_t} \dB_r^\nu 
W_{\mu\nu}(T^\ast_s-T^\ast_r,  B_s-B_r),
\en
where the pair potential $W_{\mu\nu}$ is given by 
\eq{wsrpf}
W_{\mu\nu}(t,X)=
\half\int_\BR\frac{|\vp(k)|^2}{\omega(k)}\lk\delta_{\mu\nu}-\frac{k_\mu k_\nu}{|k|^2}\rk
\ee^{-ik\cdot X}\ee^{-|t|\omega(k)}
 {\rm d}k.
\en
See \kak{pair} and \kak{W} for details. 
Several limits of 
some finite volume Gibbs measures associated with models in 
quantum field theory are considered,  e.g., examples include  the Nelson  model \cite{bhlms02,os99}, 
spin-boson model \cite{hhl12} and the Pauli-Fierz model \cite{bh09}.   
In this paper we consider a  limit 
of  finite volume Gibbs measures associated with  the SRPF model.
The pair interaction  associated with a spin-boson model \cite{hhl12}, 
the Nelson model \cite{bhlms02} and the Pauli-Fierz model \cite{bh09, hir00a, spo87} are 
given by 
\begin{align}
\label{sb}
&
W^{\rm SB}=
\int_{-t}^t ds \int_{-t}^t \dr 
\int_\BR\frac{|\vp(k)|^2}{2\omega(k)}  (-1)^{N_s-N_r}\ee^{-|s-r|\omega(k)} \dk,\\
&
\label{nelson}
W^{\rm N}=\int_{-t}^t ds\int_{-t}^t \dr \int_\BR \frac{|\vp(k)|^2}{2\omega(k)} 
\ee^{-ik\cdot(B_s-B_r)}\ee^{-|s-r|\omega(k)} \dk,\\
&
\label{pf}
W^{\rm PF}=\sum_{\mu,\nu=1}^d\int_{-t}^t
\! \dB_s^\mu  
\int_{-t}^t 
\! \dB_r^\nu 
\!\!\!
\int_\BR\frac{|\vp(k)|^2}{2\omega(k)}
\lk\delta_{\mu\nu}-\frac{k_\mu k_\nu}{|k|^2}\rk \ee^{-ik\cdot(B_s-B_r)}\ee^{-|s-r|\omega(k)}
 \dk, 
\end{align}
respectively.
Let $\mu_t^\#$
be the finite volume Gibbs measure  with 
the pair interaction $W^\#$, where 
$\#$ stands for ${\rm SRPF, SB, N, PF}$. 
Note that 
$W^{\rm N}$ and $W^{\rm SB}$ are uniformly bounded  with respect to paths, i.e., 
$$W^\#\leq \int_{-t}^t ds\int_{-t}^t \dr \int_\BR\frac{|\vp(k)|^2}{2\omega(k)}e^{-|t-s|\omega(k)}\dk,\quad \#={\rm SB,N},$$
while $W^{\rm SRPF}$ and $W^{\rm PF}$ are not uniformly bounded.
In addition, $\mu_t^{\rm N}$ and 
$\mu_t^{\rm PF}$ are measures defined on 
the set of continuous paths, 
$\mu_t^{\rm SB}$ and $\mu_t^{\rm SRPF}$, however,  on the set of paths with jumps.
See Figure \ref{measure}. 
\begin{figure}[t]
\begin{center}
\arrayrulewidth=1pt
\def\arraystretch{1.5}
\begin{tabular}{l|c|c}
\       & Path without jumps & Path with jumps \\
\hline
Uniformly bounded $W^\#$   & $\mu_t^{\rm N}$ & 
$\mu_t^{\rm SB}$ \\
\hline
Non-uniformly bounded $W^\#$ & $\mu_t^{\rm PF}$ & $\mu_t^{\rm SRPF}$\\
 \end{tabular}
\end{center}
\caption{Finite volume Gibbs measures }
\label{measure}
\end{figure}%

Existence of limits of  
$\mu_t^{\rm N}$ and $\mu_t^{\rm PF}$ 
 is proven in \cite[Theorem 6.12]{lhb11} and \cite{bh09}, respectively,  
 by showing  the tightness of the family of measures $(\mu_t^{\rm N})_{t\geq0}$ and $(\mu_t^{\rm PF})_{t\geq0}$.
It is, however,  
not straightforward  
to show the convergence of  $\mu_t^{\rm SB}$,  since $(\mu_t^{\rm SB})_{t\geq0}$ 
is a measure defined on the set of paths with jumps $\pm 1$. 
Then the local weak convergence of 
$\mu_t^{\rm SB}$ is shown in \cite{hhl12} instead of a weak 
convergence. 
Since both $\mu_t^{\rm N}$ and 
$\mu_t^{\rm SB}$ include the uniformly bounded  pair interactions, 
we can fortunately easily use the limit measures  
to express the ground state expectation with some  observable, e.g. $\ee^{+\beta\rm N}$, etc. 
See \cite{hhl12} and \cite[Section 6]{lhb11}.
On the other hand 
since $\mu_t^{\rm PF}$ includes 
the non-uniformly bounded  pair interaction, 
it is unfortunately hard to apply the limit measure  
to express the ground state expectation with some concrete observable. 
See \cite[p.196-197]{spo04}.
It is however worthwhile showing the existence of limit measure itself, 
since our pair interaction is 
far singular than that of e.g. \cite{os99}.  
The family of probability measures 
$\mu_t^{\rm SRPF}$, which is our main object in this paper,  
is  defined on the set of c\'adl\'ag paths, 
and its pair interaction  is not uniformly bounded. 
We  prove  that $\mu_t^{\rm SRPF}$  converges to a probability measure $\mu_\infty^{\rm SRPF}$  in the local weak sense as $t\to \infty$ by using the existence of the ground state of $\PFF$, which is studied in \cite{hha13,kms09,kms11}.

This paper is organized as follows:
Section \ref{sec2} is devoted to defining the SRPF  Hamiltonian $\PF$  in both a Fock space and a function space to study the semigroup by a path measure. 
In Section \ref{sec3} we construct a Feynman-Kac type formula for $\PF$.
In Section \ref{sec4} we show the essential self-adjointness  and the self-adjointness of 
$\PF$.
In Section \ref{sec5} we define the self-adjoint operator $\PFK$ of the SRPF  Hamiltonian 
with a potential in the relativistic Kato-class,  
and show that some stochastic process is martingale by which a spatial decay of bound states is proven.
Section \ref{sec6} is devoted to showing  a Gaussian domination  of the ground state. 
In Section \ref{sec7}  
the existence of an infinite volume limit of  finite Gibbs measures is shown.  
In Section \ref{sec8} we give comments on a model with spin $\han$ and 
model with a fixed total momentum.
Finally in Appendix   we give fundamental tools of probability theory and proofs of some equalities used in this paper.

\section{Semi-relativistic Pauli-Fierz model}
\label{sec2}
\subsection{SRPF  model in Fock space}
Let us begin by defining fundamental tools of quantum field theory in Fock representation. 
Let 
$
\LRZ=L^2(\BR\!\times\!\{1,..,d\!-\!1\})$
 be the Hilbert space of a
single photon in the $d$-dimension Euclidean space, 
where $\BR\times\{1,..,d\!\!-\!\!1\}\ni(k,j)$ denotes the  pair of 
momentum $k$
and polarization $j$ of a single photon.
We denote the  $n$-fold symmetric tensor
product of $\LRZ$ by $\otimes_{\rm sym}^n \LRZ$ for $n\geq1$ 
and 
 set 
$\otimes_{\rm
sym}^0\LRZ=\CC$, where $\CC$ is the set of complex numbers.  
The boson Fock space describing the full photon field is
defined then as the Hilbert space
\eq{fock}
\fff=\bigoplus_{n=0}^\infty
\lk
\otimes_{\rm sym}^n \LRZ
\rk
\en
endowed with the scalar product 
$
(\Psi,\Phi)_{\fff}= \sum_{n=0}^\infty(\Psi^{(n)},\Phi^{(n)})_
{\otimes^n\LRZ}$ 
for 
 $\Psi = \bigoplus_{n=0}^\infty
\Psi^{(n)}$ and  $\Phi = \bigoplus_{n=0}^\infty \Phi^{(n)}$.
Alternatively, $\fff$ can be identified as the set of
$\ell^2$-sequences $\{\Psi^{(n)}\}_{n=0}^\infty$ with $\sum_{n=0}^\infty\|\Psi^{(n)}\|^2_{
\otimes_{\rm sym}^n \LRZ}<\infty$. 
The vector $\Omega_{\rm b}=\{1,0,0,...\}\in\fff$
is called the Fock vacuum. The finite particle subspace $\ffff$ is
defined by
\eq{fock2}
\ffff=\left\{\{\Psi^{(n)}\}_{n=0}^\infty\in\fff \big|
 \Psi^{^{(m)}} = 0 \mbox{ for } \forall m \geq M\mbox{ with some } M\right\}.
 \en
With each $f\in\LRZ$ a  creation operator and an annihilation operator are 
associated. 
The creation operator $\add(f):\fff\rightarrow \fff$ is
defined by
\eq{dreation}
(\add(f)\Psi)^{(n)}=\sqrt n
S_n(f\otimes \Psi^{(n-1)})
\en for $n\geq1$, 
where $S_n(f_1\otimes \cdots \otimes f_n)= (1/n!) \sum_{\pi \in
{\mathfrak S}_n} f_{\pi(1)} \otimes \cdots \otimes f_{\pi(n)}$ is the
symmetrizer with respect to the permutation group ${\mathfrak S}_n$ of degree
$n$. The domain of $\add(f)$ is maximally defined by
$\D(\add(f))= \left\{\{\Psi^{(n)}\}_{n=0}^\infty \in\fff\,
 \left| \,
\sum_{n=1}^\infty n\| S_n(f\otimes \Psi^{(n-1)})\|^2<\infty\right.
\right\}$.
 The annihilation operator $a(f)$ is
introduced as the adjoint of $\add(\bar f)$, i.e., 
$a(f) = (\add(\bar f))^\ast$.
Both $\add(f)$ and $a(f)$ are closable operators, their closed extensions
are denoted by the same symbols. Also, they leave $\ffff$
 invariant and obey the canonical commutation relations on
$\ffff$:
\eq{fock3}
[a(f), \add(g)]=(\bar f, g)\one,\quad 
[a(f),a(g)]=0,\quad 
 [\add(f),\add(g)]=0.
 \en
 The dispersion relation considered in this paper is chosen to be 
$
\omega(k)=|k|$ for $k\in\BR$.
We denote $\hat f$ the Fourier transformation of $f\in\LR$. 
We use the informal expression $\d \jjj \int \ass(k,j) f(k,j)\dk $ for $\ass(f)$ for convenience. 
Then the quantized radiation field smeared by $f\in\LR$ is defined by 
\begin{align}\label{radiationfield}
A_\mu(f,x)=\frac{1}{\sqrt 2} \jjj \int 
\frac{e_\mu(k,j)}{\sqrt{\omega(k)}}
\lk \add(k,j)\ee^{-ikx}{\hat f (k)}
+a(k,j)\ee^{ikx}{\hat f (-k)}\rk \dk 
\end{align}
for each $x\in\BR$ and its momentum conjugate  by 
\begin{align}\label{momentumconjugate}
\Pi_\mu(f,x)=\frac{i}{\sqrt 2}\jjj   \int e_\mu(k,j)\sqrt{\omega(k)}
\lk \add(k,j)\ee^{-ikx}{\hat f (k)}
-a(k,j)\ee^{ikx}{\hat f (-k)}\rk \dk ,
\end{align}
where 
$e(k,j)$, $k\in\BR\setminus\{0\}$, $j=1,...,d-1$, are  $d$ dimensional 
polarization vector such that 
$e(k,j)\cdot e(k,j')=\delta_{jj'}$ and $k\cdot e(k,j)=0$. 
From canonical commutation relations it  follows that 
$
[A_\mu(f, x),\Pi_\nu(g,y)]=i\int\delta_{\mu\nu}^\perp(k)\hat f(-k) \hat g(k) \ee^{ik(x-y)} \dk$,
where 
$$\delta_{\mu\nu}^\perp(k)=\delta_{\mu\nu}-
\frac{k_\mu k_\nu}{|k|^2},\quad k\ne 0,$$ denotes the 
transversal delta function. 
The
quantized radiation field with a fixed ultraviolet cutoff function $\vp$ is 
then defined by 
\eq{amu}
A_{\mu}(x)=A_\mu(\varphi,x).
\en
By $k\cdot e(k,j)=0$, the Coulomb gauge
condition
\eq{coulomb}
\nabla_x \cdot A(x)=0
\en
holds as an operator.
A standing  assumption in this paper is as follows. 
\begin{assumption}\label{ass1}
{\rm We suppose that  $\ov{\vp(k)}=\vp(-k)$ 
and 
$ \vp/\sqrt \ob\in\LR$.
}\end{assumption}
We also introduce an assumption.
\begin{assumption}\label{ass2}{\rm 
We suppose that  $\omega\sqrt\omega\vp,
\vp/\sqrt\omega\in\LR$.
}\end{assumption}

Under Assumption \ref{ass1},  
$A_\mu(x)$ is a well-defined {symmetric}
operator in $\fff$.  
By the fact that $\d \sum_{n=0}^\infty 
\frac{\|A_{\mu}(x)^n
\Phi\|t^n}{n!}<\infty$ for $\Phi\in\ffff$ and $t>0$, and Nelson's analytic vector
theorem \cite{nel59}, the symmetric operator 
$A_{\mu}(x)\lceil_{\ffff}$ is essentially self-adjoint. We denote
its closure $\ov{A_{\mu}(x)\lceil_{\ffff}}$ by the same symbol
 $A_{\mu}(x)$.
 
 Next we define the free quantum field Hamiltonian on $\fff$. 
The free quantum field Hamiltonian is defined  as the infinitesimal generator
of a one-parameter unitary group.
 This unitary
group is constructed through a functor $\Gamma$. Let ${\ms C}(X\to
Y)$ denote the set of contraction operators from  $X$ to $Y$. 
We set ${\ms C}(X)$ for ${\ms C}(X\to X)$ for simplicity. 
Functor $\Gamma: {\ms C}(\LRZ) \rightarrow 
{\ms C}(\fff)$ is defined as
$
\Gamma(T)=\bigoplus_{n=0}^\infty [\otimes^n T]$, 
where $\otimes^0T=\one$.
 For a self-adjoint operator $h$ on $\LRZ$,
$\Gamma(\ee^{ith})$, $t\in\RR$, is a strongly continuous one-parameter
unitary group on $\fff$. Then by Stone's theorem there exists a
unique self-adjoint operator $\dG (h)$ on 
$\fff$ such that $
\Gamma(\ee^{ith}) = \ee^{it\dG (h)}$, $t\in\RR. $ $\dG (h)$ is
called the second quantization of $h$. 
Let $\omega$ be regarded as 
the multiplication operator $f\mapsto \omega(k)
f(k,j)=|k|f(k,j)$. 
The operator
$
\dG (\omega)$ 
 is then the free quantum field Hamiltonian.

The Hilbert space describing a state space of a single electron is $\LR$. 
The semi-relativistic electron Hamiltonian on
$\LR$ with a real-valued external potential $V$ is given by
\begin{align}\label{particlehamiltonian}
\hp =\sqrt{\p^2  +m^2} -m+ V.
\end{align}
Here $\p^2=\sum_{\mu=1}^d \p_\mu^2$, 
$V$ acts as the  multiplication operator in $\LR$,  
and $m\geq0$ describes the mass of an electron. 
We regard $m\geq 0$ 
  as a non-negative parameter and it is allowed to be  $m=0$. 
The state space of the joint electron-field system is
\eq{statespace}
 \hhhf=\LR\otimes\fff.
 \en
To define the quantized 
radiation field $A$ we identify $\hhhf$ with the set of
$\fff$-valued $L^2$ functions on $\BR,$
 i.e.,
$\hhhf\cong \int_\BR^\oplus \fff   \dx $ and 
$A_\mu$ 
is defined by 
$
 A_{\mu}= \int_\BR^\oplus  A_{\mu}(x)  \dx 
$
  with the domain 
 $$\D(A_\mu)=\lkk
F\in \int^\oplus_\BR\!\!\! 
\fff \dx\left|
F(x)\in \D(A_\mu(x))\  a.e.\   x
\in\BR\ and \int_\BR\!\!\! \|A_\mu (x) F(x)\|^2_\fff \dx<\infty
\right.\rkk.$$
Hence $(A_{\mu}F)(x) = A_{\mu}(x)F(x)$ for $F(x)\in \D(A_{\mu}(x))$
and $A_{\mu}$ is self-adjoint. 
The Friedrichs extension of 
$\half (\p\otimes \one-\alpha A)^2\lceil_{\CCC\hat \otimes\ffff}$ is denoted by 
$\rh_A$. 
\begin{definition}
{\rm \TTT{Definition of SRPF Hamiltonian}
Suppose Assumption \ref{ass1}. The SRPF  Hamiltonian is defined by
 \begin{align}
\label{isogasi}
(2\rh_A +m^2)^\han -m  
+ V \otimes \one +\one \otimes \dG (\omega)
\end{align}
with 
the domain $
\D((2\rh_A +m^2)^\han )\cap \D(V \otimes \one)\cap 
\D(\one \otimes \hf)$.
} \end{definition}

\subsection{SRPF  model in function space}
In order to construct the Feynman-Kac type formula 
of the semigroup   generated by 
the SRPF  Hamiltonian  we prepare some probabilistic tools for the field and the particle. 
Let us  use  a $\ms Q$-space representation instead of 
the Fock representation. 
Define the field operator
$A_\mu(f)$ by
$$
A_\mu( f)=\frac{1}{\sqrt2}\jjj\int e_\mu(k,j) \left(
\hat f(k)
\add(k,j)+
\hat f(-k) a(k,j)\right) \dk 
$$
and the $d\times d$ matrix ${\rm D}(k) $ by
$
{\rm D}(k)  = \lk \delta_{\mu\nu}^\perp (k)
\right)_
{1\leq \mu,\nu\leq d}$ for $k\not=0$.
Consider the bilinear form 
$
\q: \oplus^d \LR\times \oplus^d\LR
\rightarrow \CC$ defined by 
\eq{qmq}
\q(f,g)=\half
\int_\BR \ab{ {\hat f(k)},  {\rm D}(k)  \hat g(k)} \dk,
\en
where $\ab{x,y}=\bar x \cdot y $ denotes the 
standard scalar product on  $\CC^d$. 
Then 
we have 
$\sum_{\mu,\nu=1}^{d-1}(A_\mu(f_\mu)\Omega_{\rm b}, A_\nu(g_\nu)\Omega_{\rm b})_{\fff}=\q(f,g)$.

 We introduce another
bilinear form 
$\qq:\oplus^d \LRR\times\oplus^d \LRR\rightarrow \CC$ by 
\begin{align}
\qq (F,G) = \half \int_{\RR^{d+1}}
\ab{
 {\hat F(k,k_0)}, 
{\rm D}(k)  \hat
G(k,k_0) }
\dk \dk _0.
\end{align}
Note that ${\rm D}(k) $ is independent of $k_0\in\RR$ in the definition of $\qq$.
We denote ${\rm q}_\#(K,K)$ by ${\rm q}_\#(K)$ for simplicity, where ${\rm q}_\#$ stands for $\q$ and $\qq$. 

Let $\SSS(\BR)$ be
the set of real-valued Schwarz  test functions on $\BR$.
Let 
$\Q=(\oplus^ d  \SSS(\BR))'$ and 
$ 
\QE=(\oplus^d \SSS(\BRR))'$. 
Here $X'$ denotes the dual space of a locally convex space $X$.
We  denote the pairing between elements of
$\Q$ and $\oplus^ d  \SSS(\BR)$ by $\ab{\phi,f}_{\rm M}\in\RR$ for $\phi\in\Q$ and $f\in \oplus^ d  \SSS(\BR)$.
We denote the expectation with respect to a probability path measure $P^x$ starting from $x$ at $t=0$ 
by $\Ebb_P^x[\cdots]=\int\cdots \dP^x$. 
By the Bochner-Minlos Theorem
there exists a probability space 
$(\Q, \Sigma_{\rm M}, \mu_{\rm M})$
 such that $\Sigma_{\rm M}$ is the smallest
$\s$-field generated by $\{\ab{\phi, f}_{\rm M} | f\in\ott {\SSS}(\BR) \}$ and
$\ab{\phi,f}_{\rm M}$ is a Gaussian random variable 
with mean zero and
the covariance given by 
$
\Ebb_{\mu_{\rm M}}\lkkk 
\ab{\phi,f}_{\rm M}\ab{\phi,g}_{\rm M}
\rkkk=\q(f,g)$.
Then we have 
\begin{align}\label{covariance}
\Ebb_{\mu_{\rm M}}\!\lkkk 
  \ee^{i\ab{\phi, f}_{\rm M}}
\rkkk = \ee^{-\half \q (f,f)}.
 \end{align}
Since  $\ab{\phi,\bigoplus_\mu^d \delta_{\mu\nu}f}$ is a
$\Q$-representation of the quantized radiation field
 with 
 test
  function  $f\in\SSS(\BR)$, we have to extend
$f\in {\SSS}(\BR)$ to a more general class since our cutoff is
$(\vp/\sqrt\omega)^\vee\in\LR$. 
 For any $f=\Re f+i \Im f\in \ott \ms S(\BR)$
we set 
$\ab{\phi,f}_{\rm M} = \ab{\phi, \Re f}_{\rm M}  + i\ab{\phi, \Im f}_{\rm M}$. 
Let 
\eq{WWW}
\W=\bigoplus^d  \LR .
\en
Since $\ms S (\BR)$
 is dense in
$\LR$
 and the equality
$
\int_{\Q}
 |\ab{\phi, f}_{\rm M}|^2 {\rm d}\mu_{\rm M}
=\half\|f\|_{\W}
  ^2
$
holds by \kak{covariance},
 we can define $\ab{\phi, f}_{\rm M}$ for $f\in 
 \W
 $ by
 $\ab{\phi, f}_{\rm M}=\slimn \ab{\phi, f_n}_{\rm M}$ in
$L^2(\Q)$, where
 $\{f_n\}_{n=1}^\infty \subset \ott  \ms S(\BR)$ is 
 any sequence such that
 $\slimn f_n=f$ in
$\W $.
Thus we define the multiplication
operator $\A(f)$ by 
$$\lk \A (f) F\rk (\phi) = \ab{\phi, f}_{\rm M} F(\phi),\quad f\in\W$$
in $L^2(\Q)$
with the 
domain
$
\D(\A  (f)) = \left\{F\in L^2(\Q)| 
\int_{\Q}
|\ab{\phi, f}_{\rm M} F(\phi)|^2{\rm d}\mu_{\rm M}<\infty\right\}
$. 
Denote the identity function in $L^2(\Q)$ by $\one_{\Q}$
and the function $\A (f) \one_{\Q}$ 
by $\A (f)$ unless confusion may arise. 
It is known as the Wiener-It\^o decomposition that 
$$
L^2(\Q) = \bigoplus_{n=0}^\infty L_n^2(\Q)
$$
with
$
L_n^2(\Q)= 
\ov{ 
{\rm {\rm L.H.}} \lkk \wick{\prod_{j=1}^n \A  (f_j)}
| f_j\in \W , \,
j=1,2,...,n\rkk}$.
Here $L_0^2(\ms Q) = \CC$
and $\wick{X}$ denotes Wick product recursively defined by
$
\wick{\A  (f)} = {\A  (f)}$ and 
$
  \wick{\A(f) \prod_{j=1}^n \A (f_j) 
 } = \A(f)\wick{ 
 \prod_{j=1}^n \A (f_j) 
 }
 -\sum_{j=1}^n \q(f,f_j) \wick{\prod_{i\not=j}^n \A (f_i) 
}$.
We set $\A_\mu (f)=\A(\bigoplus_{\nu=1}^d \delta_{\nu\mu} f)$ for $f\in\LR$. 

Let 
\eq{MMM}
\WW=\bigoplus^d  \LRR.
\en
Similarly we can define the Gaussian random variable 
 $\AA(f)$ labelled by $f\in\WW $ on 
 a probability space $(\QE,\Sigma_{\rm E},\mu_{\rm E})$ 
 with $\q$ replaced by 
$\qq$ in \kak{covariance}. In particular 
\begin{align}\label{covariance2}
\Ebb_{\mu_{\rm E}}
\lkkk
 \ee^{i\ab{\phi, f}_{\rm E}}
\rkkk = \ee^{-\half \qq (f,f)}
 \end{align}
and 
$
\lk \AA (f) F\rk (\phi) = \ab{\phi, f}_{\rm E}  F(\phi)$ hold for $f\in\WW$.


We define the second quantization on $L^2(\Q)$.
Let $T\in {\ms C}(\LR)$.
Then 
$\Gamma(T)\in {\ms C}(L^2(\Q))$
 is defined by
\eq{gamma}
\Gamma(T)\one _{\Q}  =\one _{\Q},\quad 
\Gamma(T) \wick{\prod_{j=1}^n \A(f_j)}
=
 \wick{\prod_{j=1}^n \A(Tf_j)}.
 \en
 For 
$T\in {\ms C}(\LRR)$ (resp. ${\ms C}(
\LR\to\LRR)$,
 $\Gamma(T)\in {\ms C}(L^2(\QE))$ (resp.
$\Gamma(T)\in {\ms C}(L^2(\Q)\to L^2(\QE))$ is 
similarly defined. 
For each self-adjoint operator $h$ in
$\LR$ (resp. $\LRR$),
 $\Gamma(\ee^{ith})$, $t\in\RR$,  is a one-parameter
unitary group on $L^2(\Q)$ (resp. $L^2(\QE)$).
 Then there exists 
a unique self-adjoint
operator $\dG  (h)$ in $L^2(\Q)$ (resp. $L^2(\QE)$)
such that 
 $\Gamma(\ee^{ith}) =
\ee^{it\dG (h)}$ for all  $t\in\RR$.
We set 
\begin{align}
\hf=\dG(\omega(\p)),
\quad 
\pf_\mu =\dG(\p_\mu),\quad 
{\rm N}=\dG(\one_\LR)
\end{align}
in $L^2(\Q)$, where $\omega(\p)=|\p|=\sqrt{\p^2}$. 
We also set 
\begin{align}
\ov{\hf}=\dG(\one\otimes\omega(\p)),\quad
\ov{\pf}_\mu =\dG(\one \otimes \p_\mu),
\quad 
\ov  {\rm N}=\dG(\one\otimes \one_\LR),
\end{align}
where we identify $\LRR=L^2(\RR)\otimes\LR$.
$\hf$ denotes the free field Hamiltonian  of $L^2(\Q)$, $\pf$ the momentum operator  
and ${\rm N}$ the number operator, 
and 
$\ov{\hf}$, $\ov{\pf}$ and $\ov {\rm N}$ the Euclidean version of $\hf$, $\pf$ and ${\rm N}$, respectively. 
The spaces $L^2(\Q)$ and $L^2(\QE)$ are  connected 
by the family of isometries. 
Let $\jj_t:\LR\to\LRR$, $t\in\RR$, 
 be the family of isometries  such that 
$(\jj_sf,  \jj_t g)_\LRR
=
(\hat f, \ee^{-|t-s|\omega}\hat g)_\LR
$, 
and then 
$\JJ_t=\Gamma(\jj_t)$, $t\in\RR$,  turns to be 
 the family of isometry transforming  
$L^2(\Q)$ to  $L^2(\QE)$ such that 
$(\JJ_s \Phi, \JJ_t \Psi)_{L^2(\QE)}=
(\Phi, \ee^{-|t-s|\hf} \Phi)
_{L^2(\Q)}$.
We have the relations:
\begin{align}\label{intertwining}
\JJ_t \hf=\ov{\hf}\JJ_t,\quad 
\JJ_t {\rm N}=\ov {\rm N} \JJ_t,
\quad 
\JJ_t \pf=\ov{\pf}\JJ_t.
\end{align}
It is known that  $\fff$, $A_\mu(f)$ and $\dG (h)$ are
isomorphic to $L^2(\Q)$, 
$\A_\mu(f)$ 
and $\dG ( h(\p) )$, respectively, where
$ h$ is the  multiplication operator by $h$.
That is, there exists a unitary operator 
${\mathbb  U}: \fff\to L^2(\Q)$ such that
(1) 
${\mathbb  U}\Omega_{\rm b} = \one_{L^2(\Q)} $,
(2) 
${\mathbb  U}\otimes_{\rm sym}^n {\cal W}=L^2_n(\Q)$,
(3) 
${\mathbb  U} A_\mu(f) {\mathbb  U}^{-1} = \A _\mu(f)$,
and 
(4)
${\mathbb  U} \dG (h) {\mathbb  U}\f = \dG (h(\p) )$.
We set 
\eq{darksideofthemoon}
\hhh=\LR\otimes L^2(\Q).
\en
Through the  unitary operator  ${\cal U}=\one\otimes{\mathbb  U}: 
\LR\otimes \fff \rightarrow \hhh$ the SRPF Hamiltonian 
is defined as
 an  operator on
$\hhh$. Let
\begin{align}
 \la=(\vp/\sqrt\ob)^\vee,
 \end{align}
where {\it \v f} denotes the inverse Fourier transform of $f$ in $\LR$. 
Set
${\A}_\mu (\la(\cdot-x)) = \A(\bigoplus_{\nu=1}^d \delta_{\mu\nu}
\la(\cdot-x))$. 
Then the quantized radiation field with cutoff function $\varphi$ is defined by 
$
{\A}_\mu  = \int_\BR^\oplus {\A} _\mu(\la(\cdot-x)) \dx $.
Then $\A$  is   a self-adjoint operator in $\hhh$ under the identification: 
 $\hhh\cong \int^\oplus_\BR L^2(\Q) \dx$. 
Let $L_{\rm fin}^2(\Q)$ be the finite particle subspace of $L^2(\Q)$, i.e., 
\eq{fp}
L_{\rm fin}^2(\Q)=\mbox{Linear hull of}
\lkk 
\left.
\wick{\prod_{j=1}^n \A(f_j)}, \one 
\right |f_j\in \ms M
, j=1,...,n, n\geq 1\rkk.
\en
Then the Friedrichs extension of 
$\half(\p-\alpha \A)^2\lceil_{\CCC\hat\otimes L_{\rm fin}^2(\Q)}$ is denoted by 
$\rh_{\A}$. 
\begin{definition}
{\rm 
\TTT{Definition of $\PFFF$}
Suppose Assumption \ref{ass1}. 
The SRPF  Hamiltonian 
in the function space $\hhh$ is defined by 
\begin{align}
\label{definitionofPF}
\PFFF
& =
\T +V+\hf,\\
\T&=(2\rh_{\A}+m^2)
^\han -m  \end{align}
with 
the domain $\D(\PFFF)=\D(\T)
\cap \D(V)\cap \D(\hf)$.
}\end{definition}
  We investigate $\PFFF$ instead of \kak{isogasi} in what follows.

\section{Feynman-Kac type formula}
\label{sec3}
\subsection{Markov properties}
Let 
$\mathcal O\subset \RR$ and we set 
$$U_{\mathcal O}=\ov{{\rm L.H.}
\{f\in L^2_\RR(\BRR)|f\in {\rm Ran}\ \jj_t \mbox{ with some } t\in \mathcal O\}}$$
 and define 
the sub-$\s$-field 
$
\Sigma_{\mathcal O}$ by the minimal 
$\s$-filed generated  by 
$\AA(f), f\in U_{\mathcal O}$, i.e., 
$\Sigma_{\mathcal O}=\s\lk \AA(f)|f\in U_{\mathcal O}\rk$.
We also set 
$\Sigma_{\{s\}}=\Sigma_s$.
Let  $e_{{\mathcal O}}: L^2_\RR(\BRR)\to U_{{\mathcal O}}$ be the projection and the second quantization 
$\Gamma(e_{{\mathcal O}}):L^2(\Q)\to L^2(\QE)$ is denoted by 
$E_{{\mathcal O}}$. 
Hence $E_{{\mathcal O}}L^2(\Q)$ is the set of $\Sigma_{\cal O}$-measurable functions in $L^2(\QE)$. 
Moreover 
we set $E_s=\JJ_s \JJ_s^\ast$. 
Then $E_s=E_{\{s\}}$ follows. 
Let 
$\Ebb_{\mu_{\rm E}}\lkkk \Phi|\Sigma_{{\mathcal O}}\rkkk $ be the conditional expectation of $\Phi\in L^2(\QE)$  with respect to 
$\Sigma_{{\mathcal O}}$, i.e., 
By the Jensen inequality 
$\rho=\Ebb_{\mu_{\rm E}}\lkkk \Phi|\Sigma_{{\mathcal O}}\rkkk $ is the 
unique $L^2$-function such that 
it is $\Sigma_{{\mathcal O}}$-measurable and 
$\Ebb_{\mu_{\rm E}}\lkkk 
\Psi \Phi\rkkk=\Ebb_{\mu_{\rm E}}\lkkk\Psi\rho\rkkk$ for all 
$ 
\Sigma_{{\mathcal O}}$-measurable function $\Psi$. 
\bl{conditionalexpectation}
Let $\Phi\in L^2(\QE)$. Then 
$
E_{{\mathcal O}}\Phi=
\Ebb_{\mu_{\rm E}}\lkkk \Phi|\Sigma_{{\mathcal O}}\rkkk$.
\el
\proof
We see that 
$\varrho=E_{{\mathcal O}}\Phi$  
is measurable with respect to $\Sigma_{\mathcal O}$ and 
$\Ebb_{\mu_{\rm E}}\lkkk \Psi\varrho\rkkk=(\Psi, E_{\mathcal O}\Phi)=(\Psi,\Phi)=\Ebb_{\mu_{\rm E}}\lkkk \Psi \Phi\rkkk$ for all $\Sigma_{\mathcal O}$-measurable function $\Psi$. 
Thus the lemma follows.
\qed
The property below is known as Markov property \cite{sim74}:
let $a\leq b\leq t\leq c\leq d$, then 
$
E_{[a,b]} E_t E_{[c,d]}=
E_{[a,b]} E_{[c,d]}
$
follows. 
From this property we can see the corollary  below:
\bc{markovproperty2}
It follows that 
$\Ebb_{\mu_{\rm E}}\lkkk \Phi|\Sigma_{(-\infty,s]}\rkkk=
\Ebb_{\mu_{\rm E}}\lkkk \Phi|\Sigma_s\rkkk$ for all $\Sigma_{[s,\infty)}$-measurable 
function $\Phi$.
\ec
\proof
We note that 
$E_{(-\infty, s]}E_{[s,\infty)}\Phi=
E_{(-\infty, s]}E_s E_{[s,\infty)}\Phi=E_sE_{[s,\infty)}\Phi$ by the Markov property. 
Then the lemma follows from Lemma \ref{conditionalexpectation} and 
$E_s=E_{\{s\}}$.
\qed
\subsection{Euclidean groups}
We introduce the second quantization of Euclidean group $\{u_t,r\}$  on $\LRR$,
where the time shift operator  $u_t$ is defined by
$u_t f(x_0,{\bf x})=f(x_0-t,\bf x)$ and
the time reflection  $r$ by
$r f(x_0,{\bf x})=f(-x_0,{\bf x})$.
The second quantization of $u_t$ and $r$ are denoted by ${\rm U}_t=\Gamma(u_t):L^2(\QE)\to L^2(\QE)$ and 
${\rm R}=\Gamma(r):L^2(\QE)\to L^2(\QE)$, respectively.
Note that $r^\ast=r$, $rr=r^\ast r=\one$, 
$u_t^\ast=u_{-t}$ and $u_t^\ast u_t=\one$ 
and that
${\rm U}_t$ and ${\rm R}$ are  unitary.
The time shift $u_t$, the time reflection $r$  and isometry $\jj_t$
satisfy the algebraic relations:
$u_t \jj_s=\jj_{s+t}$ and 
$r \jj_s=\jj_{-s}r$. From these relations it follows that  
$
{\rm U}_t \JJ_s=\JJ_{s+t}$ and 
$
{\rm R}{\rm U}_s={\rm U}_{-s}{\rm R}$
as operators. 

\subsection{Feynman-Kac type formula and time-shift}
Let $(\Omega_{\P}, \calb _{\P}, \P^{x})$ be a probability space, and 
$(B_t)_{t\in\RR}$ the $d$-dimensional 
Brownian  motion on whole real line 
$\RR$ on 
$(\Omega_{\P}, \calb _{\P}, \P^{x})$ starting from $x$ at $t=0$.
See Appendix A for the detail of 
the Brownian  motion on whole real line 
$\RR$.
We also introduce a  subordinator $\pro T$ on a probability space 
$(\Omega_\nu, \calb _\nu, \nu)$ such that 
\eq{characteristicfunction}
\Ebb_\nu^0 \lkkk {\ee^{-uT_t}}\rkkk =
\ee^{-t(\sqrt{2u+m^2}-m)},\quad t\geq0,\quad u\geq0.
\en
The  subordinator  
$\pro T$ is one-dimensional L\'evy process and 
indeed given by 
$T_t=\inf\{s>0| B_s^1+ms=t\}$, where $\pro {B^1}$ denotes the one-dimensional Brownian motion. 
Path $[0,\infty)\ni t\to T_t\in [0,\infty)$ is nondecreasing and   
right continuous,  and the left limit exists almost surely in $\nu$.
The distribution $\rho_t$ of $T_t$, $t\geq0$,  on $\RR$ is given by 
\begin{align}
\label{distribution}
\rho_t(s)=\frac{t}{\sqrt{2\pi}}\ee^{tm}s^{-3/2}\exp\lk -\half\lk\frac{t^2}{s}+m^2s\rk\rk
1_{[0,\infty)}(s)
\end{align}
and thus $\Ebb_\nu^x[f(T_t)]=\int_\RR f(s+x) \rho_t(s)\ds $. 
Notice that 
$\Ebb_\nu^0[T_t]<\infty$ if and only if $m>0$.
We need to define a  self-adjoint extension of $\PFFF$, 
which is constructed through a functional integration. 
The idea is a  combination of Proposition \ref{fkf1} below and 
a   subordinator $\pro T$. 
In quantum mechanics, the path integral representation of the heat semigroup generated by 
the semi-relativistic Schr\"odinger operator 
$\sqrt{(\p-a)^2+m^2}-m +V$ is given by 
\begin{align}
(f, \ee^{-t(\sqrt{(\p-a)^2+m^2}-m +V)} g)
=\IIXX  \EX\lkkk
\ov{f(B_{T_0}) } g(B_{T_t}) 
\ee^{-\int_0^t V(B_{T_s})\ds } 
\ee^{-i\int_0^{T_t} a(B_s) \circ \dB_s}
\rkkk.
\end{align}
Here 
$\int_0^{T_t} a(B_s) \circ \dB_s$ is defined by 
$\int_0^T a(B_s) \circ \dB_s$ evaluated at $T=T_t$. 
Although the SRPF Hamiltonian is of a similar form 
of $\sqrt{(\p-a)^2+m^2}-m +V$, it is not
 straightforward to construct the Feynman-Kac type formula of 
 $\ee^{-t\PFFF}$.
The Feynman-Kac type formula for the  case of  $\alpha=0$ is however immediately given by 
\begin{align}
(F, \ee^{-t(\hp+\hf)}G)_{\hhh}=
\IIXX   \EX\lkkk
(\JJ_0 F(B_{T_0}), 
\JJ_t G(B_{T_t}))_{L^2(\QE)} \ee^{-\int_0^t V(B_{T_s})\ds  }
\rkkk.
\end{align}
We shall extend this formula for an arbitrary value of $\alpha$. 
The self-adjoint operator 
$\rh_\A$ is defined by the Friedrichs extension. 
In general self-adjoint extensions are 
not unique, and it is also 
 not trivial to signify an  operator core 
  of  $\rh_{\A}$. 
As is shown in the proposition below we can  however show the essential self-adjointness of $\rh_{\A}$ by means of functional integral  approach under some conditions.  
Let $C^\infty({\rm N})=\cap_{n=1}^\infty  \D({\rm N}^n)$, where we recall that ${\rm N}$ denotes the number operator. 
We define 
the $\LR$-valued stochastic integral
$ \int_0^t \la(\cdot-B_s) \dB_s^\mu$ by 
$$\int_0^t \la(\cdot-B_s) \dB_s^\mu=
 \slimn \sum_{j=1}^{2^n} \la(\cdot-B_{\tt{j-1}})(B_{\tt j}^\mu-B_{\tt{j-1}}^\mu)$$ in 
 $L^2(\BR\times \Omega_{\rm P}, \dx\otimes \dP^x)
 $ with $\tt j=tj/2^n $.
\bp{invariant}
Let $h$ be closed and the generator of a contraction  semigroup on a Banach space. 
Let $D$ be dense and $D\subset D(h)$, 
so that $\ee^{-th}D\subset D$. 
Then $D$ is a core of $h$, i.e., $\ov{h\lceil_D}=h$.  
\ep
\proof 
See  \cite[Theorem X.49]{rs2}.
\qed
To prove an essential self-adjoint of $\rh_\A$ we apply Proposition \ref{invariant}. 
\bp{fkf1}
Suppose Assumptions \ref{ass1} and \ref{ass2}.
Then $\rh_{\A}$ is essentially self-adjoint on $\D(\p^2)\cap C^\infty({\rm N})$,
and it follows that 
\begin{align}
\label{pathintegral1}
(F, \ee^{-t\rh_{\A}} G)=\IIXX  \Ebb_{\P}^{x}\lkkk
({F(B_0)},  \ee^{-i \alpha \A(\tilde K[0,t])}
G(B_t))\rkkk,
\end{align}
where $\tilde K[0,t]=\ott
 \int_0^t \la(\cdot-B_s) \dB_s^\mu$. 
\ep
\proof
See Appendix \ref{fkf1p}.
\qed

The   path integral representation of the semigroup generated by 
the semi-relativistic Schr\"odinger operator can be constructed by 
a combination of the $d$-dimensional Brownian motion $\pro B$ 
and a subordinator $\pro T$. 
In a similar manner 
we can see the lemma below:
\bl{fkfcomputation}
Suppose Assumptions \ref{ass1} and  \ref{ass2}.
Then  \begin{align}
\label{pathintegral2}
(F, \ee^{-t\T}G)=
\IIXX  \EX\lkkk
({F(B_{T_0})},  \ee^{-i \alpha \A(K[0,t])}
G(B_{T_t}))\rkkk,
\end{align}
where $K[0,t]=\ott
 \int_0^{T_t}\la(\cdot-B_s) \dB_s^\mu$ 
is defined by 
$\ott
 \int_0^T\la(\cdot-B_s) \dB_s^\mu$ evaluated at $T={T_t}$.
\el
\proof
Since 
$\d (F, \ee^{-t\T}G)
=\Ebb_\nu^0\lkkk (\Psi, \ee^{-T_t \rh_{\A}}\Phi)\rkkk
$, 
by 
Proposition \ref{fkf1} and 
\kak{characteristicfunction}, 
we see that 
$\d (F, \ee^{-t\T}G)
=
\Ebb_\nu^0\lkkk
\IIXX  \Ebb_{\P}^{x}\lkkk
({\Psi(B_{T_0})},  \ee^{-i \alpha \A(\tilde K[0,t])}
\Phi(B_{T_t}))\rkkk\rkkk$. 
We can exchange $\int {\rm d}\nu$ and $\int \dx$ 
by Fubini's lemma. Then 
the lemma follows.
\qed

By Lemma \ref{fkfcomputation} we see that $D(\T)\cap D(\hf)$ is dense. 
Then 
we can define the quadratic form sum  
$\T\, \dot+ \, \hf$. 
Let $V$ be bounded. 
Then by the Trotter-Kato product formula \cite{km78} we have 
\begin{align}
\label{tkformula}
\ee^{-t(\K+ V)}=\slimn \lk
\ee^{-\nt \T}\ee^{-\nt\hf}\ee^{-\nt V}\rk^{2^n},\quad t\geq 0.
\end{align}
Using this formula we construct a  Feynman-Kac type formula of 
$\ee^{-t(\K+V)}$ for a bounded $V$. 
We define an $\LRR$-valued stochastic integral 
$\int_S^T \jj_s\la(\cdot-B_s) \dB_s^\mu$ by 
the strong limit: 
\eq{stochasticintegral}
\int_S^T \jj_s\la(\cdot-B_s) \dB_s^\mu
=
\slimn\sum_{j=1}^{2^n}\int_{S+\Delta_{j-1}}^{S+\Delta_j}\jj_{S+\Delta_{j-1}}\la(\cdot-B_s) \dB_s^\mu
\en
in $L^2(\BRR\times \Omega_{\rm P},\dx\otimes \dP^x)$,
where
$\Delta_j=(T-S)\frac{j}{2^n}$. We give a remark on notation.
Notation $\la(\cdot-B_r)$ denotes the  function $\la=\la(\cdot)$ shifted by 
$B_r$.
We denotes the image of $\la(\cdot-B_r)$ by the isometry $\jj_t$ by 
$\jj_t \la(\cdot-B_s)$. More precisely 
$$\d \widehat{\jj_t \la(\cdot-B_s)}(k_0,k)=
\frac{\ee^{-itk_0}}{\sqrt\pi}
\frac{\sqrt{\omega(k)}}{\sqrt{\omega(k)^2+|k_0|^2}}\hat \la(k)\ee^{-ik B_s},\qquad
(k_0,k)\in\RR\times\BR.$$
Let us recall  the family of projections: $E_t=\JJ_t\JJ_t^\ast$, $t\in\RR$. 
\bl{lemma1}
Suppose Assumptions  \ref{ass1},  \ref{ass2}, 
and that 
 $V\in \CCC$. 
 Then 
  \begin{align}
&\lk F, 
\lk
\ee^{-\nt \T}\ee^{-\nt\hf}\ee^{-\nt V}\rk^{2^n}G\rk \non \\
&\label{fkfcomputation2}=
\IIXX   \EX\lkkk
\lk \JJ_0F(B_{T_0}), \ee^{-i\alpha \AA(\KKKK_n[0,t])}\JJ_t G(B_{T_t})
\rk
\ee^{-\sum_{j=0}^{2^n}{\nt} V(B_{T_{t_j}})}
\rkkk,
\end{align}
where 
\begin{align}
\label{phase}
\d \KKKK_n[0,t]
=\ott
 \sum_{j=1}^{2^n}
\int_{T_{t_{j-1}}}^{T_{t_j}}\!\!\jj_{t_{j-1}}
\la(\cdot-B_s) \dB_s^\mu 
  \end{align}
 with  $t_j=tj/2^n$, 
 and 
 $\int_{T_{t_{j-1}}}^{T_{t_j}}\!\!\jj_{{t_{j-1}}}
\la(\cdot-B_s) \dB_s^\mu $ 
 denotes $\LRR$-valued stochastic integral 
 $\int_T^S 
 \!\!\jj_{{t_{j-1}}} \la(\cdot-B_s) \dB_s^\mu $ evaluated 
at $T= {T_{t_{j-1}}}$ and $S={T_{t_j}}$.
\el
\proof
By the formula $\JJ_t^\ast \JJ_s=\ee^{-|t-s|\hf}$, we have 
  \begin{align*}
\lk F, 
\lk
\ee^{-\nt \T}\ee^{-\nt\hf}\ee^{-\nt V}\rk^{2^n}G\rk 
=\IIXX   \EX\lkkk
U_n
\ee^{-\sum_{j=0}^{2^n}{\nt} V(B_{T_{t_j}})}
\rkkk,
\end{align*}
where 
 $$
U_n=\lk
\JJ_0F(B_{T_0}), 
\prod_{j=1}^{2^n} 
\lk 
\JJ_\tt{j-1} 
\ee^{-i\alpha\A\lk 
\ott
 \int_\TT{j-1}^\TT j \la(\cdot-B_r) \dB_r^\mu\rk}
\JJ_\tt{j-1} ^\ast \rk  \JJ_t G(B_{T_t})
\rk,$$
and we see that  
\begin{align}
\label{markovproperty3}
\JJ_\tt{j-1} 
\ee^{ 
-i\alpha\A\lk 
\ott \int_\TT{j-1}^\TT j \la(\cdot-B_r) \dB_r^\mu\rk
}
\JJ_\tt{j-1} ^\ast
=E_\tt{j-1} 
\ee^{
-i\alpha\A_{\rm E}\lk 
\ott \int_\TT{j-1}^\TT j\!\! \jj_\tt{j-1}\la(\cdot-B_r) \dB_r^\mu\rk
}
E_\tt{j-1}
\end{align}
by the definition of $\JJ_t$ and $E_t$. 
Then by the Markov property of $E_{\mathcal O}$, 
 $E_t's$ can be removed  in \kak{markovproperty3} 
and thus the lemma follows.  
\qed
$(\KKKK_n[0,t])_{t\geq0}$ can be regarded as 
an  $\WW$-valued stochastic process on 
the product probability space 
$(\Omega_{\P}\times\Omega_\nu, \calb _\P\times \calb _\nu,\P^x\otimes\nu)$. 
By the It\^o isometry we have 
\eq{itoisometry}
\Ebb_{\rm P}^{x}\lkkk \|\KKKK_n[0,t]\|^2_\WW \rkkk=
d\sum_{j=1}^{2^n}\Ebb_{\rm P}^x\!\lkkk
\int_{\TT{j-1}}^{\TT j} \|\jj_{\tt {j-1}}\la(\cdot-B_s)\|^2_\LRR \ds
\rkkk
=d T_t \|\vp/\sqrt\omega\|^2.
\en
We will show that $\KKKK_n[0,t]$ has a limit as $n\to \infty$ in some sense. 
Let $\ms N_\nu\in \calb _\nu$ be a null set, i.e., $\nu(\ms N_\nu)=0$,  such that 
for arbitrary $w\in \Omega_\nu\setminus\ms N_\nu$, 
the path  $t\mapsto T_t(w)$ is nondecreasing and  right-continuous, and 
has the left-limit.

\bl{cauchysequence}
For each $w\in\Omega_\nu \setminus \ms N_\nu$ 
the sequence $\{\KKKK_n[0,t]\}_n$ strongly converges 
in $ L^2( \Omega_{\P},\P^x)\otimes \WW$ 
as $n\to
\infty$, i.e,. there exists an $\KKKK [0,t]\in L^2(\Omega_\P,\d\P^x)\otimes\WW$ such that 
$\d \limn 
\Ebb_\P^x\lkkk \|\KKKK_n[0,t]-\KKKK [0,t]\|^2_\WW\rkkk=0$.
\el
\proof
Set $\KKKK_n=\KKKK_n[0,t]$. 
It is enough to show that $\{\KKKK_n\}_n$ is a Cauchy sequence in 
$L^2( \Omega_{\P},\P^x)\otimes \WW$. 
We have 
$\KKKK_{n+1}-\KKKK_n=\ott \sum_{m=1}^ {2^n}\int_\TT{2m-1}^\TT{2m}(\jj_\tt{2m-1}-\jj_\tt{2m-2})\la(\cdot-B_s) \dB_s^\mu$, 
where $t_j=tj /2^{n+1}$.
Thus 
\begin{align*}
 \Ebb_{\P}^{x}[\|\KKKK_{n+1}-\KKKK_n\|_\WW^2]
=
d
\sum_{m=1}^{2^n}
\Ebb_{\P }^{x}
\lkkk
\int_\TT{2m-1}^\TT{2m} \|
(\jj_\tt{2m-1}-\jj_\tt{2m-2})\la(\cdot-B_s)\|_\LRR^2
\ds\rkkk
\end{align*}
by the It\^o isometry \kak{itoisometry}. 
Notice that 
$\|(\jj_t-\jj_s)f\|^2=2(\hat f, (\one-\ee^{-|t-s|\omega})\hat f)$. 
Thus
\begin{align*}
 \Ebb_{\P}^{x}[\|\KKKK_{n+1}-\KKKK_n\|_\WW^2]
\leq
d
\sum_{m=1}^{2^n}
2 (\vp/\sqrt\omega, (\one-\ee^{-\frac{t}{2^{n+1}}\omega})\vp/\sqrt\omega)
(\TT{2m} -\TT{2m-1}).
\end{align*}
Since $T_t=T_t(w)$ is not decreasing  in $t$ for $w\in \Omega_\nu\setminus\ms N_\nu$,
$\sum_{m=1}^{2^n} (\TT{2m} -\TT{2m-1})\leq T_t$ follows. 
Thus 
$
 \Ebb_{\P}^{x}[\|\KKKK_{n+1}-\KKKK_n\|_\WW^2]
\leq 
d T_t 
\frac{t}{2^{n}}
\|\vp/\sqrt\omega\|^2$.
Hence we have 
$$
  \Ebb_{\P}^{x}[
\|\KKKK_m-\KKKK_n\|_\WW^2]
\leq
\lk
\sqrt{dt T_t}\|\vp/\sqrt\omega\|  \sum_{j=n+1}^m \lk\frac{1}{\sqrt2}\rk^j 
\rk^2
$$
for $m>n$. 
The right-hand side above converges to zero as $n,m\to\infty$. Then the sequence 
$\KKKK_n$ is a Cauchy sequence for almost surely $\nu$.
Then the lemma follows. 
\qed
\begin{remark}
\label{remark1}
{\rm Integral $\KKKK [0,t]$ is  informally written as 
\eq{informally}
\KKKK [0,t] =\ott\int_0^{T_t}\jj_{T^\ast_s}\la(\cdot-B_s) \dB_s^\mu.
\en
Here $T^\ast_s=\inf\{t|T_t=s\}$ is the first hitting time of $\pro T$ at $s$.
}\end{remark}
In a similar way to $\KKKK [0,t]$  we define $\KKKK  [s,t]$ by 
the limit of 
\begin{align}
\KKKK_n[s,t]
=\ott
 \sum_{j=1}^{2^n}
 \int_{T_{s+(t-s)_{j-1}}}^{T_{s+(t-s)_j}}\jj_{s+(t-s)_{j-1}}
 \la(\cdot-B_r) \dB_r^\mu 
\end{align}
 with  $(t-s)_j=(t-s)j/2^n$
in $L^2(\Omega_\P,\P^x)\otimes \WW$.
Moreover it can be straightforwardly seen that 
$\KKKK [s,t]$ coincides with the limit of 
subdivisions  
\begin{align}
\KKKK_n[s,t]
=\ott
 \sum_{j=1}^{a 2^n}
 \int_{T_{s+\frac{(t-s)_{j-1}}{a}}}^{T_{s+\frac{(t-s)_j}{a}}}
 \jj_{s+\frac{(t-s)_{j-1}}{a}}
 \la(\cdot-B_s) \dB_s^\mu 
\end{align}
for arbitrary $a\in{\mathbb  N}$. 
We show some properties of $\KKKK[a,b]$ in Appendix B.
\bl{fkffundamental}
Suppose Assumptions \ref{ass1} and \ref{ass2}. 
Then 
\begin{align}
\label{pathintegral0}
(F, \ee^{-t(\T\, \dot +\, \hf)}G)_\hhh=
\IIXX   \EX\lkkk
\lk
\JJ_0F(B_{T_0}), \ee^{-i\alpha\AA(\KKKK [0,t])}
\JJ_t G(B_{T_t})
\rk
\rkkk.
\end{align}
\el
\proof
The proof is similar to that of 
Theorem \ref{fkf2} below for $V=0$.
We omit it.
\qed
The immediate consequence of Lemma \ref{fkffundamental} is the diamagnetic inequality.
\bc{dia}
Suppose Assumptions \ref{ass1} and \ref{ass2}. 
Let $F,G\in\hhh$. 
Then it follows that 
\bi
\item[(1)] 
$|(F, \ee^{-t(\T\,\dot +\, \hf)}G)|\leq 
(|F|, \ee^{-t(\sqrt{\p^2+m^2}-m+\hf)}|G|)$
\item[(2)]
$|(F, \ee^{-t(\T\,\dot +\, \hf)}G)|\leq 
(\|F\|_{L^2(\Q)}, 
\ee^{-t(\sqrt{\p^2+m^2}-m)} \|G\|_{L^2(\Q)})_\LR$.
\ei
\ec
\proof 
Since $|\JJ_tG|\leq \JJ_t|G|$, 
it is straightforward to see that 
\begin{align*}
|(F, \ee^{-t(\T\,\dot +\, \hf)}G)|
&\leq 
\IIXX   \EX\lkkk
\lk
|F(B_{T_0})|, 
\ee^{-t\hf} |G(B_{T_t})|
\rk
\rkkk\\
&=
(|F|, \ee^{-t(\sqrt{\p^2+m^2}-m+\hf)}|G|).
\end{align*}
Then (1) follows. (2) is similarly proven. 
\qed

We introduce a class of potentials.
\begin{definition}
\label{vsa}
{\rm
$V$ is in $\vsa$ if and only if 
$V$
is  relatively bounded with respect to $\sqrt{\p^2+m^2}$ with a 
 relative bound
 strictly smaller  than one.
}
\end{definition}

\bl{relativebound}
Suppose Assumptions 
\ref{ass1} and 
\ref{ass2}. 
Let $V\in \vsa$.
Then 
$V$ is also relatively form bounded 
(resp. bounded) with respect to 
$\K$ 
with a relative bound smaller than $a$.
\el
\proof
Let  ${\rm sgn}F(x)=\frac{F(x)}{\|F(x)\|_{L^2(\Q)}}$ for 
$\|F(x)\|_{L^2(\Q)}\not=0$ and 
$=0$ for $\|F(x)\|_{L^2(\Q)}=0$.
Let $z>0$ be sufficiently large. 
Let $\psi\in C_0^\infty(\BR)$ and $\psi(x)\geq0$. 
Substituting the vector $F={\rm sgn}((\K+z)^{-\han}G)\cdot \psi\in\hhh$ in the  inequality 
$$|(F, (\K+z)^{-\han } G)_\hhh|\leq 
(\|F\|, (\sqrt{\p^2+m^2}-m+z)^{-\han} \|G\|)_\LR$$derived from Corollary \ref{dia} (2),  
we see that 
$$(\psi, \|(\K+z)^{-\han} G)(\cdot)\|_{L^2(\Q)})\leq (\psi, (\sqrt{\p^2+m^2}-m+z)^{-\han}\|G(\cdot)\|_{L^2(\Q)}).$$
Thus 
$\|((\K-z)^{-\han} G)(x)\|_{L^2(\Q)}\leq (\sqrt{\p^2+m^2}-m -z)^{-\han} \|G(x)\|_{L^2(\Q)}$
follows for almost every $x\in\BR$, 
and 
$$\||V|^\han (\K-z)^{-\han} G\|_\hhh
\leq 
\||V|^\han (\sqrt{\p^2+m^2}-m -z)^{-\han}G\|_\hhh$$ are  derived.  
Then $V$ is also form bounded with respect to $\K$.  
$$\||V| (\K-z)^{-1} G\|_\hhh
\leq 
\||V|(\sqrt{\p^2+m^2}-m -z)^{-1}G\|_\hhh$$
is similarly derived. 
\qed

If $V\in L_{\rm loc}^1(\BR)$, 
then 
 $\D(\T)\cap \D(\hf) \cap \D(V)$ is dense. 
Let $V=V_+-V_-$, where   $V_+=\max\{V,0\}$ is the positive part of $V$ 
and $V_-=\max\{-V,0\}$ the negative part. 
We introduce a class of potentials:
\begin{definition}{\rm 
\label{vqf}
$V=V_+-V_-$ is in $\vqf $ if and only if 
$V_+\in L_{\rm loc}^1(\BR)$ and $V_- $ 
relatively form bounded with respect to 
$(\p^2+m^2)^\han$ with relative bound strictly smaller than one.
}\end{definition}
Let $V=V_+-V_-\in \vqf$. Define the quadratic form $t$ on $\hhh$ by 
\eq{quadratic form}
t(F, G)=
(\T^\han F, \T^\han G)
+(\hf^\han F, \hf ^\han G)
+(V_+^\han F, V_+^\han G)
-(V_-^\han F, V_-^\han G)
\en 
with the form domain 
$Q(t)=
D(\T^\han)\cap 
D(\hf^\han)\cap 
D(V_+^\han)$. 
By Lemma \ref{relativebound} $t$ is semibounded and closed. 

\begin{definition}
{\rm 
\TTT{Definition of $\PF$}
Suppose  Assumptions \ref{ass1} and  \ref{ass2}.
Let $V\in \vqf $. 
Then 
the self-adjoint operator associated with 
the quadratic form $t$ is 
 denoted 
by $\PF$ and 
written as 
\begin{align}\label{selfadjointextension}
\PF=\T\, \dot + \, \hf\, \dot +\, V_+\, \dot - \, V_- .
\end{align}  
}\end{definition}
Note that the form domain of $\PF$ coincides with $Q(t)$.

We now construct a  Feynman-Kac type formula of $\ee^{-t\PF}$. 
\bt{fkf2}
Suppose  Assumptions \ref{ass1} and \ref{ass2}.
Let  $V\in \vqf$.
Then 
\begin{align}
\label{pathintegral}
(F, \ee^{-t\PF}G)_\hhh=
\IIXX   \EX\lkkk
\lk
\JJ_0F(B_{T_0}), \ee^{-i\alpha\AA(\KKKK [0,t])}
\JJ_t G(B_{T_t})
\rk
\ee^{-\int_0^t V(B_{T_s})\ds}
\rkkk.
\end{align}
\et
\proof
By the Trotter product formula \kak{fkfcomputation2}
we have
\begin{align*}
(F, \ee^{-t\PF}G)&=
\limn \lk F, 
\lk
\ee^{-\nt \T}\ee^{-\nt\hf}\ee^{-\nt V}\rk^{2^n}G
\rk\\
&=
\limn 
\IIXX   \EX\lkkk
(\JJ_0F(B_{T_0}), \ee^{-i\alpha \AA(\KKKK_n[0,t])}\JJ_t G(B_{T_t}))
\ee^{-\sum_{j=0}^{2^n}{\nt} V(B_{T_{t_j}})}
\rkkk.
\end{align*}
Suppose that $V$ is in $\CCC$. 
By Lemma \ref{lemma1} and 
the dominated convergence theorem 
we can show that the right-hand side above converges to 
that of \kak{pathintegral}.
For general $V$, by    
monotone convergence theorems for both integrals and quadratic forms,  
we can establish 
\kak{pathintegral}. See \cite[Theorem 6.2]{sim05} and \cite[Theorem 3.31]{lhb11}.
\qed

We can shift the time in the Feynman-Kac type formula.
We see it in the corollary  below. 
\bc{fkfshift}
Suppose  Assumptions  \ref{ass1} and \ref{ass2}. 
Let  $V\in \vqf $.
Then 
\begin{align}
&(F, \ee^{-2t\PF}G)_\hhh\non \\
&=
\IIXX   \EX\lkkk
\lk
\JJ_{-t}F(B_{-T_t}), \ee^{-i\alpha\AA(\KKKK [-t,0]+\KKKK [0,t])}
\JJ_t G(B_{T_t})
\rk
\label{shift2}
\ee^{-\int_{-t}^0  V(B_{-T_{-s}})\ds-\int_0^t V(B_{T_s})\ds}
\rkkk,
\end{align}
where 
$\KKKK [-t,0]$ is defined by 
\eq{k}
\KKKK [-t,0]=\bigoplus_{\mu=1}^d \limn \sum_{j=1}^{2^n}
\int_{-T_{-(t_{j-1}-t)}}^{-T_{-(t_j-t)}} \jj_{-(t_{j-1}-t)}\la(\cdot-B_s)\dB_s^\mu.\en
\ec
\proof
This is proven by means of 
the shift  ${\rm U}_t$ in the field
 and the facts that $T_s-T_t=T_{s-t}$ in law. 
By Theorem \ref{fkf2} we have 
$$
(F, \ee^{-2t\PF}G)_\hhh=
\IIXX   \EX\lkkk
\lk
\JJ_0F(B_{T_0}), \ee^{-i\alpha\AA(\KKKK [0,2t])}
\JJ_{2t} G(B_{T_{2t}})
\rk
\ee^{-\int_0^{2t} V(B_{T_s})\ds}
\rkkk
$$
and 
$$
=
\IIXX   \EX\lkkk
\lk
\JJ_{-t}
F(B_{T_0}), 
{\rm U}_t \ee^{-i\alpha\AA(\KKKK [0,2t])}
{\rm U}_{-t} \JJ_t G(B_{T_{2t}})
\rk
\ee^{-\int_0^{2t} V(B_{T_s})\ds}
\rkkk.
$$
By the shift of  the Brownian motion, $B_t\to B_{t-T_t}$,  we have 
$$
=
\IIXX   \EX\lkkk
\lk
\JJ_{-t}
F(B_{-T_t}), 
\ee^{-i\alpha \AA(S)}
\JJ_t G(B_{T_{2t}-T_t})
\rk
\ee^{-\int_0^{2t} V(B_{T_s-T_t})\ds}
\rkkk,
$$
 where 
 $\d S=\limn \bigoplus_{\mu=1}^ d \sum_{j=1}^{2\cdot 2^n}\int_{T_{t_{j-1}}-T_t}^{T_{t_j}-T_t}\jj_{t_{j-1}-t}\la(\cdot-B_s) \dB_s^\mu$
and, since $T_s-T_t=T_{s-t}$ for $s\geq t$ in law, we can check that 
$$\int_0^{2t}\!\!\! V(B_{T_s-T_t})\ds=\int_0^t \!\!\! V(B_{-(T_{t-s})})\ds +\int_t^{2t}
V(B_{T_{s-t}})\ds
=
\int_{-t}^0 \!\!\! V(B_{-T_{-s}})\ds +\int_0^t \!\!\!V(B_{T_s})\ds .$$
Furthermore we have 
\begin{align*}
& \sum_{j=1}^{2\cdot 2^n}
\int_{T_{t_{j-1}}-T_t}^{T_{t_j}-T_t}\jj_{t_{j-1}-t}\la(\cdot-B_s) \dB_s^\mu\\
&=
 \sum_{j=1}^{2^n}
\int_{-T_{-(t_{j-1}-t)}}^{-T_{-(t_j-t)}}
\jj_{-(t_{j-1}-t)}\la(\cdot-B_s) \dB_s^\mu
+
 \sum_{j=2^n+1}^{2\cdot 2^n}
\int_{T_{t_{j-1}-t}}^{T_{t_j-t}}\jj_{t_{j-1}-t}\la(\cdot-B_s) \dB_s^\mu.
\end{align*}
Then the theorem follows. 
\qed
\begin{remark}
{\rm \label{remark2}
For the notational convenience 
we denote 
 $\KKKK [-t,0]+\KKKK [0,t]$ by 
 $\KKKK [-t,t]=\bigoplus_{\mu=1}^d
 \int_{-T_{t}}^{T_t}
 \jj_{T^\ast_s}\lambda(\cdot-B_s)\dB_s^\mu$, 
 and 
$\int_{-t}^0V(B_{-T_{-s}})\ds+
\int_0^t V(B_{T_s})\ds 
$ by
$
\int_{-t}^ tV(B_{T_s}) \dB_s$. 
}\end{remark}
For later use we construct a functional integral representation  of the Green function of the form: 
\eq{green}
(F_0, \ee^{-(t_1-t_0)\PF}F_1\ee^{-(t_2-t_1)\PF}\cdots 
F_{n-1}\ee^{-(t_n-t_{n-1})\PF}F_n)_\hhh.
\en
\bc{green1}
Suppose  Assumptions \ref{ass1} and \ref{ass2}. 
Let  $V\in \vqf$.
Let $-\infty<t_0<t_1<\cdots<t_n<\infty$. For  $F_0,F_n\in\hhh$ and 
$F_j=F_j(x,\A(\rho_j))\in L^\infty(\BR)\otimes 
L^\infty(\Q)$, 
it follows that
\begin{align}
&(F_0, \ee^{-(t_1-t_0)\PF}F_1\ee^{-(t_2-t_1)\PF}\cdots 
F_{n-1}\ee^{-(t_n-t_{n-1})\PF}F_n)_\hhh\non \\
&=
\IIXX   \EX\lkkk
\lk
\JJ_0F_0(B_{T_{t_0}}), \lk \prod_{j=1}^{n-1}\tilde{F_j}\rk  
\ee^{-i\alpha\AA(\KKKK [t_0,t_n])}
\JJ_t F_n(B_{T_{t_n}})
\rk
\ee^{-\int_{t_0}^{t_n} V(B_{T_s})\ds}
\rkkk.
\end{align}
Here 
$\tilde{F_j}={F_j}(B_{T_{\tt j}}, \AA(\jj_{\tt j}(\rho_j)))$, $j=1,...,n-1$, and $T_s=-T_{-s}$ for $s<0$. 
In particular 
\begin{align}
&(f\otimes \one, \ee^{-(t_1-t_0)\PF}\one_{A_1}\ee^{-(t_2-t_1)\PF}\cdots 
\one_{A_{n-1}}\ee^{-(t_n-t_{n-1})\PF}g)_\hhh\non \\
&=
\IIXX   \EX\lkkk
\ov{f(B_{T_{t_0}})}
\lk \prod_{j=1}^{n-1}\one_{A_j}(B_{T_{t_j}})\rk
g(B_{T_{t_n}})\lk
\one,   
\ee^{-i\alpha\AA(\KKKK [t_0,t_n])}
\one
\rk
\ee^{-\int_{t_0}^{t_n} V(B_{T_s})\ds}
\rkkk.
\end{align}
\ec
\proof
Note that $F_j$, $j=1,...,n-1$, can be regarded as bounded operators. 
Thus the corollary can be proven in a similar manner to 
Theorem \ref{fkf2} and 
Corollary \ref{fkfshift}.
\qed

\section{Self-adjointness}
\label{sec4}
\subsection{Burkholder type inequalities}
In this section by using the functional integral  representation derived in 
Theorem \ref{fkf2}
we show the essential self-adjointness of $\PF$ for arbitrary values of coupling constants.  
To prove this  we find an invariant domain $D$ 
so that $D\subset \D(\PF)$ and 
$\ee^{-t\PF} D\subset D$. 
Then 
 $\PF$  is essentially self-adjoint on 
 $D$ by Proposition \ref{invariant}. 
Let $T$ be a self-adjoint operator. 
 The strategy is to estimate 
 the scalar product $(T F, \ee^{-t\PF} G)$ as 
$|(TF, \ee^{-t\PF} G)|\leq c(G,T)\|F\| $
for all $F,G\in \D(T)$ with some constant $c(G,T)$, which implies that 
$\ee^{-t\PF}G\in \D(T)$  for $G\in \D(T)$.

By the It\^o isometry  we have 
\begin{align}
\EX\lkkk
\|\one \otimes \omega(\p)^{\alpha/2}
\KKKK [0,t]\|^2_{\WW }
\rkkk
=d \EX\lkkk \int_0^{T_t}\|\omega(\p)^{\alpha/2}
\la(\cdot-B_r)\|_\LR^2\dr  \rkkk.
\end{align}
In particular 
\begin{align}
\EX\lkkk
\|\one\otimes \omega(\p)^{\alpha/2}\KKKK [0,t]\|^2_{\WW }
\rkkk
\leq d
\Ebb_\nu^0[T_t]\|\omega^{(\alpha-1)/2}
\vp\|_\LR^2
\end{align}
and the right-hand side above is finite in the case of $m>0$, 
since $\Ebb_\nu^0[T_t]<\infty$. 
We can also estimate 
$
\EX\lkkk
\|\one\otimes\omega(\p)^{\alpha/2}\KKKK [0,t]\|^{4}_{\WW}
\rkkk$.
\bl{bdg}
Suppose $m>0$. 
Then the Burkholder type inequalities  hold:
\begin{align}
\EX\lkkk
\|\one\otimes \omega(\p)^{\alpha/2}
\KKKK [0,t]\|^4_{\WW }\rkkk\leq C  
\|\omega^{(\alpha-1)/2}\vp\|^4_{\LR},
\end{align}
where $C$ is a constant.
 \el
\proof
It is known that by \cite[Theorem 4.6]{hir00b}
\begin{align}
\label{bdgformula}
\Ebb_{\P}^{x}
\lkkk
\left\|
\one\otimes \omega(\p)^{\alpha/2}\int_0^t \jj_s\la(\cdot-B_s) \dB_s^\mu
\right\|_{\LRR}^{2m}
\rkkk
\leq
\frac{(2m)!}{2^m}
t^m\|\omega^{(\alpha-1)/2}\vp\|_\LR ^{2m}.
\end{align}
Notice that 
$\d \KKKK [0,t] =\slimn \ott \sum_{j=1}^{2^n} a_j^\mu $ with 
$a_j^\mu=
 \int_{T_{t_{j-1}}}^{T_{t_j}}\!\!\jj_{t_{j-1}}
 \la_{\alpha}(\cdot-B_s) \dB_s^\mu\in \LRR$, and 
 $\la_{\alpha}= \omega(\p)^{\alpha/2}\la$ and 
 $\hat{\la}_{\alpha}=\omega^{(\alpha-1)/2}\vp$.
   We fix a $\mu$ and set   $a_j^\mu=a_j$ for simplicity. 
$a_j$ and $a_i$ are independent for $i\not=j$ and 
then 
we have 
\begin{align*}
&\EX
\lkkk 
\left\| \sum_{j=1}^{2^n} a_j
\right\|_\LRR^4 
\rkkk\\
&=\sum_{j,j'}\sum_{i,i'}
\EX\lkkk
\lk
\int_\BRR\!\!\!  a_j(x) a_{j'}(x)  \dx  
\rk
\lk
\int_\BRR\!\!\!  a_i(y)  a_{i'}(y) \dy  
\rk
\rkkk\\
&=
\sum_{j=1}^{2^n} \EX\lkkk
\lk
\int_\BRR\!\!\!  a_j(x)^2  \dx \rk^2
\rkkk
+
\sum_{j=1}^{2^n} \EX\lkkk
\int_\BRR\!\!\!  a_j(x)^2  \dx 
\rkkk
\sum _{i\not =j}
 \EX\lkkk
\int_\BRR\!\!\!  a_i(x)^2  \dx 
\rkkk\\
&
\quad +
\sum_{j=1}^{2^n} 
\sum_{i\not=j}
\EX\lkkk
\int_\BRR\!\!\!  a_j(x)a_i(x)   \dx 
\int_\BRR\!\!\!  a_j(y) a_i(y) \dy  
\rkkk.
\end{align*}
We estimate the first term of the right-hand side above. 
We have by \kak{bdgformula}
\begin{align*}
&\sum_{j=1}^{2^n} \EX
\lkkk
\lk
\int a_j(x)^2  \dx 
\rk^2
\rkkk
=
\sum_{j=1}^{2^n} \EX
\lkkk
\|a_j\|^4
\rkkk
\leq 
6\|\omega^{(\alpha-1)/2}\vp\|^4
  \sum_{j=1}^{2^n} \Ebb_\nu^0\lkkk \left|
 T_{\frac{t}{2^{n+1}}}\right|^2\rkkk.
 \end{align*}
By using the distribution \kak{distribution} of $T_t$ and the assumption 
$m>0$ 
we have 
 \begin{align*}
&\sum_{j=1}^{2^n} \EX
\lkkk
\lk
\int a_j(x)^2  \dx 
\rk^2
\rkkk\\
&\leq 
6\|\omega^{(\alpha-1)/2}\vp\|^4 
 \frac{t}{2\sqrt{2\pi}} 
 \ee^{\frac{mt}{2^{n+1}}}
 \int_0^\infty 
  \sqrt s\exp\lk
-\half
\lk\frac{(\frac{t}{2^{n+1}})^2}{s}+m^2 s\rk\rk 
\ds.
\end{align*}
The right-hand side  converges to 
\begin{align*}
\frac{3t}{\sqrt{2\pi}}
\|\omega^{(\alpha-1)/2}\vp\|^4 
 \int _0^\infty 
 \sqrt s
 \exp
 \lk
-\half  m^2 s  \rk \ds
\end{align*}
as $n\to\infty$. 
The second term is estimated as 
\begin{align*}
\sum_{j=1}^{2^n} \EX\lkkk
\int a_j(x)^2  \dx 
\rkkk
\sum _{i\not =j}
 \EX\lkkk
\int a_j(x)^2  \dx 
\rkkk
\leq
\lk
\sum_{j=1}^{2^n} \EX\lkkk
\int a_j(x)^2  \dx 
\rkkk
\rk^2.
\end{align*}
By the It\^o isometry we have 
$$ \sum_{j=1}^{2^n}
 \EX\lkkk
\int a_j(x)^2  \dx 
\rkkk
=\EX 
\lkkk
\int_0^{T_t}\|\jj_s\la_{\alpha}(\cdot-B_s)\|^2\ds
\rkkk
\leq 
\Ebb_\nu^0[T_t]\|\omega^{(\alpha-1)/2}\vp\|^2.$$
Hence 
\begin{align*}
&\sum_{j=1}^{2^n} \EX\lkkk
\int a_j(x)^2  \dx 
\rkkk
\sum _{i\not =j}
 \EX\lkkk
\int a_j(x)^2  \dx 
\rkkk
\leq
(\Ebb_\nu^0[T_t])^2 
\|\omega^{(\alpha-1)/2}\vp\|^4.
\end{align*}
Finally 
we estimate the third term. 
We see that 
\begin{align*}
\sum_{j=1}^{2^n} 
\sum_{i\not=j}
\EX
\lkkk
\int \! a_j(x)a_i(x)   \dx 
\int \! a_j(y) a_i(y) \dy  
\rkkk
\leq
\int_\BRR\!\!\! \!\!  \dx
\int_\BRR\!\!\! \!\! \dy  
\left|
 \sum_{j=1}^{2^n}
\EX
\lkkk
a_j(x)
a_j(y)
\rkkk
\right|^2. 
\end{align*}
Note 
that 
$\EX
\lkkk
a_j(x)a_j(y)
\rkkk
=
\EX\lkkk
\int_\TT{j-1}^\TT{j}
A_s(x,j) A_s(y,j)\ds \rkkk$, where 
we set $A_s(x,j)=
 (\jj_\tt {j-1}
 \la_{\alpha}(\cdot-B_s))(x)$.
By the Schwarz  inequality we have 
\begin{align*}
&
\leq
\int_\BRR\!\!\! \!\!  \dx
\int_\BRR\!\!\! \!\! \dy  
\EX
\lkkk
\lk
 \sum_{j=1}^{2^n}
\int_{\TT{j-1}}^{\TT{j}}A_s(x,j) A_s(y,j)\ds \rk^2\rkkk\\
&\leq
\int_\BRR\!\!\! \!\!  \dx
\int_\BRR\!\!\! \!\! \dy  
\EX
\lkkk
\lk
 \sum_{j=1}^{2^n}
\int_{\TT{j-1}}^{\TT{j}}A_s(x,j)^2 \ds\rk
\lk
 \sum_{j=1}^{2^n}
\int_{\TT{j-1}}^{\TT{j}}A_s(y,j)^2 \ds\rk\rkkk
\end{align*}
and the Fubini's lemma yields that 
\begin{align*}
&
=
\EX
\lkkk
\int_\BRR\!\!\! \!\!  \dx
\lk
 \sum_{j=1}^{2^n}
\int_{\TT{j-1}}^{\TT{j}}A_s(x,j)^2 \ds\rk
\int_\BRR\!\!\! \!\! \dy  
\lk
 \sum_{j=1}^{2^n}
\int_{\TT{j-1}}^{\TT{j}}A_s(y,j)^2 \ds\rk\rkkk\\
&
=
\EX
\lkkk
T_t^2\|\omega^{(\alpha-1)/2}\vp\|^4
\rkkk
=
\Ebb_\nu^0[T_t^2]\|\omega^{(\alpha-1)/2}\vp\|^4.
\end{align*}
Note that 
$\Ebb_\nu^0[T_t^n]=\frac{te^{tm}}{\sqrt{2\pi}}\int_0^\infty \frac{s^n}{s^{3/2}}\exp\lk -\half\lk \frac{t^2}{s}+m^2s\rk\rk ds<\infty$ for $n\geq0$. 
Then the lemma follows. 
\qed

\subsection{Invariant domain and  self-adjointness}
Let $\tot_\mu=\p_\mu\otimes\one+\one\otimes\pf_\mu$ be the total momentum 
operator in $\hhh$. 
\bl{invariant2}
Let $V=0$. Then 
$
\ee^{-it\tot_\mu }\ee^{-s \PF}\ee^{it\tot_\mu }=\ee^{-s\PF}$.
\el
\proof
By the Feynman-Kac type formula we have 
\begin{align*}
\!\!(F, \ee^{-it\tot_\mu }\ee^{-s \PF}\ee^{it\tot_\mu }G)
\!=
\!\!\!\IIXX \Ebb_{\P\times\nu}^{x,0}
\!\!\lkkk (\JJ_0 F(B_{T_0}), 
\ee^{-it\p_\mu}
\ee^{-it\ov{\pf}_\mu}
\ee^{-i\AA(\KKKK[0,t])}
\ee^{it\ov{\pf}_\mu}
\ee^{it\p_\mu}
\JJ_t G(B_{T_t}))
\rkkk.
\end{align*}
Since $\ee^{-it\p_\mu}
\ee^{-it\ov{\pf}_\mu}
\ee^{-i\AA(\KKKK[0,t])}
\ee^{it\ov{\pf}_\mu}
\ee^{it\p_\mu}
=\ee^{-i\AA(\KKKK[0,t])}
$, the lemma follows. 
\qed

\bl{lemmamomentuminvariance}
Suppose Assumptions \ref{ass1} and \ref{ass2}. 
Let $V=0$. 
For 
 $F\in \D(\p_\mu)$
and $G\in \D(\p_\mu)\cap \D(\hf^\han)$ 
it follows that
 \begin{align}\label{pbound}
(\p_\mu F, \ee^{-t\PF} G)\leq 
C
\lk
(\|\sqrt{\omega} \vp\|+\| \vp  \|)
\|(\hf+\one)^\han G\|
+
 \|\p_\mu G\|\frac{}{}
 \rk
 \|F\|.
\end{align}
\el
\proof
Notice that 
 $(\ee^{is \p_\mu} F, \ee^{-t\PF}G)=
(\ee^{-is \pf _\mu} F, \ee^{-t\PF}\ee^{-is \tot_\mu}G)$.  
Then 
\begin{align}\label{fkfp}
(\ee^{is\p_\mu} F, \ee^{-t\PF} G)=
\IIXX   \EX\lkkk
\lk
\JJ_0 
F(B_{T_0}), 
\ee^{+is \ov{\pf}_\mu} \ee^{-i\alpha\AA(\KKKK )}
\ee^{-is \ov{\pf}_\mu} 
\JJ_t \ee^{-is\p_\mu} G(B_{T_t})
\rk
\rkkk.
\end{align}
Here and in what follows in this proof we set $\KKKK =\bigoplus_\mu ^ d \KKKK ^\mu=\KKKK [0,t]$. 
We see that 
$\ee^{+is \ov{\pf}_\mu} \ee^{-i\alpha\AA(\KKKK )}
\ee^{-is \ov{\pf}_\mu} =
\ee^{-i\alpha\AA(\ee^{is(\one\otimes \p_\mu)}\KKKK )}$. 
Take the derivative at $s=0$ on both sides of \kak{fkfp}.
We  have 
\begin{align}
(i\p_\mu F, \ee^{-t\PF} G)
&=
\IIXX   \EX\lkkk
\lk
\JJ_0  
F(B_{T_0}), -i\alpha{\AA}_\mu (i\p_\mu \KKKK ^\mu )
\ee^{-i\alpha\AA(\KKKK )}\JJ_t G(B_{T_t})
\rk
\rkkk\non\\
&
\label{fkfa}
+
\IIXX   \EX
\lkkk
\lk
\JJ_0 
F(B_{T_0}), \ee^{-i\alpha\AA(\KKKK )}
\JJ_t (-i\p_\mu  G)(B_{T_t})
\rk
\rkkk.
\end{align}
It is trivial to see that 
\begin{align*}
\left|
\IIXX   \EX\lkkk
\lk
\JJ_0  
F(B_{T_0}), 
\ee^{-i\alpha\AA(\KKKK )}\JJ_t 
(-i\p_\mu
G)(B_{T_t})
\rk
\rkkk\right|
\leq
\|F\|\|\p_\mu G\|.
\end{align*}
We  can estimate the first  term on the right-hand side of \kak{fkfa} as 
\begin{align*}
&\left|
\IIXX   \EX\lkkk
\lk
\JJ_0  
F(B_{T_0}), {\AA}_\mu (i\p_\mu \KKKK ^\mu )
\ee^{-i\alpha{\AA}_\mu (\KKKK ^\mu )}\JJ_t G(B_{T_t})
\rk
\rkkk\right|\\
&\leq
\IIXX   \EX\lkkk
\|
{\AA}_\mu (i\p_\mu \KKKK ^\mu)\JJ_0  
F(B_{T_0})\|
\|
\JJ_t G(B_{T_t})\|
\rkkk
\end{align*}
By the bound 
$\|{\AA}_\mu (f) \Phi\|\leq C (\|\hat f\|+\| \hat f/\sqrt\omega \|)\|(\hf+\one)^\han \Phi\|$ with some constant $C>0$, 
we have 
\begin{align*}
&\leq
C\IIXX   \EX\lkkk
(\|\p_\mu \KKKK \|+\|\omega(\p)^{-\han}\p_\mu \KKKK \|)
\|
(\hf+\one)^\han  
F(B_{T_0})\|
\|
 G(B_{T_t})\|
\rkkk
\end{align*}
and by the Schwarz  inequality,   
\begin{align*}
&\leq
C\lk
\IIXX   
\EX
\lkkk
(\|\omega(\p)\KKKK ^\mu \|+\|\omega(\p)^{\han} \KKKK ^\mu \|)
^2
\rkkk 
\|(\hf+\one)^\han  F(x)\|^2
\rk^\han
\|G\|\\
&\leq
C(\|\omega^\han \vp\|+\|\vp\|)
\|(\hf+\one)^\han  F\|
\|G\|.
\end{align*}
Then the lemma follows. 
\qed
We define the momentum conjugate of $\AA(f)$ by 
$\Pi_{\rm E}(f)=i [\ov{\hf}, \AA(f)]$ in the function space.

\bl{freefieldinvariance}
Suppose Assumptions \ref{ass1} and \ref{ass2}. 
Let $V=0$. 
Then for  $F,G\in \D(\hf)$ it follows that 
\begin{align*}
&(\hf F, \ee^{-t\PF} G)\\
&\leq 
\lk\frac{}{}
\|\hf G\|+|\alpha|(\|\sqrt\omega \vp\|+\|\vp\|) 
\|(\hf+\one)^\han G\|
+|\alpha|^2\|\vp/\sqrt\omega\|^2\|G\|\rk
 \|F\|.
\end{align*}
\el 
\proof
By the Feynman-Kac type formula we have 
\begin{align*}
(\hf F, \ee^{-t\PF} G)
=
\IIXX   \EX\lkkk
\lk
\JJ_0 F(B_{T_0}), \ee^{-i\alpha\AA(\KKKK[0,t] )}
S \JJ_t G(B_{T_t})
\rk
\rkkk,
\end{align*}
where 
$
S=\ee^{i\alpha\AA(\KKKK[0,t] )}
\ov{\hf} 
\ee^{-i\alpha\AA(\KKKK [0,t])}=\ov{\hf} -\alpha \Pi_{\rm E}(\KKKK[0,t] )+
\alpha^2 g
$
 with the constant 
 $g=\qq(\KKKK [0,t])$.
 It is trivial to see that 
\begin{align}
&
\left|
\IIXX   \EX\lkkk
\lk
\JJ_0 F(B_{T_0}), \ee^{-i\alpha\AA(\KKKK[0,t] )}
\hf 
 \JJ_t G(B_{T_t})
\rk
\rkkk
\right|
\leq \|F\| \|\hf G\|.
\end{align}
In the same way as the estimate of the first term of the right-hand side of \kak{fkfa} 
we can see that 
\begin{align*}
&\left|
\IIXX   \EX\lkkk
\lk
\JJ_0 F(B_{T_0}), \ee^{-i\alpha\AA(\KKKK[0,t] )}
\Pi_{\rm E}(\KKKK[0,t] )
 \JJ_t G(B_{T_t})
\rk
\rkkk
\right|\\
&
\leq  C(\|\sqrt\omega \vp\|\!+\!\|\vp\|)\|F\|\|(\hf+\one)^\han G\|
\end{align*}
with some constant $C>0$. 
Here we used the fundamental bound 
$\|{\Pi_{\rm E}}_\mu(f)\Phi\|
\leq C \lk\|\sqrt \omega \hat f\|+\|\hat f\|\rk
\|(\hf+\one)^\han\Phi\|$ and Lemma \ref{bdg}.
Finally 
we see that $g\leq  C \|\KKKK [0,t]\|_\WW^2$ and 
by Lemma \ref{bdg} again,  
\begin{align}
&\left|
\IIXX   \EX\lkkk
\lk
\JJ_0 F(B_{T_0}), \ee^{-i\alpha\AA(\KKKK[0,t] )}
g 
 \JJ_t G(B_{T_t})
\rk
\rkkk
\right|\non\\
&\leq 
C \lk
\IIXX   \EX\lkkk
\|\KKKK [0,t]\|_\WW^4\rkkk \| F(x)\|^2 
\rk^\han
\|G\|
\leq 
C  \|\vp/\sqrt\omega\|^2 \|F\|\|G\|.
\end{align}
Then the lemma follows. 
\qed

\bt{essentialselfadjointnesstheorem}
\TTT{Essential self-adjointness}
Let $V\in \vsa$. 
Suppose  
that 
$m>0$, and 
Assumptions \ref{ass1} and \ref{ass2} hold. 
Then 
 $\PF$ is essentially self-adjoint on $\D(|\p|)\cap \D(\hf)$.
\et
\proof
Suppose $V=0$. 
Let $F\in \CCC\otimes\ffff$. Then we see that
\begin{align*}
\|(\K)F\|^2
\leq C_1\| |\p|  F\|^2+C_2\|\hf F\|^2+ C_3\|F\|^2
\end{align*}
with some constants $C_1,C_2$ and $C_3$. 
Since $\CCC\otimes \ffff$ is a core of $|\p|+\hf$, 
\eq{one}
\D(\K)\supset \D(|\p|)\cap \D(\hf)
\en
 follows from a limiting argument.
By Lemmas \ref{lemmamomentuminvariance} and \ref{freefieldinvariance}, 
we also see that 
\eq{two}
\ee^{-t(\K)}
\lk
 \D(|\p|)\cap \D(\hf)\rk
\subset 
\lk
 \D(|\p|)\cap \D(\hf)\rk.
 \en
\kak{one} and \kak{two} 
imply that 
$\K$  is essentially self-adjoint on $\D(|\p|)\cap \D(\hf)$ by 
Proposition \ref{invariant}. 
Next we suppose that $V$ satisfies assumptions in the theorem.
By Lemma \ref{relativebound}, 
 $V$ is also relatively bounded with respect to $\K$ with a relative bound strictly smaller than one. 
Then the theorem follows by the Kato-Rellich theorem. 
\qed
Furthermore in Hidaka and Hiroshima \cite {hh13} 
the self-adjointness of $\PF$ for arbitrary 
$m\geq 0$ is established. 
The key inequality is as follows. 
\bl{hidaka}
Suppose  that $m>0$, and 
Assumptions 
\ref{ass1} and \ref{ass2} hold.
Let $V=0$. 
Then 
there exists a constant $C$ such that 
\eq{hidaka1}
\||\p | F\|^2+\|\hf F\|^2\leq C\|(\K+\one)F\|^2
\en
for all $F\in D(|\p|)\cap D(\hf)$.
\el
\proof See \cite[Lemma 2.7]{hh13}.
\qed

\bt{sasa}\TTT{Self-adjointness \cite{hh13}}
Suppose  that $m\geq0$, and 
Assumptions 
\ref{ass1} and \ref{ass2} hold.
Let $V\in \vsa$.
Then $\PF$ is self-adjoint on 
$D(|\p|)\cap D(\hf)$.
\et
\proof
We show an outline of the proof. See \cite{hh13} for detail. 
Suppose that $V=0$ and $m>0$.
We write $\PFF _m$ for $\PF$ to emphasize  $m$-dependence. 
By \kak{hidaka1}, $\PFF _m\lceil_{D(|\p|)\cap D(\hf)}$ is closed on $D(|\p|)\cap D(\hf)$. 
Then $\PFF _m$ is self-adjoint on 
$D(|\p|)\cap D(\hf)$. Note that 
$\PFF _0=\PFF _m+(\PFF _0-\PFF _m)$ and $\PFF _0-\PFF _m$ is bounded. Then 
$\PFF _0$ is also self-adjoint on 
$D(|\p|)\cap D(\hf)$ for $V=0$.
Finally  
let $V\in \vsa$.
Then $V$ is also 
relatively bounded with respect to 
 $\PFF _m$ with a 
 relative bound
 strictly smaller  than one.
 Then the theorem follows from Kato-Rellich theorem. 
\qed

\begin{example}
\label{hydrogen}
\TTT{Hydrogen like atom}{\rm 
Let $d=3$.
A  spinless hydrogen like atom is defined by 
introducing the Coulomb potential $V_{\rm Coulomb}(x)=-g/|x|$, $g>0$,  
which is relatively form bounded with respect to $\sqrt{\p^2+m^2}$ 
with a relative bound strictly smaller  than one if  $g\leq 2/\pi$ by \cite{her77} 
(see also 
 \cite[Theorem 2.2.6]{be11}).  
Furthermore if $g<1/2$, $V_{\rm Coulomb} $ is relatively bounded with respect to 
$\sqrt{\p^2+m^2}$ with  a relative bound strictly smaller  than one.   
 Let $\A_\Lambda $ be the quantized radiation field with  
the cutoff function $\vp(k)=\one_{|k|<\Lambda}(k)/\sqrt{(2\pi)^3}$, where $\Lambda>0$ 
describes  a UV cutoff parameter. 
By Lemma \ref{relativebound}, when $g<2/\pi$, $V$ is relatively form bounded with respect to $\K$ and 
$\PF$ is well defined as a self-adjoint operator. 
 Furthermore 
by Theorem \ref{sasa}
when $g<1/2$,  
 $\PF$ is self-adjoint on $D(|\p|)\cap D(\hf)$.
 All the statements mentioned above are true for arbitrary values of $\alpha\in\RR$ and $\Lambda>0$.}
\end{example}

\section{Martingale properties and fall-off of bound states}
\label{sec5}
\subsection{Semigroup and  relativistic Kato-class potential}
In this subsection we define the self-adjoint operator $\PFK$ 
with a potential $V$ in the so-called relativistic Kato-class through the Feynman-Kac type formula.
Let us define the relativistic Kato-class.
\begin{definition}
{\rm 
\label{VK}
\TTT{Relativistic Kato-class}
(1)
Potential $V$ is in the  relativistic Kato-class 
if and only if 
\begin{align}
\label{katoclass}
\sup_x\EX\lkkk \ee^{\int_0^t V(B_{T_s}) \ds }\rkkk<\infty.
\end{align}
(2)
$V=V_+-V_-$ is in $\vk $ if and only if 
$V_+\in L_{\rm loc}^1(\BR)$ and $V_-$ is in relativistic Kato-class.
}\end{definition}
The property \kak{katoclass} is used in the proofs of Lemmas \ref{sup} and  \ref{bound2}, and Corollary \ref{martingale2}. 
When $V\in \vk $,
 we can see that 
$$r_t(F,G)=\IIXX  \EX
\lkkk\lk \JJ_0 F(B_{T_0}), \ee^{-i\alpha\AA(\KKKK [0,t])}\ee^{-\int_0^t V(B_{T_r})\dr} \JJ_t G(B_{T_t})\rk\rkkk $$ 
is well defined for all $F,G\in \hhh$, and 
$|r_t(F,G)|\leq c_t \|F\| \|G\|$ follows with some constant $c_t$. 
Then the Riesz representation theorem yields that 
there exists a bounded operator $S_t$ such that 
$r_t(F,G)=(F, S_t G)$ for $F,G\in\hhh$ and $\|S_t\|\leq c_t$. 
By the Feynman-Kac type formula \kak{fkf2} 
we indeed see that 
$
(S_t G)(x)=\Ebb_{\P\time\nu}^{x,0}\lkkk 
\QTT  
G(B_{T_t})\rkkk$, 
where 
\eq{qzt}
\QTT =
\JJ_0^\ast \ee^{-\int_0^t V(B_{T_r}) \dr} \ee^{-i\alpha\AA(\KKKK [0,t])}\JJ_t.\en
\bt{semigroup}
Let $V\in \vk $.
Suppose Assumption \ref{ass1}. 
Then 
$S_t$, $t\geq0$, is a  strongly continuous  one-parameter symmetric semigroup. 
\et
\begin{definition}
{\rm 
\TTT{Definition of $\PFK$}
\label{vk}
Let $V\in \vk $.
Suppose Assumption \ref{ass1}. 
The unique self-adjoint generator of $S_t, t\geq 0$,  is  denoted by 
$\PFK$, i.e., $S_t=\ee^{-t\PFK}$, $t\geq0$.
}\end{definition}
\begin{remark}
{\rm 
Note that 
\eq{comp}
\vsa\subset\vqf,\quad \vk \subset \vqf.
\en
It is easy to see that  $\vsa\subset \vqf$. 
See Appendix \ref{appK} for 
the inclusion $\vk \subset \vqf$.  
We give a remark on the difference between 
$\PF$ and $\PFK$. 
In order to define $\PF$ 
we need Assumptions \ref{ass1} and 
\ref{ass2}, 
an extra Assumption \ref{ass2} 
is, however,  not needed to define 
$\PFK$.
}\end{remark}
In order to prove  Theorem \ref{semigroup} we need several lemmas:

\bl{semigroup1}
Let $V\in \vk $.
Suppose Assumption \ref{ass1}.
Then $S_t, t\geq 0$, satisfies the  semigroup property, i.e., 
$S_sS_t=S_{s+t}$ for all $s,t\geq0$.
\el
\proof 
We have 
$(F, S_sS_t G)=\lk 
F, 
\EX\lkkk 
\QSS 
\Ebb_{\P\times\nu}^{B_{T_s},0}
\lkkk \QTT  
G(B_{T_t})
\rkkk
\rkkk \rk$. 
By  Lemma  \ref{mar} 
we show that 
\eq{mar1}
\EX\lkkk 
\QSS \Ebb_{\P\times\nu}^{B_{T_s},0}
\lkkk \QTT  G(B_{T_t})\rkkk\rkkk 
=\EX\lkkk 
\QSS 
\JJ_0^\ast \ee^{-\int_s^{s+t} V(B_{T_r})\dr}
 \ee^{-i\alpha \AA(\ima [s,s+t])} \JJ_t  G(B_{T_{s+t}}) 
\rkkk,
\en
where 
\eq{k2-1}
\ima [s,s+t]=
\slimn 
\ott \sum_{j=1}^{2^n}\int _{T_{\frac{t}{2^n}(j-1)+s}}
^{T_{\frac{t}{2^n}j+s}}
\jj_{\frac{t}{2^n}(j-1)}\la(\cdot-B_r) \dB_r^\mu.
\en
Since 
it is obtained that 
\begin{align*}
&\QSS 
\JJ_0^\ast \ee^{-\int_s^{s+t} V(B_{T_r})\dr}
 \ee^{-i\alpha \AA(\ima [s,s+t])} \JJ_t  G(B_{T_{s+t}}) \\
&=
\JJ_0^\ast 
 \ee^{-\int_0^{s+t} V(B_{T_r})\dr}
\ee^{-i\alpha \AA(\KKKK [0,s])} 
\JJ_s
\JJ_0^\ast 
\ee^{-i\alpha \AA(\ima [s,s+t])} 
\JJ_t  G(B_{T_{s+t}}) 
\end{align*}
and 
$\JJ_s
\JJ_0^\ast 
={\rm U}_s \JJ_0\JJ_0^\ast={\rm U}_s E_s$, we have 
\begin{align*}
(F, S_s S_t G)
=
\lk F, 
\EX\lkkk 
\JJ_0^\ast 
\ee^{-i\alpha \AA(\KKKK [0,s])} 
{\rm U_s}E_s
\ee^{-i\alpha \AA(\ima [s,s+t])} 
 \ee^{-\int_0^{s+t} V(B_{T_r})\dr}
\JJ_t  G(B_{T_{s+t}}) \rkkk\rk.
\end{align*}
By the Markov property of projection $E_s$,  $E_s$ can be deleted, and 
${\rm U}_s$ satisfies that 
${\rm U_s}
\ee^{-i\alpha \AA(\ima [s,s+t])} 
\JJ_t  G(B_{T_{s+t}}) =
\ee^{-i\alpha \AA(\KKKK [s,s+t])} 
 \JJ_{s+t}  G(B_{T_{s+t}})$. Then 
by Proposition \ref{linearity}  we have 
\begin{align*}
&
(F, S_s S_t G)=
\lk F, 
\EX\lkkk 
\JJ_0^\ast 
 \ee^{-\int_0^{s+t} V(B_{T_r})\dr}
\ee^{-i\alpha \AA(\KKKK [0,s+t])} 
\JJ_{s+t}  G(B_{T_{s+t}}) \rkkk\rk=
(F, S_{s+t} G).
\end{align*}
 Then the semigroup property, $S_sS_t=S_{s+t}$, follows. 
\qed
\bl{semigroup2}
Let $V\in \vk $.
Suppose Assumption \ref{ass1}.
Then $S_t$, $t\geq 0$,  is strongly continuous in  $t$ and 
$\d \slim_{t\to0} S_t=\one$.
\el
\proof
It  is enough to show that $(F, S_t G)\to (F,G)$ as $t\to 0$ for 
 $F,G\in \CCC\otimes \ffff$.
 Let $F=f\otimes\Psi$ and $G=g\otimes \Phi$. Since $V\in \vk$,
 we have 
 \begin{align*}
| (F, (S_t-\one)  G)|
\leq
C\|F\|_\hhh 
\lkk 
\int \dx
\EX
\lkkk\|(\ee^{-i\alpha\AA(\KKKK [0,t])}-1)g(B_{T_t})\Phi\|^2\rkkk\rkk^\han.
\end{align*}
Since $g\in \CCC$, $|g(x)|\leq a\one_K(x)$ with some $a$ and 
a compact domain $K\subset\BR$, we have
\begin{align*}
|(F, (S_t-\one) G)|\leq 
aC\|F\|_\hhh 
\lkk 
\int_K \dx
\EX\lkkk\|(\ee^{-i\alpha\AA(\KKKK [0,t])}-1)\Phi\|^2\rkkk\rkk^\han.
\end{align*}
By 
the bound 
$
\Ebb\lkkk\|(\ee^{-i\alpha\AA(\KKKK [0,t])}-1)\Phi\|^2\rkkk
\leq
|\alpha|\|\KKKK [0,t]\| \|({\rm N}+\one)^\han \Phi\|$,
we have 
\begin{align*}
(F, (S_t-\one) G)|
&\leq
|\alpha| aC\|F\|_\hhh 
\lkk 
\int_K \dx
\EX
\lkkk \|\KKKK [0,t]\|^2 \rkkk 
\|({\rm N}+\one)^\han\Phi\|
\rkk^\han\\
&\leq
\sqrt 
t
 a|\alpha|
 C
 \|\vp/\sqrt\omega\|
\|F\|_\hhh 
\lk
\int_K \dx
\rk^\han 
\|({\rm N}+\one)^\han\Phi\|.
\end{align*}
Then 
$|(F, (S_t-\one) G)|
\to 0$ as $t\to 0$ follows. 
\qed
\bl{semigroup3}
Let $V\in \vk $.
Suppose Assumption \ref{ass1}.
Then
$S_t$, $t\geq 0$,  is symmetric, i.e., $S_t^\ast=S_t$ for all $t\geq0$.
\el
\proof
Recall that 
 ${\rm R}=\Gamma(r)$ is the second quantization of  the reflection $r$.
We have 
\begin{align*}
(F, S_tG)
=
\int \dx\EX\lkkk 
\ee^{-\int_0^t V(B_{T_r})\dr}
\lk  \JJ_0 F(B_{T_0}),\ee^{-i\alpha\AA(r \KKKK [0,t])}
\JJ_{-t} G(B_{T_t})
\rk
\rkkk,
\end{align*}
and by the time-shift ${\rm U}_t=\Gamma(u_t)$,
\begin{align*}
=
\int \dx\EX\lkkk 
\ee^{-\int_0^t V(B_{T_r})\dr}
\lk 
 \JJ_t  F(B_{T_0}),
  \ee^{-i\alpha\AA(u_t r \KKKK [0,t])}
 \JJ_{0} G(B_{T_t})
\rk
\rkkk.
\end{align*}
Notice that 
$\d u_t r \KKKK [0,t]=\limn  \sum_{j=1}^{2^n} \int_{\TT{j-1}}^{\TT{j}} \jj_{t-\tt{j-1}} 
\la(\cdot-B_s)\dB_s^\mu$.
Exchanging integrals $\int \dP^0$ and $\int \dx$ and 
changing the variable $x$ to $y-B_{T_t}$, we can have 
\begin{align}
\label{toshiba}
= \EZ \lkkk
\int \dy
\ee^{-\int_0^t V(B_{T_r}-B_{T_t}+y)\dr}
\lk
\JJ_t F(y-B_{T_t}), 
\ee^{-i\alpha\AA(u_t r \tilde \KKKK [0,t])}\JJ_0 G(y)\rk
\rkkk,
\end{align}
where 
$u_t r \d \tilde \KKKK [0,t]
=\limn  \sum_{j=1}^{2^n} \int_{\TT{j-1}}^{\TT{j}} \jj_{t-\tt{j-1}} 
\la(\cdot-(B_s-B_{T_t}+y))d B_s^\mu$.
By Lemma \ref{lemma0},
we can see that 
\begin{align}
\label{toshiba2}
\kak{toshiba}=\int \dy
\Ebb_{\P\times\nu}^{0,y}
\lkkk
\ee^{-\int_0^t V(B_{T_r})\dr}
\lk
\JJ_0^\ast \ee^{-i\alpha\AA(\KKKK [0,t])}\JJ_t F(y+B_{T_t}), 
G(y)\rk
\rkkk
=(S_t F, G).\non
\end{align}
Then the lemma follows. 
\qed

{\it Proof of Theorem \ref{semigroup}}

Lemmas \ref{semigroup1}-\ref{semigroup3} yield that $S_t$ is 
symmetric and strongly continuous one-parameter  semigroup. 
Then there exists the unique self-adjoint operator such that $S_t=\ee^{-t\PFK}$ by a semigroup version of the Stone theorem \cite[Proposition 3.26]{lhb11}.
\qed

\subsection{Martingale properties}
Let $\bou$ be a bound  state of $\PFK$ and $E\in\RR$ the eigenvalue 
associated with $\bou$:
$$\PFK \bou=E\bou.$$
In this section we study  the spatial decay of $\|\bou(x)\|_{L^2(\Q)}$ as $|x|\to\infty$. 
In order to do that we show the martingale property of 
the stochastic process $(\xt )_{t\geq 0}$: 
\begin{align}
\xt =\ee^{tE}\ee^{-\int_0^t V(B_{T_s}+x)\ds }\ee^{-i\alpha\AA(\KKKK ^x[0,t])}
\JJ_t \bou(B_{T_t}+x),\quad t\geq 0,
\end{align}
on $\Omega_{\P}
\times \Omega_\nu\times\QE$.
Here $\KKKK ^x[0,t]$ is defined by $\KKKK [0,t]$ with $B_s$ replaced by $B_s+x$, i.e., 
$\KKKK ^x[0,t]=\bigoplus_{\mu=1}^d \int_0^{T_t}
\jj_{T^\ast _s}\la(\cdot-B_s-x)\dB_s^\mu$. 
Using the stochastic process $( \xt )_{t\geq 0}$, 
bound state $\bou$ 
can be  represented as 
\begin{align}
\bou(x)=\EZ\lkkk \JJ_0^\ast \xt \rkkk
\end{align}
for arbitrary  $t\geq0$. 
We can also obtain 
that 
$\d (u\otimes \Phi, \bou)
=(u\otimes \Phi, \ee^{-t(\PFK-E)}\bou)
\!=
\!\int_\BR \dx \ov{u(x)}
\EZ\Ebb_{\mu_{\rm E}}
\lkkk \JJ_0\Phi\cdot 
\xt \rkkk$. Then   we have 
$(\Phi, \bou(x))_{L^2(\Q)}=
\EZ\Ebb_{\mu_{\rm E}}\lkkk 
\JJ_0\Phi\cdot 
\xt \rkkk$.
\bl{sup}
Let $V\in \vk $.
Suppose Assumption \ref{ass1}.  
Then 
$\|\bou(\cdot)\|_{L^2(\Q)}\in L^\infty(\BR)$. 
\el
\proof
By $\bou(x)=\EZ\Ebb_{\mu_{\rm E}}
\lkkk \JJ_0^\ast \xt \rkkk$ 
for arbitrary $t>0$, we have 
\begin{align*}
\|\bou(x)\|_{L^2(\Q)}
\leq
\ee^{tE}
\lk
\EZ\lkkk \ee^{-2\int_0^tV(B_{T_s}+x)\ds}\rkkk\rk
^\han 
\lk
\EZ
\lkkk \| \bou(B_{T_t}+x)\|^2 
\rkkk\rk^\han.  
\end{align*}
We have 
$\sup_{x\in\BR}
\EZ\lkkk \ee^{-2\int_0^tV(B_{T_s}+x)\ds}\rkkk
<\infty$, 
 since $V$ is relativistic Kato-class, and 
$$\EZ
\lkkk  
\| \bou(B_{T_t}+x)\|^2 \rkkk 
=\int_\BR \dy
\int_0^\infty {\rm d}s  \frac{\rho_t(s)\ee^{-|y|^2/(2s)}}{(2\pi s)^{d/2}}  \|\bou(x+y)\|^2 
\leq C\|\bou\|^2_\hhh.$$
 Then $\sup_{x\in\BR}\|\bou(x)\|^2\leq C\|\bou\|^2_\hhh$ follows. 
 \qed
\bc{martingale2}
It follows that 
$
\EZ\Ebb_{\mu_{\rm E}}
\lkkk
\|\xt \|_{L^2(\Q)}
\rkkk<\infty$ for all $x\in\BR$.
\ec
\proof
This follows from Lemma \ref{sup}.
\qed
We define a filtration under which $(\xt )_{t\geq 0}$ is martingale.  
Let 
\eq{f1}
{\cal F}_{[0,t]}^{(1)}=\lkk \left. \bigcup_{w_1\in \Omega_\nu}
(A(w_1),w_1)\right|A(w_1)\in \s(B_r, 0\leq r\leq T_t(w_1))
\rkk\subset \calb _\P\times \calb _\nu
\en
and
\eq{f2}
{\cal F}_{[0,t]}^{(2)}=\lkk\left. \bigcup_{w_2\in \Omega_{\P}
}(w_2, B(w_2))
\right|B(w_2)\in \s(T_r, 0\leq r\leq t)\rkk
\subset \calb _{\P}\times \calb _\nu.
\en
Then we set  
$
{\cal F}_{[0,t]}={\cal F}_{[0,t]}^{(1)}\cap {\cal F}_{[0,t]}^{(2)}$, 
$t\geq0$,
and define a filtration in  $\calb _\P\times\calb _\nu\times \Sigma_{\rm E}$ by 
\begin{align}
\pro {{\cal M}}=\lk {\cal F}_{[0,t]}\times \Sigma_{(-\infty, t]}\rk_{t\geq0}.
\end{align}
\bt{martingale}
\TTT{Martingale property of $\pro \Y$}
Let $V\in \vk $.
Suppose Assumption \ref{ass1}. 
Then the  stochastic process 
$(\xt )_{t\geq 0}$ is martigale with respect to the filtration $\pro{{\cal M}}$. I.e., 
$
\EZ\Ebb_{\mu_{\rm E}}
\lkkk
\xt |{\cal M}_s
\rkkk=\xs $ for $t\geq s$. 
\et
\proof
By Proposition  \ref{linearity} we have 
$\AA(\KKKK^x[0,t])=\AA(\KKKK^x[0,s])+\AA(\KKKK^x[s,t])$ for $s\leq t$. 
Since 
$\ee^{-i\alpha\AA(\KKKK^x[0,s])}
\ee^{-\int_0^s V(B_{T_r}) \dr  }
$ is ${\cal M}_s$-measurable, 
we have 
\begin{align*}
&\EZ\Ebb_{\mu_{\rm E}}
\lkkk
\xt |{\cal M}_s
\rkkk=
\ee^{tE}
\ee^{-i\alpha\AA(\KKKK^x[0,s])}
\ee^{-\int_0^s V(B_{T_r}+x) \dr }\\
&\times 
\EZ\Ebb_{\mu_{\rm E}}\!\!\!\lkkk
\ee^{-i\alpha \AA(\KKKK^x[s,t]) } \ee^{-\int_s^t V(B_{T_r}+x)\dr }
\JJ_t \bou(B_{T_t}+x)|{\cal M}_s
\rkkk.
\end{align*}
By the definition of $\KKKK [s,t]$ 
it is seen that 
\begin{align*}
&\EZ\Ebb_{\mu_{\rm E}}\lkkk
\ee^{-i\alpha \AA(\KKKK^x[s,t]) } \ee^{-\int_s^t V(B_{T_r}+x)\dr }
\JJ_t \bou(B_{T_t}+x)|{\cal M}_s
\rkkk\\
&=
\limn 
\EZ\Ebb_{\mu_{\rm E}}\lkkk
\ee^{-i\alpha \AA(\KKKK^x_n [s,t]) } \ee^{-\int_s^t V(B_{T_r}+x)\dr }
\JJ_t \bou(B_{T_t}+x)|{\cal M}_s
\rkkk, 
\end{align*}
and then 
\begin{align*}
&
\EZ\Ebb_{\mu_{\rm E}}\lkkk
\ee^{-i\alpha \AA(\KKKK_n^x [s,t]) } \ee^{-\int_s^t V(B_{T_r}+x)\dr }
\JJ_t \bou(B_{T_t}+x)|{\cal M}_s
\rkkk\\
&=
\Ebb_{\mu_{\rm E}}
\lkkk
\Ebb_{\nu}^0\lkkk
\Ebb_{\P}^{0}\lkkk
\ee^{-i\alpha \AA(\KKKK_n^x [s,t]) } \ee^{-\int_s^t V(B_{T_r}+x)\dr }
\JJ_t \bou(B_{T_t}+x)|{\cal F}_{[0,s]}^{(1)}
\rkkk
|
{\cal F}^{(2)}_{[0,s]}\rkkk
|\Sigma_{[-\infty,s]}
\rkkk.
\end{align*}
By the Markov property of the  Brownian motion we see that 
\begin{align*}
&
\Ebb_{\P}^{0}\lkkk
\ee^{-i\alpha \AA(\KKKK^x_n [s,t]) } \ee^{-\int_s^t V(B_{T_r}+x)\dr }
\JJ_t \bou(B_{T_t}+x)|{\cal F}_{[0,s]}^{(1)}
\rkkk\\
&=
\limn \Ebb_{\P}^{B_{T_s}}
\lkkk
\ee^{-i\alpha \AA(\KKKK_n ^{(1),x}[s,t]) } \ee^{-\int_s^t V(B_{T_r-T_s}+x)\dr }
\JJ_t \bou(B_{T_t-T_s}+x)
\rkkk,
\end{align*}
where $\Ebb_\P^{B_{T_s}}$ menas $\Ebb_\P^y$ evaluated at 
$y=B_{T_s}$ and 
$$\KKKK_n ^{(1),x}[s,t]=\ott  \sum_{j=1}^{2^n}
\int _{T_{\frac{(t-s)}{2^n}(j-1)+s}-T_s}
^{T_{\frac{(t-s)}{2^n}j+s}-T_s}
\jj_{\frac{(t-s)}{2^n}(j-1)+s}\la(\cdot-B_r-x) \dB_r^\mu.
$$
Since the subordinator $\pro T$ is also 
a  Markov process, 
 we have 
 \begin{align*}
&
\Ebb_\nu^0
\lkkk
\Ebb_{\P}^{B_{T_s}}
\lkkk
\ee^{-i\alpha \AA(\KKKK_n ^{(1),x}[s,t]) } 
\ee^{-\int_s^t V(B_{T_r-T_s}+x)\dr }
\JJ_t \bou(B_{T_t-T_s}+x)\rkkk 
|{\cal F}^{(2)}_{[0,s]}
\rkkk\\
&=
\Ebb_\nu^{T_s}
\Ebb_{\P}
^{B_{T_0}}
\lkkk
\ee^{-i\alpha \AA(\KKKK_n ^{(2),x}[s,t]) } 
\ee^{-\int_s^t V(B_{T_{r-s}-T_0}+x)\dr }
\JJ_t \bou(B_{T_{t-s}-T_0}+x)
\rkkk, 
\end{align*}
where 
$\Ebb_\nu ^{T_s}$ also means $\Ebb_\nu^y$ evaluated at $y=T_s$ and 
$$\KKKK_n ^{(2),x}[s,t]=\ott  \sum_{j=1}^{2^n}\int _{T_{\frac{(t-s)}{2^n}(j-1)}-T_0}
^{T_{\frac{(t-s)}{2^n}j}-T_0}
\jj_{\frac{(t-s)}{2^n}(j-1)+s}\la(\cdot-B_r-x) \dB_r^\mu.$$
Again the  Markov property of the Euclidean field yields that 
\begin{align*}
&
\Ebb_{\mu_{\rm E}}\lkkk
\Ebb_\nu^{T_s}
\Ebb_{\P}^{B_{T_0}}
\lkkk
\ee^{-i\alpha \AA(\KKKK_n ^{(2),x}[s,t]) } 
\ee^{-\int_s^t V(B_{T_{r-s}-T_0}+x)\dr }
\JJ_t \bou(B_{T_{t-s}-T_0}+x)\rkkk 
|\Sigma_{[-\infty,s]}
\rkkk\\
&=
\Ebb_{\mu_{\rm E}}\lkkk
\Ebb_\nu^{T_s}
\Ebb_{\P}^{B_{T_0}}
\lkkk
\ee^{-i\alpha \AA(\KKKK_n ^{(2),x}[s,t]) } 
\ee^{-\int_s^t V(B_{T_{r-s}-T_0}+x)\dr }
\JJ_t \bou(B_{T_{t-s}-T_0}+x)
\rkkk |\Sigma_s
\rkkk.
\end{align*}
The right-hand side above  equals to 
\begin{align}
&
=
E_s
\Ebb_\nu^{T_s}
\Ebb_{\P}^{B_{T_0}}
\lkkk
\ee^{-i\alpha \AA(\KKKK_n ^{(2),x}[s,t]) } 
\ee^{-\int_s^t V(B_{T_{r-s}-T_0}+x)\dr }
\JJ_t \bou(B_{T_{t-s}-T_0}+x)
\rkkk \non \\
&\label{martingale3}
=
\JJ_s \JJ_0^\ast {\rm U}_{-s}
\Ebb_\nu^{T_s}
\Ebb_{\P}^{B_{T_0}}
\lkkk
\ee^{-i\alpha \AA(\KKKK_n ^{(2),x}[s,t]) } 
\ee^{-\int_s^t V(B_{T_{r-s}-T_0}+x)\dr }
\JJ_t \bou(B_{T_{t-s}-T_0}+x)
\rkkk.
\end{align}
Since ${\rm U}_{-s}$ is the shift by $-s$, we have 
\begin{align*}
&
=
\JJ_s \JJ_0^\ast 
\Ebb_\nu^{T_s}
\Ebb_{\P}^{B_{T_0}}
\lkkk
\ee^{-i\alpha \AA(\KKKK_n ^{(3),x}[s,t]) } 
\ee^{-\int_s^t V(B_{T_{r-s}-T_0}+x)\dr }
\JJ_{t-s} \bou(B_{T_{t-s}-T_0}+x)
\rkkk,
\end{align*}
where
$$\KKKK_n ^{(3),x}[s,t]=\ott  \sum_{j=1}^{2^n}\int _{T_{\frac{(t-s)}{2^n}(j-1)}-T_0}
^{T_{\frac{(t-s)}{2^n}j}-T_0}
\jj_{\frac{(t-s)}{2^n}(j-1)}\la(\cdot-B_r-x) \dB_r^\mu.$$
We notice that
the random variable $T_t+ y$ under $\nu$ has the same law 
as $T_t$ under $\nu^y$, i.e., $\Ebb_\nu^y[f(T_t)]=\Ebb_\nu^0[f(T_t+y)]$, 
we can  see that 
\begin{align*}
&=
\left.
\JJ_s \JJ_0^\ast 
\Ebb_\nu^0
\Ebb_{\P}^{B_{u+T_0}}
\lkkk
\ee^{-i\alpha \AA(\KKKK_n ^{(3),x}[s,t]) } 
\ee^{-\int_s^t V(B_{T_{r-s}-T_0}+x)\dr }
\JJ_{t-s} \bou(B_{T_{t-s}-T_0}+x)
\rkkk\right\lceil_{u=T_s}\\
&=
\JJ_s \JJ_0^\ast 
\Ebb_\nu^0
\Ebb_{\P}^{B_{T_s}}
\lkkk
\ee^{-i\alpha \AA(\KKKK_n ^{(4),x}[s,t]) } 
\ee^{-\int_s^t V(B_{T_{r-s}}+x)\dr }
\JJ_{t-s} \bou(B_{T_{t-s}}+x)
\rkkk,
\end{align*}
where 
$$\KKKK_n ^{(4),x}[s,t]=\ott  \sum_{j=1}^{2^n}\int _{T_{\frac{(t-s)}{2^n}(j-1)}}
^{T_{\frac{(t-s)}{2^n}j}}
\jj_{\frac{(t-s)}{2^n}(j-1)}\la(\cdot-B_r-x) \dB_r^\mu.$$
Taking the limit $n\to\infty$, we 
finally obtain that 
\begin{align*}
\EZ
\Ebb_{\mu_{\rm E}}
\lkkk
\xt |{\cal M}_s
\rkkk
&=
\ee^{sE}
\ee^{-i\alpha\AA(\KKKK^x [0,s])}
\ee^{-\int_0^s V(B_{T_r}+x) \dr }
\JJ_s\\
&\hspace{-1cm}\times 
\ee^{(t-s)E}
\Ebb_{\P \times \nu}^{B_{T_s}, 0}
\lkkk
\JJ_0^\ast 
\ee^{-i\alpha \AA(\KKKK^x [0, t-s]) } 
\ee^{-\int_0^{t-s} V(B_{T_r}+x)\dr }
\JJ_{t-s} \bou(B_{T_{t-s}}+x)
\rkkk.
\end{align*}
Notice that 
$$
\ee^{(t-s)E}
\Ebb_{\P \times \nu}^{B_{T_s}, 0}
\lkkk
\JJ_0^\ast 
\ee^{-i\alpha \AA(\KKKK^x [0, t-s]) } 
\ee^{-\int_0^{t-s} V(B_{T_r}+x)\dr }
\JJ_{t-s} \bou(B_{T_{t-s}}+x)
\rkkk=
\bou(B_{T_s}+x)
$$
and hence 
\begin{align*}
\EZ\Ebb_{\mu_{\rm E}}
\lkkk
\xt |{\cal M}_s
\rkkk=
\ee^{sE}
\ee^{-i\alpha\AA(\KKKK^x[0,s])}
\ee^{-\int_0^s V(B_{T_r}+x) \dr }
\JJ_s 
\bou(B_{T_s}+x)=\xs .
\end{align*}
Then the proof is complete.
\qed

Since we show that $(\xt )_{t\geq 0}$ is a martingale, 
for an arbitrary  stopping time $\tau$ with respect to $({\cal M}_t)_{t\geq 0}$,  
$(\xtt )_{t\geq 0}$ is also a martingale. 
By using this fact we can show a spatial decay of bound state $\bou$ of $\PFK$. 

\subsection{Fall-off of bound states}
Let us recall that $\pro z$ is the $d$-dimensional L\'evy   process on  
a probability space $(\Omega_\rmw, \calb _\rmw , \rmw ^x)$ such that 
$\Ebb_\rmw ^x \lkkk \ee^{-iu \cdot z_t} \rkkk =\ee^{-t (\sqrt{|u|^2+m^2}-m)}\ee^{-iu\cdot x}$. 
Hence the generator of $\pro z$ is given by 
$\sqrt{\p^2+m^2}-m$, and the distribution 
$k_{t,m}(x)$ of 
$z_t$ by 
\begin{align*}
k_{t,m}(x)&=2\lk \frac{m}{2\pi}\rk^{\frac{d+1}{2}}\frac{t \ee^{tm}K_{\frac{d+1}{2}}(m\sqrt{t^2+|x|^2})}
{(t^2+|x|^2)^{\frac{d+1}{4}}},&\quad m>0,\\
k_{t,0}(x)&=
\frac{\Gamma(\frac{d+1}{2})}{(2\pi)^{\frac{d+1}{2}}}\frac{t}
{(t^2+|x|^2)^{\frac{d+1}{2}}},&\quad m=0.
\end{align*}
Here $\Gamma(m)$ denotes the Gamma function, 
$K_\nu(z)$ is the modified Bessel function of
the third kind of order $\nu$, and 
it is known that $K_\nu(z) \sim  \half \Gamma(\nu)(\half z)^{-\nu}$ as $z\sim 0$.

\bl{bound2}
Let $V\in \vk $.
Suppose Assumption \ref{ass1}.  
Let   $\tau$  be a stopping time with respect to the filtration $\pro{{\cal M}}$.
Then 
\eq{expdecay}
\|\bou(x)\|
\leq
\|\bou\|_{\hhh}
\Ebb_\rmw ^x\lkkk
\ee^{-\int_0^{t\wedge \tau} (V(z_r)-E)\dr }
\rkkk.
\en
\el
\proof
Since 
$({\JJ_0\Phi\cdot \xt })_{t\geq0}$ is a  martingale with respect to 
the filtration $\pro {{\cal M}}$, 
also is $(\JJ_0\Phi\cdot \xtt )_{t\geq0}$. 
Then 
$\EZ\Ebb_{\mu_{\rm E}}
\lkkk \JJ_0\Phi\cdot \xt \rkkk=
\EZ\Ebb_{\mu_{\rm E}}
\lkkk \JJ_0\Phi\cdot \xtt \rkkk$ follows.
 It is immediate  to see by Lemma \ref{sup} that 
\begin{align*}
|
\EZ\Ebb_{\mu_{\rm E}}
\lkkk \JJ_0\Phi\cdot \xtt \rkkk
|
\leq C
\|\Phi\|
\EZ
\lkkk \ee^{-\int_0^{t\wedge \tau} (V(B_{T_r}+x)-E)\dr }
\rkkk,
\end{align*}
where $\d C=\sup_{x\in\BR}\|\bou(x)\|$. 
Since $B_{T_t}=
z_t$ in law, 
we then have 
\eq{law}
|\EZ\Ebb_{\mu_{\rm E}}
\lkkk \JJ_0\Phi\cdot \xt \rkkk
|\leq
\|\Phi\|
\Ebb_\rmw ^x\lkkk
\ee^{-\int_0^{t\wedge \tau} (V(z_r)-E)\dr }
\rkkk.
\en
From 
$\|\bou(x)\|_{L^2(\Q)}=\sup_{\Phi\in L^2(\Q),\Phi\not=0}
\EZ\Ebb_{\mu_{\rm E}}\lkkk \JJ_0\Phi\cdot \xt \rkkk/{\|\Phi\|}
$,
the lemma follows. 
\qed

\bt{falloff1}
\TTT{Fall-off of bound states}
Let $V=V_+-V_-\in \vk$.
Suppose Assumption \ref{ass1}.  
\bi
\item[(1)]
Suppose that $\d \lim_{|x|\to\infty}V_-(x)+E=a<0$. 
Then 
\bi
\item[Case $m=0$]: 
there exists $C>0$  such that 
$\d \frac{\|\bou(x)\|_{L^2(\Q)}}{\|\bou\|_\hhh}\leq \frac{C}{1+|x|^{d+1}}$;
\item[Case $m>0$]: 
there exist $C>0$  and $c>0$ such that 
$\d \frac{\|\bou(x)\|_{L^2(\Q)}}{\|\bou\|_\hhh}
\leq C \ee^{-c|x|}$.
\ei
\item[(2)]
Suppose that $\d \lim_{|x|\to\infty}V(x)=\infty$. 
Then there  exist $C>0$  and $c>0$ such that 
$\|\bou(x)\|_{L^2(\Q)}\leq C \ee^{-c|x|}\|\bou\|_\hhh$.
\ei
\et
\proof
(1) Suppose  that $V_-(x)+E<a+\epsilon<0$ for all $x$ such that $|x|>R$, 
and $\tau_R=\inf \{s||z_s|<R\}$ is a stopping time with respect to the filtration 
$\pro{{\cal M}}$.
By \kak{expdecay}
we have 
$
\|\bou(x)\|
\leq
\|\bou\|_{\hhh}
\Ebb_\rmw ^x
\lkkk
\ee^{+2(\epsilon+a) ({t\wedge \tau_R})} 
\rkkk$ for $|x|>R$.
In a similar way to  \cite[Proposition IV.1]{cms90} we have 
\eq{array}
\begin{array}{ll}
\d \Ebb_\rmw ^x\lkkk
\ee^{+2(\epsilon+a) ({t\wedge \tau_R})}
\rkkk\leq \frac{C}{1+|x|^{d+1}},&  m=0,\\
\d \Ebb_\rmw ^x\lkkk
\ee^{+2(\epsilon+a) ({t\wedge \tau_R})}
\rkkk\leq C \ee^{-c|x|},& m>0.
\end{array}
\en
Thus (1)  follows. 

(2)
Let $\tau_R=\inf\{s| |z_s|>R\}$, which is the stopping time with respect to 
the filtration $\pro {{\cal M}}$.
Let $W(x)=\inf\{V(y)||x-y|<R\}$.
Then it can be shown in \cite[Theorem 4.7]{hil13} and \cite[Proposition IV.4]{cms90} 
that 
\eq{seen}
\Ebb_\rmw ^x\lkkk \ee^{(t\wedge \tau_R)E } \ee^{-\int_0^{t\wedge \tau_R}V(z_r)\dr}\rkkk
\leq \ee^{-t(W(x)-E)}+C \ee^{-\alpha R}\ee^{ct}
\en
with some constants $\alpha,c$ and $C$. 
Inserting $R=p|x|$ with any $0<p<1$, we see that $W(x)\to\infty$ as $|x|\to\infty$. 
Substituting $t=\delta|x|$ for sufficiently small $\delta>0$ and $R=p|x|$ with some $0<p<1$, (2)  follows. 
\qed

\section{Gaussian domination  of ground \label{sec6}
states}
\begin{figure}[t]
\centering
\includegraphics[width=200pt]{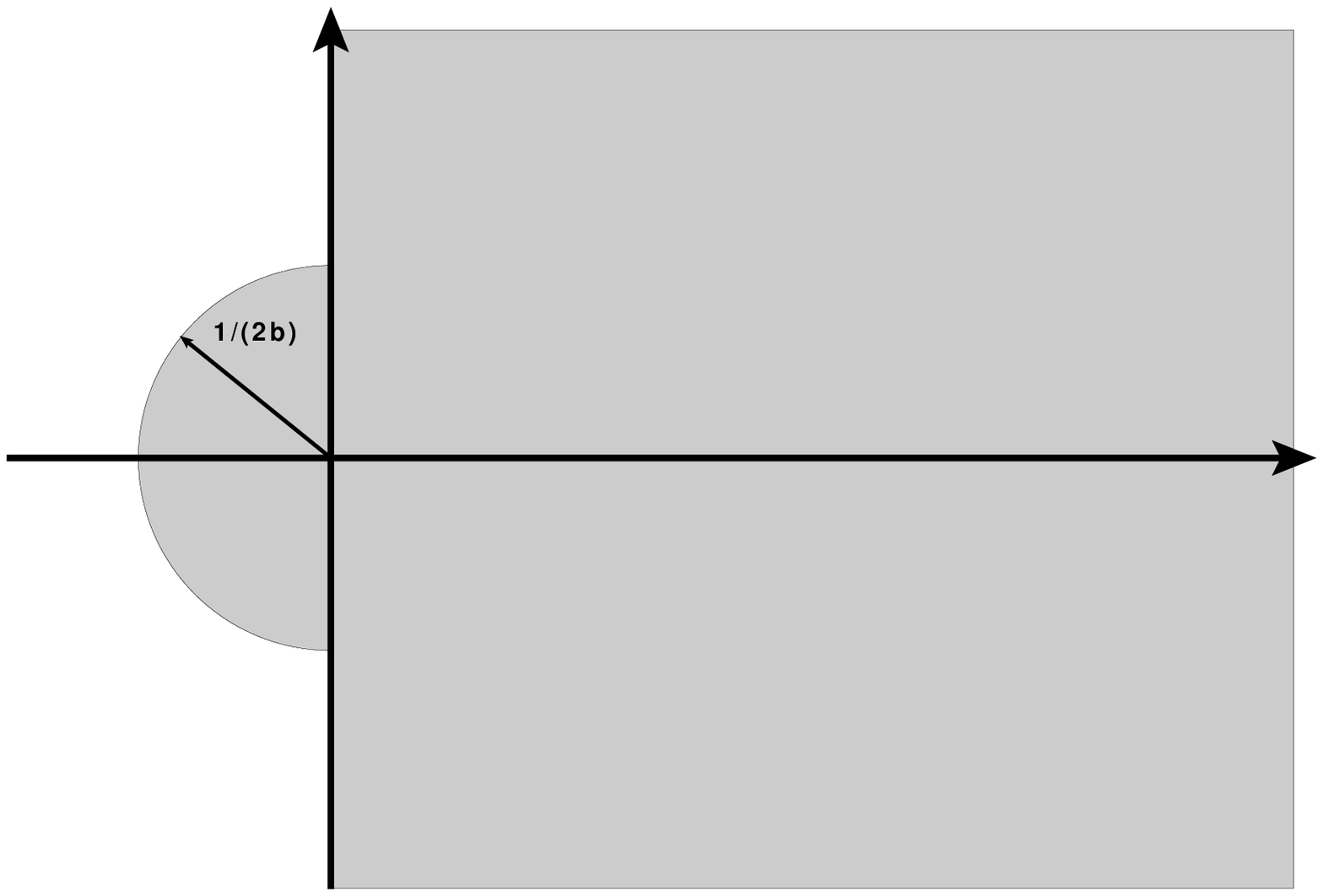}
\caption{Domain $D$}
\label{domain}
\end{figure}

Let $\PFF =\PF$ or $\PFK$ in this section. 
Throughout this section, when we consider $\PF$ we suppose 
Assumptions \ref{ass1} and \ref{ass2}, and when we consider 
$\PFK$ we suppose 
 Assumption \ref{ass1}.
A fundamental assumption in  this section  is that $\PFF $ has a ground state $\gr$. 
\begin{assumption}
\label{gr}
Suppose that  $m\geq 0$ and  $\PFF $ has a ground state $\gr$
, i.e., 
\eq{groundstate}
\PFF \gr=E\gr,\quad E=\is(\PFF ).
\en
\end{assumption}
The existence of ground state is studied in \cite{hha13,kms09,kms11}. 
\bc{positivityimproving}
The operator $\ee^{i\frac{\pi}{2} {\rm N}}\ee^{-t\PFF}\ee^{-i\frac{\pi}{2} {\rm N}} $ is 
positivity improving for $t>0$, i.e., 
 $(F, \ee^{i\frac{\pi}{2} {\rm N}}\ee^{-t\PFF}\ee^{-i\frac{\pi}{2} {\rm N}} G)>0$ for any $F\geq0$ and $G\geq 0$ $(F\not\equiv 0, G\not\equiv 0)$.
 In particular  $\ee^{i\frac{\pi}{2} {\rm N}}\gr$ is strictly positive and then the ground state of $\PFF $ is unique up to multiplication constants.
  \ec
 \proof
 It is established in \cite{hir00a} that $\JJ_0 ^\ast \ee^{i\frac{\pi}{2} {\rm N}}  \ee^{-i\alpha \AA(f)}
\ee^{-i\frac{\pi}{2} {\rm N}}\JJ_t $ is positivity improving for arbitrary 
$f\in \bigoplus^d  L_\RR^2(\BR)$. 
Thus the first statement follows. 
Since 
$\ee^{i\frac{\pi}{2} {\rm N}}$ is unitary, the statement on the uniqueness also follows from the Perron-Frobenius theorem. 
 \qed
For an arbitrary fixed $0\leq \phi\in \LR$ but $\phi\not\equiv 0$, 
we define 
 \begin{align}
\grtt =
 \ee^{-t(\PFF -E)} 
(\phi\otimes \one),\quad 
 \grt=
\grtt /\|\grtt \|. 
\end{align}
Then it follows that 
$\grt\to \grn$ strongly as $t\to\infty$, since $(\phi\otimes\one, \gr)\not=0$.
Let
\begin{align}
\ms L_t=
\phi(B_{-T_{t}})\phi(B_{T_t}) \ee^{-\frac{\alpha^2}{2}
\qq(\KKKK [-t,t])}\ee^{-\int_{-t}^t V(B_{T_s})\ds},\quad t\geq0.
\end{align}
\begin{remark}
{\rm We formally write the pair interaction 
$W^{\rm SRPF}=\qq(\KKKK [-t,t])$
by 
\begin{align}
\label{pair}
\qq(\KKKK [-t,t])=-\frac{\alpha^2}{2}\sum_{\mu,\nu=1}^d
\int_{-T_t}^{T_t} \dB_s^\mu 
\int_{-T_t}^{T_t} \dB_r^\nu 
W_{\mu\nu}(T^\ast_s-T^\ast_r,  B_s-B_r), 
\end{align}
where the pair potential, $W_{\mu\nu}(t,X)$,  is given by 
\eq{W}
W_{\mu\nu}(t,X)=\half\int_\BR\frac{|\vp(k)|^2}{\omega(k)}\lk\delta_{\mu\nu}-\frac{k_\mu k_\nu}{|k|^2}\rk \ee^{-ik\cdot X}\ee^{-\omega(k)|t|}
 {\rm d}k.
\en
}\end{remark}
\bd{64}
{\rm 
Define the probability measure $\mu_t^{\rm SRPF}=\mu_t$ on the measurable space 
$(\Omega_\P\times\Omega_\nu, \calb _\P\times\calb _\nu)$
by 
\eq{M}
\calb _\P\times\calb _\nu\ni A\mapsto \mu_t(A)=\frac{1}{Z_t}
\IIXX\EX\lkkk \one_A \ms L_t\rkkk,\quad t\geq0.
\en
Here $Z_t$ is the normalizing constant such that $\mu_t(\Omega_\P\times\Omega_\nu)=1$. 
}\ed
We define  the self-adjoint operator $\A_\xi$ in $\hhh$ by  
$\A_\xi= \int_\BR^\oplus \A(\xi(\cdot-x))\dx$, where 
$\xi\in \ottt L_\RR^2(\BR)$.
Then we have 
\eq{convgr}
(\gr, \ee^{-i\beta \A_\xi}\gr)=\limt 
\frac{(\ee^{-t\PFF}\phi\otimes \one, \ee^{-i\beta \A_\xi} \ee^{-t\PFF}\phi\otimes \one)}
{(\ee^{-t\PFF}\phi\otimes \one,\ee^{-t\PFF}\phi\otimes \one)},\quad \beta\in\RR. 
\en
\bl{expectationA}
Let $\beta\in\RR$. Then it  follows that 
\eq{expa}
\frac{(\ee^{-t\PFF}\phi\otimes \one, \ee^{-i\beta  \A_\xi} 
\ee^{-t\PFF}\phi\otimes \one)}
{(\ee^{-t\PFF}\phi\otimes \one,\ee^{-t\PFF}\phi\otimes \one)}
=
\Ebb_{\mu_t}[\ee^{-\half(2\alpha\beta \Re 
\qq(\KKKK[-t,t],\jj_0\xi)+
\beta^2 \qq(\jj_0\xi))}]
\en
\el
\proof
This follows from Corollary \ref{green1}.
\qed
Note that both $ \qq(\KKKK[-t,t],\jj_0\xi)$ and $\qq(\jj_0\xi)$ do not depend on $x$. 
\bc{zero}
Let $\xi=\bigoplus_{\nu=1}^d \delta_{\mu\nu} \xi_\mu$ and 
 $\A_\mu=\int_\BR^\oplus \A (\xi(\cdot-x))\dx$. 
We suppose that 
${\rm supp} \hat \xi_\mu\cap {\rm supp} {\vp}=\emptyset$.
Then 
\begin{align}
(\gr, \A_\mu^n \gr)_\hhh
&=(\one, \A_\mu(0)^n \one)_{L^2(\Q)}\non \\
&=\lkk\begin{array}{ll}(-1)^m(2m-1)!! \lk \half \int _\BR|\hat\xi_\mu(k)|^2(1-\frac{k_\mu^2}{|k|^2})\dk\rk ^m&n=2m\\
0&n=2m-1,\end{array}\right.
\label{kokoro}
\end{align}
 where 
 $\A_\mu(0)= \A (\xi)$.
 \ec
\proof
Formally we see that 
$$
\qq(\KKKK[-t,t],\jj_0\xi)=\half \sum_{\nu=1}^d 
\int_{-T_t}^ {T_t} 
\!\!\!
\dB_s^\nu 
\lk \int _\BR 
\hat \xi_\mu(k) \frac{\vp(k)}{\sqrt{\omega(k)}} \ee^
{-T^\ast _s \omega(k)} \ee^{-ikB_s} \lk\delta_{\mu\nu}-\frac{k_\mu k_\nu}{|k|^2}\rk \dk \rk 
=0.$$
This is proven rigorously from the definition of $\KKKK[-t,t]$.
By \kak{expa} and taking the limit $t\to\infty$, we have 
$(\gr, \ee^{-i\beta \A_\mu}\gr)=
\ee^{-\beta^2 \qq(\jj_0\xi))/2}$.
Since $\gr\in D(\A_\mu^n)$ by Theorem \ref{gaussiandecay} below, 
we derive \kak{kokoro} by taking $n$-times  derivative at $\beta=0$. 
\qed
\bl{guss1}
Suppose that  $\beta<(2\qq(\jj_0 \xi))\f$.
Then 
$\grt\in \D(\ee^{\beta\A_\xi^2/2})$ and 
\eq{expaa}
\|\ee^{\beta\A_\xi^2/2}\grt\|^2=
(1-2\beta \qq(\jj_0\xi))^{-1/2}
\Ebb_{\mu_t}\lkkk
\ee^{\frac{-\beta \alpha^2  \qq(\KKKK [-t,t],\jj_0\xi)^2}
{(1-2\beta \qq(\jj_0\xi))}}\rkkk.
\en
\el
\proof
We have 
$
(\grt, \ee^{-ik \A_\xi}\grt)=
\Ebb_{\mu_t}\lkkk 
\ee^{-\alpha k \qq(\KKKK [-t,t], \jj_0\xi)} \rkkk
\ee^{-\half k^2 \qq(\jj_0\xi)}$. 
By the Gaussian transformation with respect to $k$, 
we see that 
\begin{align*}
&(\grt, \ee^{-\A_\xi^2/2}\grt)=
(2\pi)^{-1/2} \int_\RR  \ee^{-\frac{k^2}{2}}
\Ebb_{\mu_t}\lkkk 
\ee^{-\alpha k \qq(\KKKK [-t,t], \jj_0\xi)} \rkkk
\ee^{-\half k^2 \qq(\jj_0\xi)}\dk, 
\end{align*}
and by  Fubini's lemma, we can exchange $\int \dk$ and $\int {\rm d}\mu_t$. 
Then 
\eq{66}
(\grt, \ee^{-\A_\xi^2/2}\grt)=
\frac{1}{\sqrt{1+\qq(\jj_0\xi)}}
\Ebb_{\mu_t}\lkkk 
\ee^{\frac{\alpha^2 \qq(\KKKK [-t,t], \jj_0\xi)^2}{2(1+\qq(\jj_0\xi))}} \rkkk.
\en
Replacing $\xi$ with $\sqrt{-2\beta}\xi$ for  $\beta<0$, 
we have \kak{expaa} with $\beta<0$.
We can extend this to $\beta<(2\qq(\jj_0\xi))\f$ by an analytic continuation.
For notational simplicity we set 
$\www =\qq(\jj_0\xi)$.
Let 
\begin{align*}
\chi(z)=(\grt, \ee^{-z \A_\xi^2}\grt),\quad 
\rho(z)=
\Ebb_{\mu_t}\lkkk 
\exp\lk z  \alpha^2 \frac{\qq(\KKKK [-t,t], \jj_0\xi)^2}{2\www }\rk \rkkk,\quad
 \theta (z)=\frac{2z \www }{1+2z\www }.
 \end{align*}
 Then \kak{expaa} is realized as 
\eq{ana1}
\chi(z)=
\frac{1}{\sqrt{1+2z \www }}
\rho\circ \theta (z)
\en
 for $z\geq 0$. 
Notice that 
$\Ebb_{\mu_t}\lkkk 
\exp\lk z  \alpha^2 \frac{\qq(\KKKK [-t,t], \jj_0\xi)^2}{2\www }\rk \rkkk<\infty$ 
for all $z>0$.
Then 
we know that 
\eq{expbb}
\rho(z)=\sum_{n=0}^\infty \frac{1}{n!}\Ebb_{\mu_t}
\lkkk \lk\frac{\alpha^2 \qq(\KKKK [-t,t], \jj_0\xi)^2}{2\www }\rk^n
\rkkk
z^n
\en
 for $z\geq0$, and hence 
 $\rho(z)$ can be analytically continued to the whole complex plane $\CC$, which is denoted by 
 $\bar \rho(z)$ and it follows that 
 $\bar \rho(z)=
\Ebb_{\mu_t}\lkkk 
\exp\lk z  \alpha^2 \frac{\qq(\KKKK [-t,t], \jj_0\xi)^2}{2\www }\rk \rkkk$ for $z\in\CC$.
Then 
$\frac{1}{\sqrt{1+2z \www }}
\rho\circ \theta (z)
$ can be analytically continued to 
the domain: (Fig.\ref{domain})
$$D=\{z\in\CC||z|<(2\www )\f\}\cup\{z\in\CC|  \Re z>0\}.$$
In particular the radius of convergence $r$ of 
$\frac{1}{\sqrt{1+2z \www }}
\bar \rho\circ \theta (z)$ 
at $z=0$ 
satisfies that 
$1-\epsilon<r<1$ for an arbitrary $\epsilon>0$.
By the equality \kak{ana1}, 
$\chi$ can be also analytically continued to the domain $D$, which is denoted by 
$\bar \chi$. 
Let $\epsilon>0$. 
Then 
\eq{67}
\chi(z)=\sum_{n=0}^\infty 
\lk
\frac{(-1)^n}{n!}\int_0^\infty \lambda^n \ee^{-\epsilon \lambda}
\dE(\lambda)\rk (z-\epsilon)^n
\en
 for $0<\epsilon-z$, 
where $\dE(\lambda)$ denotes the spectral resolution of the self-adjoint operator 
$\A_\xi^2$ with respect to $\grt$. 
Since we have 
\eq{expcc}
\frac{1}{\sqrt{1+2z \www }}
\bar \rho\circ \theta (z)=\sum_{n=0}^\infty a_n (z-\epsilon)^n
\en
for $z\in\CC$ such that $|z-\epsilon|<\sqrt{\frac{1}{(2\www )^2}+\epsilon^2}$.
Comparing both expansions \kak{67} and \kak{expcc} we see that 
$a_n= 
\frac{(-1)^n}{n!}\int_0^\infty \lambda^n \ee^{-\epsilon \lambda}\dE(\lambda)$
and by \kak{expcc}
we have 
\eq{expdd}
\frac{1}{\sqrt{1+2z \www }}
\bar \rho\circ \theta (z)=
\sum_{n=0}^\infty 
\lk
\frac{(-1)^n}{n!}\int_0^\infty \lambda^n \ee^{-\epsilon \lambda}\dE(\lambda)\rk
(z-\epsilon)^n.
\en
In particular it follows that 
for $-\delta<0$ with  $\epsilon+\delta<\sqrt{\frac{1}{(2\www )^2}+\epsilon^2}$,
$$
\bar\chi(\delta)=\sum_{n=0}^\infty 
\lk
\frac{1}{n!}\int_0^\infty \lambda^n \ee^{-\epsilon \lambda}\dE(\lambda)\rk
(\delta+\epsilon)^n<\infty.$$
Thus 
$$
\sum_{n=0}^\infty 
\lk
\frac{1}{n!}
\int_0^N \lambda^n \ee^{-\epsilon \lambda}\dE(\lambda)\rk (\delta+\epsilon)^n=
\int_0^N  \ee^{\delta  \lambda}\dE(\lambda)$$
and take $N\to\infty$ on both sides we have 
$\bar \chi(\delta)=\int_0^\infty  \ee^{\delta  \lambda}\dE(\lambda)<\infty$.
Since $\epsilon>0$ is arbitrary, then 
it follows that $(\grt, \ee^{z \A_\xi^2}\grt)<\infty$ for $\beta>(2\www )\f$. 
\qed

\bt{gaussiandecay}
\TTT{Gaussian domination  of the ground state}
Let $\beta<(2\qq(\jj_0\xi))\f$. Then 
$\gr\in \D(\ee^{\beta \A_\xi^2/4})$ follows.
\et
\proof
By Lemma \ref{guss1} we have 
the uniform bound 
$
\|\ee^{\beta\A_\xi^2}\grt\|^2\leq \frac{1}{\sqrt{1-2\beta\qq(\jj_0\xi)^2}}
$
in $t$. Thus 
there exists a subsequence $t'$ such that $\|\ee^{\beta\A_\xi^2/4}\gr^{t'}\|^2$ 
converges  to some $c$ as $t'\to\infty$. 
We reset $t'$ as $t$. 
We claim that $\{\ee^{\beta\A_\xi^2/4}\grt\}_t$ is a Cauchy sequence. 
Directly we have 
$
\|\ee^{\beta\A_\xi^2/4}\grt-
\ee^{\beta\A_\xi^2/4}\grn^s\|^2
=
\|\ee^{\beta\A_\xi^2/4}\grt\|^2+
\|\ee^{\beta\A_\xi^2/4}\grn^s\|^2-
2(\grn^s,
\ee^{\beta\A_\xi^2/2}\grt)$.
Note that $\grt$ strongly converges to $\gr$ as $t\to\infty$. 
Since 
the uniform bound of $\|\ee^{\beta\A_\xi^2}\grt\|^2$ 
 implies that 
$$
(\grn^s, \ee^{\beta\A_\xi^2/2}\grt)
=
(\grn^s-\grt, \ee^{\beta\A_\xi^2/2}\grt)
+\|\ee^{\beta\A_\xi^2/4}\grt\|^2
\to c$$
as $t,s\to\infty$, we obtain that
$\lim_{t,s\to\infty}
\|\ee^{\beta\A_\xi^2/4}\grt-
\ee^{\beta\A_\xi^2/4}\grn^s\|=0$
and 
$\ee^{\beta\A_\xi^2/4}\grt$, $t>0$,  is a convergent sequence. 
Hence 
the closedness of $\ee^{\beta\A_\xi^2/4}$ yields the desired results. 
\qed

\section{Measures associated with the ground state}
\label{sec7}
Similar to Section \ref{sec6} 
in this section let $\PFF =\PF$ or $\PFK$, 
and we 
suppose 
 that $\PFF $ has a ground state $\gr$. 
\subsection{Outline}
\label{621}We set 
$\ms X=\Omega_\P \times \Omega_\nu$ and  
$\msw ^x=\P^x\otimes \nu$ in what follows. 
Let $X_t=B_{T_t}$ for $t\geq0$ and $X_{-t}=B_{-T_t}$ for $-t<0$. 
Thus 
$t\mapsto X_t(\omega_1,\omega_2)=B_{T_t(\omega_2)}(\omega_1)$ for $(\omega_1,\omega_2)\in \ms X$ is a 
c\'adl\'ag path,  i.e., 
paths are  right continuous and 
the left limits exist. 
Let 
${\cal F}_{[-s,s]}=\s(X_r;r\in[-s,s])$.
Then \eq{life1}
\d\ms G_t=\bigcup_{0\leq s\leq t}
{\cal F}_{[-s,s]},\quad
\d\ms G=\bigcup_{0\leq s}{\cal F}_{[-s,s]}
\en
are  finitely additive families of sets.
We  define the correction of probability spaces by  
\eq{x1}
(\ms X, \s(\ms G), \mu_t),\quad t >0,
\en
where $\mu_t$ is given by \kak{M}.  
We show in this section that there exists a probability measure $\mu_\infty$  on $(\ms X,\s(\ms G))$ such that 
$\mu_t\to \mu_\infty$ as $t\to\infty$ in the local weak sense.

The outline of the idea to show the convergence is as follows. 
First by using 
$\gr^t$
we  define the family of finitely additive set  functions $\rho_t$ 
on $(\ms X, \ms G_t)$, $t>0$,   
and we denote the  extension to the probability measure on $(\ms X, \s(\ms G_t))$ by $\bar \rho_t$. 
Thus we define the probability space 
\eq{x2}
(\ms X, \s(\ms G_t)), \bar \rho_t).
\en
We show in Lemma \ref{equality} by using functional integrations that 
\eq{e1}
\bar \rho_t(A)=\rho_t(A)=\mu_t(A)
\en
for $A\in \ms G_s$ for all   $s\leq t$. 
Next by using the ground state $\gr$ 
we define a finitely additive set function $\mu$ on $(\ms X,\ms G)$
and denote the  extension to the probability measure on $(\ms X, \s(\ms G))$ by 
$\mu_\infty$.
Thus we define the probability space 
\eq{x3}
(\ms X, \s(\ms G)), \mu_\infty).
\en
By applying the fact that $\gr^t$ strongly converges to $\gr$ as $t\to\infty$, 
we prove that 
\eq{e2}
\rho_t(A)\to \mu(A),\quad t\to\infty,
\en
for $A\in \ms G$ in Lemma \ref{doko}, which, together with \kak{e1},  implies that 
\eq{x4}
\mu_t(A)\to \mu_\infty(A),\quad A\in \ms G
\en
and $\mu_t$ converges to 
the measure $\mu_\infty$ in the sense of local weak.
By the construction of $\mu_\infty$ we can show an explicit form of $\mu_\infty(A)$ for $A\in\ms G$. 
See Figure \ref{lwc}. 
\begin{figure}[t]
$$\mu_t\stackrel{\rm Lemma \ref{equality}}{=}\rho_t\subset\bar\rho_t\xrightarrow {\rm Lemma \ref{doko}}\mu\subset\mu_\infty$$
\caption{Local weak convergence of $\mu_t$ to $\mu_\infty$}
\label{lwc}
\end{figure}%

\subsection{Local weak convergences}
Let 
us define 
\eq{qtzz}
\QT = 
\JJ_{-t}^\ast \ee^{-\int_{-t}^tV(X_s)\ds } 
\ee^{-i\alpha\AA(\KKKK [-t,t]) }
\JJ_t .
\en
Note that for a.s.  $(\omega_1,\omega_2)\in \ms X$, 
$\QT:L^2(\Q)\to L^2(\Q)$ is a bounded linear operator. 
Define an additive set function $\mu:\ms G\to \RR$ 
by 
\begin{align}
\mu(A)=\ee^{2Et}\int _\BR 
\dx \EW\lkkk \one_A
( \grn(X_{-t}), \QT  \grn(X_t))\rkkk,\quad 
A\in {\cal F}_{[-t,t]}.
\end{align}
\bl{positive}
It follows that $\mu(A)\geq 0$ for $A\in{\cal F}_{[-t,t]}$.
\el
\proof
We note that $\ee^{i\frac{\pi}{2}\N}\gr>0$ and 
$\ee^{i\frac{\pi}{2}\N}\QT \ee^{-i\frac{\pi}{2}\N}$ is positivity improving by Corollary  \ref{positivityimproving}. 
Then 
\begin{align*}
\mu(A)=\ee^{2Et}
\int _\BR 
\dx \EW\lkkk \one_A
( \ee^{i\frac{\pi}{2}\N}\grn(X_{-t}), \ee^{i\frac{\pi}{2}\N}\QT  \ee^{-i\frac{\pi}{2}\N}\ee^{i\frac{\pi}{2}\N}\grn(X_t))\rkkk\geq0, 
\end{align*}
the lemma follows.
\qed

\bl{welldefined}
The set function $\mu$ is well defined, i.e., 
for $A\in {\cal F}_{[-t,t]}\subset {\cal F}_{[-s,s]}$
\begin{align*}
\mu(A)&=\ee^{2Et}
\int _\BR 
\dx \EW\lkkk \one_A
( \grn(X_{-t}), \QT  \grn(X_{t}))
\rkkk\\
&=
\ee^{2Es}\int _\BR 
\dx \EW\lkkk \one_A
( \grn(X_{-s}), \QS  \grn(X_{s}))
\rkkk 
\end{align*}
\el
\proof
Let $\mu_{(t)}=\mu\lceil_{{\cal F}_{[-t,t]}}$. Then $\mu_{(t)}$ is a probability measure on 
$(\ms X, {\cal F}_{[-t,t]})$. Let $-s< -t=t_0<t_1<\cdots <t_n=t<s$. 
Then by Corollary \ref{green1} 
the finite dimensional distribution is given by 
\begin{align*}
&\mu_{(t)}^{t_0,...,t_n}(A_0\times\cdots \times A_n)=\mu(X_{t_0}\in A_0,\cdots,X_{t_n}\in A_n)\\
&=
\ee^{2Et}\int_\BR \dx\EW \lkkk 
\lk \prod _{j=0}^n \one_{A_j} (X_{t_j})\rk (\grn(X_{-t}),\QT\grn(X_{t}))\rkkk\\
&=
(\grn, \one_{A_0}\ee^{-(t_1-t_0)(\PFF -E)}\cdots \ee^{-(t_n-t_{n-1})(\PFF -E)}\one_{A_n}\grn).
\end{align*}
By $\ee^{-(t_0+s)(\PFF -E)}\grn=\grn$ we have
\begin{align*}
&=
(\grn, \ee^{-(t_0+s)(\PFF -E)}
\one_{A_0}\ee^{-(t_1-t_0)\PFF }\cdots \ee^{-(t_n-t_{n-1})\PFF }\one_{A_n}\ee^{-(s-t_n)(\PFF -E)}\grn)\\
&=
\ee^{2Es}\int_\BR \dx\EW \lkkk 
\lk \prod _{j=0}^n \one_{A_j}(X_{t_j}) \rk (\grn(X_{-s}),\QS\grn(X_s))\rkkk
\\
&=\mu_{(s)}^{t_0,...,t_n}(A_0\times\cdots \times A_n).
\end{align*}
It can be also seen that the finite dimensional distributions 
$\mu_{(t)}^{\Lambda}$, $\Lambda\subset [-t,t], \#\Lambda<\infty$,  satisfy the consistency condition, i.e., 
$$\mu_{(t)}^{t_0,...,t_n}(A_0\times\cdots \times A_n)
=\mu_{(t)}^{t_0,...,t_n,t_{n+1},...,t_{n+l}}
(A_0\times \cdots \times A_n\times \prod^l\BR).$$
By the Kolmogorov extension  theorem 
there exists a unique probability space 
$({\cal Y}, \calb _\pq, \pq )$ 
and  a stochastic process $(Y_s)_{s\in[-t,t]}$ up to isomorphisms (e.g., \cite[Theorem 2.1]{sim05})
such that $\calb _\pq$ is the minimal $\s$-field, $\calb _\pq=\s(Y_s,s\in[-t,t])$, and 
$\mu^{t_0,...,t_n}_{(t)}(A_0\times\cdots \times A_n)=\pq (Y_{t_0}\in A_0,\cdots,Y_{t_n}\in A_n)$. 
By the uniqueness,  
$({\cal Y}, \calb _\pq, \pq )$ and $({\ms X}, {\cal F}_{[-t,t]}, \mu_{(t)})$ are isomorphic, and 
also 
is $({\cal Y}, \calb _\pq, \pq )$ and $({\ms X}, {\cal F}_{[-t,t]}, \mu_{(s)}\lceil_{{\cal F}_{[-t,t]}})$.
Hence $\pq (A)=\mu_{(s)}(A)=\mu_{(t)}(A)$ for $A\in {\cal F}_{[-t,t]}$ follows. 
\qed
Clearly $\mu$ is a completely  additive set function on $(\ms X, \ms G)$.
There exists a unique probability measure $\mu_\infty$ on $(\ms X, \s(\ms G))$ such that 
$\mu_\infty (A)=\mu(A)$ for 
$A\in \ms G$
 by the Hopf theorem. 
\bt{gibbs}\TTT{Local weak convergence and uniqueness}
The probability measures $\mu_t$ converges to $\mu_\infty$ in the local weak sense, 
i.e., $\mu_t(A)\to\mu_\infty(A) $ as $t\to\infty$ for each $A\in \ms G$, 
and $\mu_\infty$ is independent of $\phi$.
\et
Before giving a proof of Theorem \ref{gibbs} we need several lemmas. 
We 
define an additive set function 
$\rho_t:\ms G_t\to\RR$  by 
\eq{seki}
\rho_t(A)=\ee^{2Es}\IIXX \EW\lkkk 
\one_{A} 
\lk
\frac{\phi_{t-s}(X_0)}{\|\phi_t\|},\QS
\frac{\phi_{t-s}(X_s)}{\|\phi_t\|}\rk
\rkkk
\en
for $A\in{\cal F}_{[-s,s]}$ with $s\leq t$. 
\bl{welldefined3}
The set function $\rho_t$ satisfies $\rho_t(A)\geq 0$ and 
is well defined, i.e., 
\begin{align}
\rho_t(A)&=\ee^{2Er}\IIXX \EW\lkkk 
\one_{A} 
\lk
\frac{\phi_{t-r}(X_{-r})}{\|\phi_t\|},\QR  
\frac{\phi_{t-r}(X_{r})}{\|\phi_t\|}\rk
\rkkk\non \\
&\label{welldefined2}
=\ee^{2Es}\IIXX \EW\lkkk 
\one_{A} 
\lk
\frac{\phi_{t-s}(X_{-s})}{\|\phi_t\|},\QS 
\frac{\phi_{t-s}(X_s)}{\|\phi_t\|}\rk
\rkkk
\end{align}
for all $r\leq s\leq t$.
\el
\proof
$\rho_t(A)\geq0$ follows in a similar way to Lemma \ref{positive}. 
The proof of the second statement is similar to that of Lemma \ref{welldefined}. 
The left-hand side of \kak{welldefined2} is denoted by $\rho_{(r)}(A)$ and the right-hand side by $\rho_{(s)}(A)$. 
The finite dimensional distribution of 
$\rho_{(r)}$ is 
given by 
\begin{align*}
&
\rho_{(r)}^{t_0,...,t_n}(A_0\times\cdots \times A_n)=
\rho_{(r)}(X_{t_0}\in A_0,...,X_{t_n}\in A_n)\\
&=
\frac{\ee^{2Er}}{\|\phi_t\|^2}\int_\BR \dx\EW \lkkk \lk
\prod _{j=0}^n \one_{A_j}(X_{t_j})
\rk({\phi_{t-r}(X_{-r})},\QR {\phi_{t-r}(X_{r})}\rkkk.
\end{align*}
By Corollary \ref{green1} the right-hand side above can be represented  as
\begin{align*}
&=
\frac{1}{\|\phi_t\|^2}
{\lk
{\phi_{t-r}}, \ee^{-(t_0+r)(\PFF -E)} \one_{A_0}\ee^{-(t_1-t_0)(\PFF -E)}\cdots \ee^{
-(t_n-t_{n-1})(\PFF -E)}
\one_{A_n}\ee^{-(r-t_n)(\PFF -E)}{\phi_{t-r}}\rk}\\
&=
\frac{1}{\|\phi_t\|^2}
\lk
{\phi\otimes\one}, \ee^{-(t+t_0)(\PFF -E)} \one_{A_0}\ee^{-(t_1-t_0)(\PFF -E)}\cdots \ee^{-(t_n-t_{n-1})(\PFF -E)}
\one_{A_n}\ee^{-(t-t_n)(\PFF -E)}{\phi\otimes\one}\rk\\
&=
\frac{1}{{\|\phi_t\|^2}}
\lk
{\phi_{t-s}}, \ee^{-(t_0+s)(\PFF -E)} \one_{A_0}\ee^{-(t_1-t_0)(\PFF -E)}\cdots \ee^{-(t_n-t_{n-1})(\PFF -E)}
\one_{A_n}\ee^{-(s-t_n)(\PFF -E)}{\phi_{t-s}}\rk
\\
&=
\frac{\ee^{2Es}}{\|\phi_t\|^2}\int_\BR \dx\EW \lkkk \lk
\prod _{j=0}^n \one_{A_j}(X_{t_j})\rk
({\phi_{t-s}(X_{-s})},\QS {\phi_{t-s}(X_s)})\rkkk\\
&=
\rho_{(s)}^{t_0,...,t_n}(A_0\times\cdots \times A_n).
\end{align*}
Note that $\rho_{(r)}^{\Lambda}$ and $\rho_{(s)}^{\Lambda}$, $\Lambda\subset [-t,t]$, $\#\Lambda<\infty$,  satisfy  the consistency condition. 
Note that  $\rho_{(r)}\lceil_{{\cal F}_{[-r,r]}}$ and $\rho_{(s)}\lceil_{{\cal F}_{[-r,r]}}$ are  probability measures on 
$(\ms X, {\cal F}_{[-r,r]})$. 
By the Kolmogorov extension   theorem 
we see that $\rho_{(r)}(A)=\rho_{(s)}(A)$
 for $A\in {\cal F}_{[-r,r]}\subset {\cal F}_{[-s,s]}$.
Then the lemma follows.
\qed
By the Hopf theorem 
 there exists a probability measure $\bar \rho_t$
on $(\ms X, \s(\ms G_r))$ such that 
$\rho_t=\bar \rho_t\lceil_{\ms G_t}$.

\bl{equality}
Let $s\leq t$ and $A\in \ms G_s$. 
Then 
$
\bar \rho_t(A)=
\mu_t(A)
$.
\el
\proof
For 
 $\Lambda=\{t_0,t_1,\cdots,t_n\}\subset [-s,s]$ and 
$\d A_0\times\cdots\times A_n\in \times_{j=0}^n \ms B (\BR)$, 
we define 
\begin{align*}
\rho_t^\Lambda(A_0\times\cdots\times A_n)
&=
\rho_t(X_{t_0}\in A_0,..., X_{t_n}\in  A_n)\\
&=
\frac{\ee^{2Es}}{{\|\phi_t\|}^2}
\IIXX \EW\lkkk 
\lk\prod_{j=0}^n  \one_{A_j}(X_{\tt j})\rk 
\lk
{\phi_{t-s}(X_{-s})},\QS 
{\phi_{t-s}(X_{s})}\rk
\rkkk
\end{align*}
and 
\begin{align*}
\mu_t^\Lambda(A_0\times\cdots\times A_n)
=
\mu_t(X_{t_0}\in A_0,..., X_{t_n}\in  A_n)
=
\frac{1}{Z_t}
\int _\BR \!\!\! \dx\EW
\lkkk 
\lk\prod_{j=0}^n \one_{A_j}(X_{\tt j})\rk 
\ms L_t
\rkkk. 
\end{align*}
Both $\rho_t^\Lambda$ and 
$\mu_t^\Lambda$ 
are  probability measures on 
$((\BR)^\Lambda, \calb (\BR)^\Lambda)$.  
We have 
\begin{align*}
\mu_t^\Lambda(A_0\times\cdots\times A_n)
&=
\frac{(\phi\otimes\one, \ee^{-(t_0+t)\PFF }\one_{A_0}\ee^{-(t_1-t_0)\PFF }\one_{A_1}\cdots\one_{A_n}\ee^{-(t-t_n)\PFF }\phi\otimes\one)}{\|\phi_t\|^2}\\
&=\frac{\ee^{2Es}(\phi_{t-s}, \ee^{-(t_0+s)\PFF }\one_{A_0}\ee^{-(t_1-t_0)\PFF }\one_{A_1}\cdots\one_{A_n}\ee^{-(s-t_n)\PFF }\phi_{t-s})}{\|\phi_t\|^2}
\end{align*}
by the definition of $\phi_{t-s}$. 
The right-hand side above can be expressed as 
\begin{align*}
=\ee^{2Es}
\IIXX \EW\lkkk 
\lk\prod_{j=0}^n \one_{A_j}(X_{\tt j})\rk 
\lk
\frac{\phi_{t-s}(X_0)}{\|\phi_t\|},\QS 
\frac{\phi_{t-s}(X_{s})}{\|\phi_t\|}\rk
\rkkk.
\end{align*}
Then 
$
\rho_t^\Lambda(A_0\times\cdots\times A_n)=
\mu_t^\Lambda(A_0\times\cdots\times A_n)
$ follows.
The probability measures $\mu_t^\Lambda$ and $\rho_t^\Lambda$ 
 satisfy the  consistency condition. 
Then by the Kolmogorov extension theorem 
there exists a  unique probability space 
$({\cal Y}, \calb _\pq, \pq )$ and stochastic process $Y_s$ such that 
$\calb _\pq=\s(Y_s,s\in[-t,t])$ 
and $\pq (Y_{t_0}\in A_0,\cdots, Y_{t_n}\in A_n)=
\mu_t^{t_0,...,t_n}(A_0\times\cdots\times A_n)=
\rho_t^{t_0,...,t_n}(A_0\times\cdots\times A_n)$.
On the other hand 
it holds that 
$\mu_t^{t_0,...,t_n}(A_0\times\cdots\times A_n)=
\rho_t^{t_0,...,t_n}(A_0\times\cdots\times A_n)=
\bar \rho_t(A_0\times\cdots\times A_n)=
\mu_t\lceil_{{\ms G}_t}(A_0\times\cdots\times A_n)$. 
Hence 
$\bar \rho_t=\pq =\mu_t\lceil_{{\ms G}_t}$ follows by the uniqueness of extensions.
\qed
\bl{doko}
Let $A\in \ms G$. Then 
$\d \lim_{t\to\infty}\mu_t(A)=\mu_\infty(A)$.
\el
\proof
Suppose that $A\in \ms G_s$ with some $s$. 
By Lemma \ref{equality} we have 
\begin{align*}
\lim_{t\to\infty} \mu_t(A)
=\lim_{t\to\infty} \bar \rho_t(A)
=\lim_{t\to\infty}
\ee^{2Es}\IIXX \EW\lkkk 
\one_{A} 
\lk
\frac{\phi_{t-s}(X_{-s})}{\|\phi_t\|},\QS
\frac{\phi_{t-s}(X_{s})}{\|\phi_t\|}\rk
\rkkk.
\end{align*}
Since $\phi_t\to\gr$ strongly as $t\to\infty$, we have 
$$
\lim_{t\to\infty} \mu_t(A)
=\ee^{2Es}\IIXX \EW\lkkk 
\one_{A} 
\lk
\gr(X_{-s})
,\QS  
\gr(X_{s})\rk
\rkkk=\mu_\infty(A).
$$
Then the lemma follows. 
\qed

Now we state the proof of Theorem \ref{gibbs}. 

{\it Proof of Theorem \ref{gibbs}}:
By Lemma \ref{doko} it follows that $\mu_t(A)\to\mu_\infty(A)$ for $A\in \ms G$. 
Next we show that 
$\mu_\infty$ is  independence of 
the choice of $\phi$.
Suppose that $\mu_\infty'$ is a local weak limit of $\mu_t'$ defined by $\mu_t$ with $\phi$ replace by $\phi'$ such that $0\leq \phi'\in\LR$. 
By the construction of $\mu_\infty$, 
$\mu_\infty(A)=\mu_\infty'(A)$ for $A\in \ms G$. 
The uniqueness of Hopf's extension implies $\mu_\infty=\mu_\infty'$. 
Thus $\mu_\infty$ is independent of the choice  of $\phi$.
Then the theorem follows.  
\qed

\section{Concluding remarks}
\label{sec8}
\subsection{Translation invariant models}
Let $\PFF =\PFK$ or $\PF$. 
Suppose that $V=0$. Then we already see that 
$\ee^{-it\tot} \ee^{-t\PFF} \ee^{it \tot}=\ee^{-t\PFF}$.
Then 
  $\PFF$ can be decomposable with respect to the spectrum of $\tot$. Thus we have 
\begin{align}
\PFF=\int _\BR^\oplus \PFF(p) dp.
\end{align}
Here  $\PFF (p)$ is defined by 
\eq{hp}
\PFF (p)=\sqrt{L(p)+m^2}-m\ \dot +\ \hf
\en
and 
\eq{hpp}
L(p)=\ov{(p-\pf -\alpha\A(0))^2\lceil_{D(\pf^2)\cap D(\hf)}}.
\en
It is established that 
$(p-\pf -\alpha\A(0))^2$ is essentially self-adjoint on
$D(\pf^2)\cap D(\hf)$ in \cite[Theorem 2.3]{hir07}. 
We can construct the functional integral  representation of $\ee^{-t\PFF(p)}$ for each $p\in\BR$ in a similar manner to \cite{hir07}. 
\bt{pathintegralrepresentationwithp}
Let $F,G\in L^2(\Q)$. 
Then it follows that 
\begin{align}
\label{psemi}
(F, \ee^{-t\PFF(p)} G)=
\EZ
\lkkk
\ee^{-ip\cdot  B_{T_t}} 
\lk \JJ_0F(B_{T_0}), \ee^{i\pf\cdot B_{T_t}} \ee^{-i\alpha\AA(\KKKK [0,t])}\JJ_t G(B_{T_t})
\rk
\rkkk.
\end{align}
\et
From this functional integral representation 
we  can show the self-adjointness of  $\PFF(p)$ in a similar manner to $\PFF $. 
\bc{essentialselfadjointnesswithp}
Suppose Assumptions \ref{ass1} and \ref{ass2}. Then 
for all $p\in\BR$, $\PFF(p)$ is self-adjoint on $ \D(|\pf|)\cap \D(\hf)$.
\ec
\proof
The proof is similar to  that of Theorems \ref{essentialselfadjointnesstheorem} and \ref{sasa}, 
i.e, it can be show that 
$ \ee^{-t\PFF (p)}$ leaves $ \D(|\pf|)\cap \D(\hf)$ invariant fo r$m>0$, and that by using the inequality 
$\||p-\pf|\Phi\|^2+\|\hf\Phi\|\leq C\|(\PFF (p)+\one)\Phi\|$ we can show the self-adjointness of $\PFF (p)$ for $m\geq0$. 
See \cite{hh13}. 
\qed

\subsection{Spin 1/2 and generalizations}
Let us assume that the space dimension $d=3$. 
The SRPF  Hamiltonian with spin $\han$ is defined by 
\eq{a1PF}
\PFF _{\rm SR}=\sqrt{(\s\cdot(\p -\alpha \A))^2+m^2}-m+V+\hf
\en
on the Hilbert space $\lk \CC^2\otimes L^2(\RR^3)\rk \otimes L^2(\Q)$.
Here $\s=(\s_1,\s_2,\s_3)$ are 
 the $2\times 2$ Pauli
matrices given by
\eq{paulimatrices}
\s_1=\mmm 0 1 1 0,\ \ \ \s_2=\mmm 0 {-i} i 0 ,\ \ \ \s_3=\mmm 1 0 0 {-1}.
\en
Let $\pro N$ be the Poisson process with the unit intensity 
on a probability space $(\Omega_\nu, \ms B_\nu, \nu)$. We define  
the stochastic process $\s_t=\s(-1)^{N_t}$, $t\geq 0$, where $\s\in \{-1, +1\}$.   
Under some condition we can construct a functional integral representation of $\ee^{-t\PFF}$ in terms of 
stochastic processes $\pro B$, $\pro T$ and $\pro \s$.
We can identify 
$\lk \CC^2\otimes L^2(\RR^3)\rk \otimes L^2(\Q)$ with $
L^2(\RR^3\times\{\pm 1\}; L^2(\Q))$. 
Under this identification we can construct the Feynman-Kac type formula of 
$e^{-t\PFF }$ with 
$$\PFF _{\rm NR}=\half(\s\cdot(\p-\alpha \A))^2+V+\hf$$
in \cite {hl08}. 
By a minor modification we can also construct the Feynman-Kac type formula for $\PFF $ in \kak{a1PF}. 
 \bt{soinFKF}
Let 
$F, G\in 
L^2(\RR^3\times\{\pm 1\}; L^2(\Q))$. 
Then 
\eq{tt}
(F, \ee^{-t \PFF _{\rm SR}}G)=
\ee^{T_t}\sum_{\s=\pm 1}\int_{\RR^3}
\dx\Ebb_{{\rm P}\times \mu\times \nu}^{x,0,\s}\lkkk \ee^{-\int_0^tV(B_{T_s})\ds}
\lk 
\JJ_0F(B_{T_0},\s_{T_0}), \ee^S \JJ_t G(B_{T_t},\s_{T_t})\rk \rkkk, 
\en
where 
\begin{align*}
S=&-i\alpha\AA(\KKKK[0,t])-\frac{\alpha}{2}\int_0^{T_t} \!\!\!
\s_s {\rm B}_3(\la(\cdot-B_s))\ds\\
&+
\int_0^{T_t+}\!\!\!\!\!\!\log\lk \frac{\alpha}{2}({\rm B}_1(\la(\cdot-B_s))-i\s_s{\rm B}_2(\la(\cdot-B_s))\rk dN_s
\end{align*}
and 
${\rm B}(x)=\nabla _x\times \AA (x)$ describes the quantized magnetic field.  
\et
We can furthermore  consider  general Hamiltonians of the form:
\eq{a3PF}
\Psi\lk 
\half (\s\cdot(\p -\alpha \A))^2
\rk+V+\hf,
\en
where $\Psi$ denotes a Bernstein function. 
The standard Pauli-Fierz Hamiltonian is realized by $\Psi(u)=u$, and 
the SRPF Hamiltonian with spin $\han$ by $\Psi(u)=\sqrt{2u+m^2}-m$.
\kak{a3PF}  can be also investigated  by path measures, 
and 
only the difference from \kak{tt} is to take the subordinator $\pro{{T^\Psi}}$ associated with Bernstein function $\Psi$ instead of $\pro T$.
See Appendix \ref{D} for relationship between Bernstein functions and subordinatos. 
We will publish details somewhere in near future.  

\begin{remark}
{\rm 
We give comments on both of semigroups \kak{psemi} and \kak{tt}. \\ 
(1) The semigroup \kak{psemi} is not positivity improving for $p\not=0$ and positivity improving for $p=0$, since the semigroup includes $e^{-ip\cdot B_{T_t}}$. \\
(2)  Let 
 $V$ and $\vp$ be  rotation invariant. 
Then in a similar manner to \cite[Corollary 7.70]{lhb11} 
it can be shown that \kak{a1PF} has degenerate ground state if it exists. 
In particular in this case \kak{tt} can not be  positivity improving.  }
\end{remark}

\subsection{Gaussian domination and local weak convergence}
We can see that 
$\qq(\KKKK[-t,t],\jj_0\xi)$ in \kak{expaa} converges as $t\to\infty$. 
\bl
{cauchy}
Sequence $\{\qq(\KKKK[-t,t],\jj_0\xi)\}_t$ is a Cauchy sequence in 
$L^2(\ms X, \msw^0)$.
\el
\proof
Let $s<t$ and we estimate 
$\Ebb_\msw^0
[\qq(\KKKK  [s,t],\jj_0\xi)^2]$.
By the definition of $\KKKK  [s,t]$ we have 
\begin{align*}
\Ebb_\msw^0[\qq(\KKKK  [s,t],\jj _0\xi)^2]
\leq \limn
\Ebb_\msw^0\lkkk \left|
\sum_{j=1}^{2^n}
\int_{T_{t_{j-1}}}^{T_{t_j}} (\jj_{t_{j-1}} \la(\cdot-B_s),\jj_0 \xi) 
 \dB_s\right|^2\rkkk.
 \end{align*}
By the independent increments of the Brownian motion we have 
 \begin{align*}
\leq 
\limn\sum_{j=1}^{2^n}
\Ebb_\msw^0\lkkk 
\int_{T_{t_{j-1}}}^{T_{t_j}}
\!\!\!
(\xi, \ee^{-2t_{j-1}\omega}  \xi)   \ds
\rkkk
\!\!
\|\la\|^2
\!\!\!
=\!\!\!
\limn\sum_{j=1}^{2^n}
\Ebb_\msw^0\lkkk 
(T_{t_j}-T_{t_{j-1}})
(\xi, \ee^{-2t_{j-1}\omega}  \xi)   \rkkk
\!\!\|\la\|^2.
 \end{align*}
Since 
 $T_{t_j-t_{j-1}}$ and $T_{t_j}-T_{t_{j-1}}$ have the same low, we see that 
 \begin{align*}
=
\lk
\xi, 
\limn\sum_{j=1}^{2^n}
\Ebb_\msw^0
\lkkk
T_{t_j-t_{j-1}}
 \ee^{-2t_{j-1}\omega}
 \rkkk
 \xi\rk
 \|\la\|^2.
\end{align*}
Using the distribution of $T_t$ we have 
 \begin{align*}
=
\lk 
\xi, 
\limn\sum_{j=1}^{2^n}
\lk 
\int_0^\infty \ds 
\frac{\Delta t_j}{\sqrt{2\pi}} 
\frac{1}{\sqrt s}\exp
\lk
-\half (\frac{(\Delta t_j)^2}{s}+m^2 s)
\rk
 \ee^{-2t_{j-1}\omega}\rk
 \xi\rk
\|\la\|^2,
\end{align*}
where $\Delta t_j=t_{j}-t_{j-1}$. Since $m>0$ we obtain that 
 \begin{align*}
&\leq
C
\lk
\xi, 
\xi 
\limn\sum_{j=1}^{2^n}
\Delta t_j
 \ee^{-2t_{j-1}\omega}
\rk=
 C
\lk\xi, 
\xi \int _s^t \ee^{-2r\omega } \dr
\rk=
C
\lk\xi, 
\frac{\ee^{-2s\omega}-\ee^{-2t\omega}}{2\omega }\xi
\rk
\end{align*}
with some constant $C$.
Then 
$\qq(\KKKK  [-t,t],\jj_0\xi)$ is a Cauchy sequence. 
\qed
By Lemma \ref{cauchy}
 there exists 
$\qq(\KKKK(-\infty, \infty),\jj_0\xi)$ such that 
$\d \limt  \qq(\KKKK[-t,t],\jj_0\xi)
=\qq(\KKKK(-\infty,\infty),\jj_0\xi)$ in $L^2(\ms X, \msw^0)$.

\begin{remark}
{\rm By Theorem \ref{gibbs} and Lemma \ref{cauchy} we  
conjecture that 
\eq{conj1}
(\gr, \ee^{\beta\A_\xi^2}\gr)=
\frac{1}{\sqrt{1-2\beta \qq(\jj_0\xi)}}
\Ebb_{\mu_\infty}\lkkk 
\ee^{
\frac
{-\beta \alpha^2\qq(\KKKK[-\infty,\infty],\jj_0\xi)^2}
{(1-2\beta\qq(\jj_0\xi))}
}\rkkk
\en
and 
$
\lim_{\beta\uparrow \qq(\jj_0\xi)/2}
\|\ee^{\beta\A_\xi^2/2}\gr\|=\infty$.
This type of results are derived for a spin-boson model \cite{hhl12}. 
}\end{remark}

\appendix
\section{Brownian motion on $\RR$}
Let $(B_t)_{t\in\RR}$ be $d$-dimensional Brownian motion on a probability space $(\Omega_\P, \ms B _\P, \P^x)$. 
The properties of Brownian motion on the whole real line can be summarized as follows.
Let $N_t$ be the Gaussian random variable with mean zero and covariance $t$. 
\begin{itemize}
\item[(1)]
$\P^x(B_0=x)=1$;
\item[(2)]
the increments $(B_{t_i}-B_{t_{i-1}})_{1\leq i\leq n}$ are independent Gaussian random
variables for any $0=t_0<t_1<\cdots<t_n$ with $B_t-B_s\stackrel{\rm d}{=}
N_{t-s}$, for
$t>s$;
\item[(3)]
the increments $(B_{-t_{i-1}}-B_{-t_i})_{1\leq i\leq n}$ are independent Gaussian random
variables for any $0=-t_0>-t_1>\cdots>-t_n$ with $B_{-t}-B_{-s}
\stackrel{\rm d}{=}N_{s-t}$,
for $-t>-s$;
\item[(4)]
the function $\RR\ni t\mapsto B_t(\omega)\in \RR$ is continuous for almost every $\omega$;
\item[(5)]
$B_t$ and $B_s$ for $t>0$ and $s<0$ are independent;
\item[(6)]
the joint
distribution of $B_{t_0},\ldots, B_{t_n}$, $-\infty<t_0<t_1<\cdots<t_n<\infty$, with
respect to $\dx \otimes \dP^x$ is invariant under time shift, i.e.,
\eq{shiftinvariant}
\int_\BR\dx\Ebb_\P^x
 \left[\prod_{i=0}^n f_i(B_{t_i})\right] =
\int_\BR\dx\Ebb_\P^x
\left[\prod_{i=0}^n f_i(B_{t_i+s})\right]
\en
for all $s \in\RR$. 
\end{itemize}

\section{Proof of Proposition \ref{fkf1}}
\label{fkf1p}
{\it Proof of Proposition \ref{fkf1}}: 
We show an outline of a proof. 
This is a modification of 
\cite[Theorem 2.7]{hir00b}  and \cite[Lemma 7.53]{lhb11}.
By the Riesz theorem the right-hand side of \kak{pathintegral1} can be expressed as 
$(F, S_t G)$ with some bounded operator $S_t$. 
We can check that $S_t$, $t\geq0$,  is symmetric and strongly continuous one-parameter  semigroup. Thus 
there exists a self-adjoint operator $K$ such that $S_t=\ee^{-tK}$. 
It is also shown \cite[the proof of Lemma 4.8]{hir97} that 
\eq{rie}
\frac{1}{t}((e^{-tK}-\one)F, G)=\int_0^1
(-\rh_\A F, e^{-ts K} G)\ds
\en
for $F, G\in C_0^\infty(\BR)\otimes L^2_{\rm fin}(\Q)$.  
By the inequality 
$\|\rh_\A  F\|\leq C(\|p^2 F\|+\|(\N+\one) F\|)$ with some positive constant $C$, 
\kak{rie} can be extended for 
$F, G\in D(\p^2)\cap 
D(\N)$. 
Thus  $K=\rh_{\A}$ on 
$\D(\p^2)\cap C^\infty({\rm N})$.
We also see that $|(U F, \ee^{-tK}G)|\leq 
C(U,K,G) \|F\| $ for $F,G\in D(U)$, where 
$C(U,K,G)$ is a positive constant, 
$U=\p^2$ and $U=\N^n$ for any $n\geq1$.
Thus 
$\ee^{-tK}$ leaves 
$\D(\p^2)\cap C^\infty({\rm N})$ invariant. 
Thus the proposition follows from 
Proposition \ref{invariant}. 
 
\section{Relativistic Kato-class}
\label{appK}
Let $m\geq0$. Set $h=\sqrt{\p^2+m^2}-m$. 
It is known that 
$V\geq 0$ is in the relativistic Kato-class if and only if 
$\d \lim_{E\to\infty} 
\sup_{x\in\BR}
|((h-E)^{-1}V)(x)|=0$.
See e.g.\cite[Proposition 4.5]{hil13}.
\bl{katoclass2}
Let $V>0$ be in the relativistic Kato-class. Then $V$ is infinitesimally small form bounded with respect to 
$h$,
i.e., for arbitrary $\e$ there exists $b_\e\geq 0$ such that
$\|V^\han f\|\leq \epsilon\|h^\han f\|+
b_\epsilon \|f\|$ for arbitrary $f\in D(h^\han)$.
In particular $\vk \subset \vqf $.
\el
\proof
Let  $\|\cdot\|_{p,p}$ 
be bounded operator
norm on $L^p(\BR)$.
By duality it is seen that 
$
\|(h+E)^{-1}V\|_{1,1}=
\|(h+E)^{-1}V\|_{\infty,\infty}$.  
By the Stein interpolation theorem 
we have 
$\|V^\han (h-E)^{-1}V^\han\|_{2,2}\leq
\|(h+E)^{-1}V\|_{1,1}
$
and 
notice 
that 
$\|(h+E)^{-1}V\|_{\infty,\infty}
=
\sup_{x\in\BR}
|((h-E)^{-1}V)(x)|
$.
Hence 
$\|V^\han (h-E)^{-1/2}\|^2_{2,2}
\leq
\sup_{x\in\BR}
|((h-E)^{-1}V)(x)|
\to 0$
as $E\to \infty$. 
From 
$\|V^\han f\|\leq 
\|V^\han (h-E)^{-1/2}\|_{2,2}
\|(h-E)^\han f\|$  
it follows that 
$V$ is form bounded with an infinitesimally small relative bound.
\qed

\section{Integral $\KKKK[a,b]$}
\bp{j}
Let 
$\d \KKKK'_n[0,t]=\ott 
\sum_{j=1}^{2^n}
\int_{T_{\tt{j-1}}}^{T_{\tt j}}\jj_{\tt j}\la(\cdot-B_s)\dB_s^\mu$. 
Then 
$\d \slimn \KKKK_n'[0,t]=\KKKK [0,t]$ in $L^2(\Omega_{\rm P},\P^x)\otimes\ms E$.
\ep
\proof
We have 
$\|\KKKK_n'[0,t]-\KKKK_n[0,t]\|^2=
d
(T_t-T_0)(\la, 2(1-\ee^{-t\omega/2^n}) \la) 
\to0$ as $n\to \infty$. 
Then the proof is complete. 
\qed


\bp{linearity}
For each $w\in \Omega_\nu\setminus\ms N_\nu$, 
 $\KKKK [0,t]=\KKKK [0,s]+\KKKK [s,t]$ for $0<s<t$ follows 
 in the 
 sense of $L^2(\Omega_\P,\P^x)\otimes \WW$, i.e., 
\begin{align}
\label{linearity1}
\Ebb_\P^x\lkkk\left \|
 \KKKK [0,t]-\KKKK [0,s]-\KKKK [s,t]\right\|^2_\WW\rkkk=0.
 \end{align}
 \ep
\proof
By a limiting argument we see that 
\eq{itoisometry2}
\Ebb_\P^x\lkkk \|\KKKK  [0,t]\|^2_\WW\rkkk=
d T_t \|\vp/\sqrt\omega\|^2
\en
for almost surely in $\nu$. 
We suppose that $s=at/2^k$ with some $a,k\in{\mathbb  N}$. 
Then by the definition of $\KKKK_n[0,t]$ we have 
$\d \KKKK  [0,t]=\limn 
\ott
 \sum_{j=1}^{2^{n+k}}
 \int_\TT{j-1}^\TT{j}
 \jj_{\tt{j-1}}
 \la(\cdot-B_s) \dB_s^\mu 
$ with $\tt j=\frac{tj}{2^{n+k}}$,
and 
\begin{align*}
& \sum_{j=1}^{2^{n+k}}
 \int_\TT{j-1}^\TT{j}
 \jj_{\tt{j-1}}
 \la(\cdot-B_s) \dB_s^\mu \\
& =\sum_{j=1}^{2^n a}
 \int_{T_{{\frac{s}{2^n a}(j-1)}}}
 ^{T_{{\frac{s}{2^n a}j }}}
   \jj
   _{{\frac{s}{2^n a}(j-1)}}
 \la(\cdot-B_r) \dB_r^\mu 
+
\sum_{j=1}^{2^n b}
 \int_{T_{s+{\frac{t-s}{2^n b}(j-1)}}}
 ^{T_{s+{\frac{t-s}{2^n b}j }}}
   \jj
   _{s+{\frac{t-s}{2^n b}(j-1)}}
 \la(\cdot-B_r) \dB_r^\mu,
\end{align*}
where $b=2^k-a$. 
Hence 
$\KKKK [0,t]=\KKKK [0,s]+\KKKK [s,t]$ follows. 
Let $0<s<t$. Then there exists $s(\epsilon)>s$ such that $s(\epsilon)=a/2^k$ 
with some $a,k\in{\mathbb  N}$ and $s(\epsilon)\downarrow s$ as $\epsilon\to 0$. 
Hence 
$\KKKK [0,t]=\KKKK [0,s(\epsilon)]+\KKKK [s(\epsilon),t]$. 
Note that $\KKKK [0,s(\epsilon)]-\KKKK [0,s]=\KKKK [s,s(\epsilon)]$
and $\Ebb_{\P}^{x}\lkkk\|\KKKK [s,s(\epsilon)]\|^2\rkkk=
(T_{s(\epsilon)}-T_s)\|\vp/\sqrt\omega\|^2$ by the It\^o isometry \kak{itoisometry2}.  
Since 
$T_s=T_s(w)$ is right continuous in $s$ for $w\in \Omega_\nu\setminus\ms N_\nu$, 
\kak{linearity1}  follows. 
\qed

\bp{itoformula}
Let $a\leq b$ and  $c\leq d$, and 
suppose that 
$[a,b]\cap [c,d]=[c,b]$.
Then for each $w\in \Omega_\nu\setminus\ms N_\nu$, 
$\Ebb_{\P}^{x}\lkkk (\KKKK [a,b],\KKKK [c,d])_\WW\rkkk
=
d (T_b-T_c)\|\vp/\sqrt\omega\|^2_\LR$.
\ep
\proof 
Suppose that  $[a,b]\cap [c,d]=\emptyset$. 
 Then 
it follows that 
$$\Ebb_{\P}^{x}\lkkk (\KKKK [a,b],\KKKK [c,d]) \rkkk
=\limn \limm \Ebb_{\P}^{x}\lkkk (\KKKK_n[a,b],\KKKK_m[c,d]) \rkkk=0.$$
Thus 
by Proposition  \ref{linearity} we see that 
\begin{align*}
\Ebb_{\P}^{x}\lkkk (\KKKK [a,b],\KKKK [c,d]) \rkkk
=
&\Ebb_{\P}^{x}\lkkk (\KKKK [a,c],\KKKK [b,d]) \rkkk
+
\Ebb_{\P}^{x}\lkkk (\KKKK [a,c],\KKKK [c,b]) \rkkk\\
&+
\Ebb_{\P}^{x}\lkkk (\KKKK [b,d],\KKKK [c,b]) \rkkk
+\Ebb_{\P}^{x}\lkkk \|\KKKK [c,b]\|^2 \rkkk.
\end{align*}
Then the lemma follows from 
$\Ebb_{\P}^{x}\lkkk \|\KKKK [c,b]\|^2 \rkkk
=d \Ebb_\P^x\lkkk \int_{T_c}^{T_b}\|\la(\cdot-B_r)\|^2 \dr\rkkk$ by the It\^o 
isometry \kak{itoisometry2}.
\qed

\section{Proofs of \kak{mar1} and \kak{toshiba2}}
\bl{mar}
\kak{mar1} follows.
\el
\proof
From the proof of Theorem \ref{martingale} and  \kak{martingale3},  
it follows that 
\begin{align}
&
\Ebb_{\P\times \nu}^{0,0}
\lkkk
\left. 
\ee^{-i\alpha \AA(\KKKK^x[s,t]) } 
\ee^{-\int_s^{t} V(B_{T_r}+x)\dr }
\JJ_t G(B_{T_t}+x)
\right|
{\cal F}_{[0,s]}
\rkkk \non \\
&\label{arb}
=
\Ebb_{\P\times\nu}^{B_{T_s},0}
\lkkk
\ee^{-i\alpha \AA(\KKKK^{(2),x}[0,t-s]) } 
\ee^{-\int_0^{t-s} V(B_{T_r}+x)\dr }
\JJ_t G(B_{T_{t-s}}+x)
\rkkk 
\end{align}
for arbitrary $G\in\hhh$.
Then 
we have 
\begin{align*}
&
\EX\lkkk 
\QSS \Ebb_{\P\times\nu}^{B_{T_s},0}
\lkkk \QTT  G(B_{T_t})\rkkk\rkkk 
=\EZ \lkkk 
\QSS (x) \Ebb_{\P\times\nu}^{B_{T_s},0}
\lkkk \QTT (x)  G(B_{T_t}+x)\rkkk\rkkk \\
&=
\EZ \lkkk 
\QSS (x) 
\EZ 
\lkkk \left. \JJ_0^\ast\ee^{-\int_s^{s+t} V(B_{T_r}+x)\dr}
 \ee^{-i\alpha \AA(\ima ^x [s,s+t])} \JJ_t  
G(B_{T_{s+t}}+x) \right| {\cal F}_{[0,s]}
\rkkk\rkkk.
\end{align*}
Here $\ima^x[s,s+t]$(resp.  $\QSS (x) 
$) denotes 
$\ima[s,s+t]$ (resp. $\QSS $) 
 with $B_r$ replaced by $B_r+x$. 
Since a conditional expectation leaves expectation invariant, we have 
\begin{align*}
&
=
\EZ \lkkk 
\QSS (x) 
\JJ_0^\ast \ee^{-\int_s^{s+t} V(B_{T_r}+x)\dr}
 \ee^{-i\alpha \AA(\ima ^x[s,s+t])} \JJ_t  G(B_{T_{s+t}}+x) 
\rkkk\\
&=
\EX\lkkk 
\QSS 
\JJ_0^\ast \ee^{-\int_s^{s+t} V(B_{T_r})\dr}
 \ee^{-i\alpha \AA(\ima [s,s+t])} \JJ_t  G(B_{T_{s+t}}) 
\rkkk
\end{align*}
and \kak{mar1} follows. 
\qed

\bl{lemma0}
\kak{toshiba2}  follows.
\el
\proof
Note that 
$B_{T_r}-B_{T_t}=B_{T_t-T_r}$ and 
$y-B_{T_t}=y+B_{T_t}$ in law.
We investigate $u_t r\tilde \KKKK_n[0,t]$. 
We see that
\begin{align*}
u_t r\tilde \KKKK_n[0,t]
&=\ott \sum_{j=1}^{2^n} \int_{\TT{j-1}}^{\TT{j}} \jj_{t-\tt{j-1}} 
\la(\cdot-(B_s-B_{T_t}+y))d B_s^\mu\\
&
=\limm \ott\sum_{i=1}^{2^m}
\sum_{j=1}^{2^n}
\jj_{t-\tt{j-1}}\la\lk 
\cdot-(B_{T_{\tt{j-1}}+(i-1)\Delta_{j-1}}-B_{T_t}+y)\rk\\
&\hspace{2cm}
\times
\lk B_{T_{\tt{j-1}}+i\Delta_{j-1}}-B_{T_{\tt{j-1}}+(i-1)\Delta_{j-1}}
\rk\\
&=
\limm \ott\sum_{i=1}^{2^m}
\sum_{j=1}^{2^n}
\jj_{t-\tt{j-1}}\la\lk 
\cdot-B_{T_t-T_{\tt{j-1}}-(i-1)\Delta_{j-1}}-y\rk\\
&\hspace{2cm}
\times
\lk B_{T_t-T_{\tt{j-1}}-i\Delta_{j-1}}-B_{T_t-T_{\tt{j-1}}-(i-1)\Delta_{j-1}}
\rk,
\end{align*}
where 
$\Delta_{j-1}=\frac{1}{2^n}\lk T_{\tt j}-T_{\tt{j-1}}\rk$.
Since $T_t-T_s$ has the same law as $T_{t-s}$, we can replace 
the right-hand side above with 
\begin{align}
&-\limm \ott\sum_{i=1}^{2^m}
\sum_{j=1}^{2^n}
 \jj_{t-\tt{j-1}}\la\lk \cdot-B_{T_{t-\tt{j-1}}-\frac{i-1}{2^n}(T_{t-\tt{j-1}}-T_{t-\tt j})}-y\rk\non \\
&\hspace{2cm}\times 
\lk B_{T_{t-\tt{j-1}}-\frac{i-1}{2^n}(T_{t-\tt{j-1}}-T_{t-\tt j})}-
B_{T_{t-\tt{j-1}-\frac{i}{2^n}(T_{t-\tt{j-1}}-T_{t-\tt j})}}\rk.
\end{align}
By the definition of $\int_S^T \jj_s\la(\cdot-B_s)\dB_s^\mu$ 
and the Coulomb gauge condition \kak{coulomb}
it follows that 
\begin{align}
=-\ott
 \sum_{j=1}^{2^n}\int_{T_{t-\tt j}}^{T_{t-\tt{j-1}}}\jj_{t-\tt{j-1}}\la(\cdot-B_s-y)\dB_s^\mu
 \label{cou}
=-\ott
 \sum_{j=1}^{2^n}\int_{T_{\tt{j-1}}}^{T_{\tt j}}
\jj_{\tt j} \la(\cdot-B_s-y)\dB_s^\mu.
\end{align}
Finally we have by Proposition  \ref{j}
\begin{align}
\label{lemma00}
-\ott
 \sum_{j=1}^{2^n}\int_{T_{\tt{j-1}}}^{T_{\tt j}}
\jj_{\tt j} \la(\cdot-B_s-y)\dB_s^\mu.
=-\ott  \sum_{j=1}^{2^n}\int_{T_{\tt{j-1}}}^{T_{\tt j}}
\jj_{\tt {j-1}} \la(\cdot-B_s-y)\dB_s^\mu.
\end{align}
Then the proof is complete. 
\qed

\section{Subordinators}
\label{D}
A subordiantor $\pro T$ is a $1$-dimensional L\'evy process 
which has  a almost surely nondecreasing path $t\mapsto T_t$.  
Subordinator may be thought as a random time, since $T_t\geq 0$ and $T_t\leq T_s$ for $t\leq s$. The subordinator $\pro T$ satisfies that 
$\Ebb[\ee^{-uT_t}]=\ee^{-t\psi(u)}$, where 
\eq{bernstein}
\psi(u)=bu+\int_0^\infty (1-\ee^{-uy}) \lambda(\dy)
\en
for $u>0$, where $b\geq0$ a constant and $\lambda(\dy)$ denotes a L\'evy measure such that $\lambda((-\infty,0))=0$ and $\int_0^\infty (y\wedge 1)\lambda(\dy)<\infty$. 
Let $f\in C^\infty((0,\infty))$ with $f\geq0$. 
$f$ is a Bernstein function if and only if $(-1)^n d^n f/dx^n\leq 0$ for all $n=1,2,3,...$.
For each  Bernstein function $\psi$ 
such that  $\lim_{u\downarrow 0}\psi(u)=0$ can be realized as 
\kak{bernstein}. The examples of Bernstein functions are $\psi(u)=u^\alpha$ with $0<\alpha<1$ and $\psi(u)=\sqrt{u^2+m^2}-m$.

\bigskip
\noindent {\bf Acknowledgments:} 
The author  acknowledges support of Grant-in-Aid for Science Research (B) 20340032,
23340032,  
and Challenging Exploratory Research 22654018 from JSPS. He  thanks
the invitation of the conference 
{\it The 24th Max Born
Symposium} at Wroclaw university at 2008, where his research started, 
and 
also thanks the hospitality of universit\'e de Paris XI, universit\'e d'Aix-Marseille-Luminy, universit\'e de
Rennes I, and Bologna university,  where part of this work has been  done.

\end{document}